*(revised version)*

# Flexo-Strain Engineering of Phonons and Ferrons in Thin Films of Van der Waals Ferrielectrics

Anna N. Morozovska[1], Eugene A. Eliseev[2], Yujie Zhu[3], Mykola Yelisieiev[4,1], Hanna V. Shevliakova[5], Yulian M. Vysochanskii[6*], Venkatraman Gopalan[7†], Long-Qing Chen[7‡], and Jia-Mian Hu[3§]

[1] Institute of Physics, National Academy of Sciences of Ukraine, 46, Nauky Avenue, 03028 Kyiv, Ukraine

[2] Frantsevich Institute for Problems in Materials Science, National Academy of Sciences of Ukraine, Omeliana Pritsaka str., 3, Kyiv, 03142, Ukraine

[3] Department of Materials Science and Engineering, University of Wisconsin-Madison, Madison, WI, 53706, USA

[4] V. Lashkaryov Institute of Semiconductor Physics, National Academy of Sciences of Ukraine, 41, pr. Nauky, 03028 Kyiv, Ukraine

[5] Department of Microelectronics, Igor Sikorsky Kyiv Polytechnic Institute, Kyiv, Ukraine

[6] Institute of Solid-State Physics and Chemistry, Uzhhorod University, 88000 Uzhhorod, Ukraine

[7] Department of Materials Science and Engineering, Pennsylvania State University, University Park, PA 16802, USA

## Abstract

The influence of the flexoelectric coupling on the fluctuations of electric polarization and elastic strains can lead to the principal changes of the dispersion law of soft optical and acoustic phonons and ferrons in a bulk van der Waals ferrielectric. Since the size, gradient and strain effects determine phase diagrams and polarization behavior in thin films, it is reasonable to assume that the flexocoupling and mismatch strains should have a strong influence on the dispersion of phonons and ferrons in thin ferroelectric films. Using the Landau-Ginzburg-Devonshire approach, in this work we reveal that the dispersion of soft optical and acoustic phonons and ferrons is strongly dependent on the sign and magnitude of elastic strains, which originate from the lattice constants mismatch in thin strained films of van der Waals ferrielectric $CuInP_2S_6$. In particular, the frequency of acoustic phonons and ferrons approaches zero at nonzero wavevectors $k > k_{cr}$, where the critical value of the wavevector $k_{cr}$ is determined by the mismatch strain, flexoelectric coupling strength and temperature. Zeroing of the acoustic phonon frequency, that appears with increase of tensile strains, indicates a dynamic instability induced by the flexo-strain effects, that may evolve into a spatially modulated incommensurate polar phase with increase in the flexocoupling strength. Analytical results, derived in this work, open the way for flexo-strain engineering of soft phonon and ferron dispersion in thin films of van der Waals ferrielectrics.

* corresponding author, e-mail: vysochanskii@gmail.com

† corresponding author, e-mail: vxg8@psu.edu

‡ corresponding author, e-mail: lqc3@psu.edu

§ corresponding author, e-mail: jhu238@wisc.edu

## 1. Introduction

Despite great fundamental scientific interest in low-dimensional van der Waals ferroelectrics and advanced application potential, their electrophysical and electronic properties, as well as electron-phonon interactions, are largely unexplored still [1]. Among the wide class of van der Waals ferroelectrics, of particular interest are $CuInP_2(S_xSe_{1-x})_6$ layered ferrielectrics [2, 3], which reveal tunable ferroelectricity, ferrielectricity, and antiferroelectricity in bulk form, thin films, freestanding and strained monolayers [4] and nanoparticles [5]. The joint action of its multi-well potential energy landscape [6, 7] and negative electrostriction coupling [8] leads to a set of interesting phenomena in nanoscale $CuInP_2S_6$, such as the anomalous dynamics of polarization in thin films [9, 10] and nanoparticles [11, 12], the temperature and strain tunability of the multiple energy-degenerate metastable polar states [13], and possible emergence of controllable negative capacitance (NC) state [14, 15, 16]. Study of $CuInP_2S_6$ thin films and multilayers is of primary interest due to the giant flexoelectric effect, which determines the domain engineering [17, 18] and governs bending-induced isostructural transitions [19], as well as opens attractive possibilities for advanced memories and NC-field effect transistors [20, 21, 22], electromechanical, electrocaloric [23] and optoelectronic [24, 25] applications of the films.

The flexoelectric effect [26, 27], originated from the linear coupling of electric polarization with the strain gradient [28], can influence significantly the electromechanical state of ferroelectric thin films [29, 30], for which the strain and/or stress gradients are very strong at the surfaces and interfaces [31], as well as near extended defects (e.g., domain walls, twins and antiphase boundaries) [32, 33, 34]. When the strength of the static flexoelectric coupling (shortly "*flexocoupling*") exceeds the "upper limit" [35], a spatially modulated state (e.g., incommensurate phase) of the polarization fluctuations could appear [36, 37]. It has been shown recently that the influence of flexocoupling on the fluctuations of electric polarization and elastic strains can lead to the principal changes of the dispersion of soft optical and acoustic phonons and ferrons in a bulk $CuInP_2S_6$ [38].

The concept of a bosonic quasiparticle – "ferron", which carry electric dipoles and emerge from the joint action of anharmonicity and broken inversion symmetry in displacive ferroelectrics, was introduced by Bauer et al. [39, 40] and Tang et al. [41]. The ferrons include both volume-type [42] and surface-type excitation modes [43, 44]. Later on, the Bauer concept of ferron was extended to the fluctuations of the spontaneous polarization vector by Yang and Chen [45]. The ferroelectric soft optical phonons, which are quanta of a coherent polarization wave with a definite phase and frequency [46, 47], are "coherent ferrons" [48, 49]. The ferrons, which are excited by the thermal fluctuations or electric noise [38, 39-41], are incoherent. Recently, a ferron-driven transfer force on ferroelectric domain walls was calculated [50], and a nonlocal ferron-driven thermoelectricity was observed [51]. Strong coupling between coherent ferrons and cavity acoustic phonons is predicted in the $CuInP_2S_6$ films [52], and ferron-driven photoferroic hysteresis was observed in $CuInP_2S_6$ flakes [53]. Note that the dispersion of phonons and ferrons is anisotropic in $CuInP_2S_6$ due to its low symmetry, and applied electric field can influence strongly on the anisotropy [54].

It was shown much earlier that finite size effects of soft phonon dispersion are principally important in nanosized ferroelectrics [55]. According to first principles calculations, interface engineering of ferroelectricity is possible in thin films and multilayers of thiophosphates $ABP_2X_6$, including $CuInP_2Se_6$ and $CuInP_2S_6$ [4]. According to phenomenological Landau approach, size and strain effects can determine phase diagrams and anomalous polarization behavior in strained thin films of $CuInP_2S_6$ [10, 10]. Thus, it is reasonable to assume that the flexocoupling and elastic strains, which originate from the lattice constants mismatch between the film and substrate, should have a strong influence on the dispersion of soft optical and acoustic phonons and ferrons in thin strained films van der Waals ferroelectrics. However, to the best of our knowledge, the question is not studied.

Using Landau-Ginzburg-Devonshire (LGD) approach, in this work we study theoretically the influence of the flexocoupling and mismatch strains on the dispersion law of soft optical and acoustic phonons and ferrons in thin strained films of $CuInP_2S_6$. In particular, the frequency of acoustic phonons and ferrons tends to zero at nonzero wavevectors $k > k_{cr}$, where the critical value of the wavevector $k_{cr}$ is determined by the mismatch strain, temperature and flexocoupling strength. Zeroing of the acoustic phonon frequency, that appears with increase of tensile strains, indicates a dynamic instability induced by the flexo-strain effects, that may evolve into a spatially modulated incommensurate polar phase with increase in the flexocoupling strength.

## 2. The Problem Statement

Let us consider an epitaxial thin $CuInP_2S_6$ film of thickness $h$. The film is covered by ideally conducting electrodes and thus the single-domain state of its spontaneous polarization is energetically preferable. The film is regarded clamped to a rigid substrate, and the mismatch strain $u_m$, which originates from the lattice constants mismatch, exists at the film-substrate interface. The most common substrates are CMOS-compatible $SiO_2$ and $MoS_2$ layers [16, 20-22].

Paraelectric and ferrielectric phases of the $CuInP_2S_6$ film have the point group symmetry $2/m$ and $m$, respectively. The symmetry axis "2" lies in the plane of layers, with spontaneous polarization pointed along the normal vector to the layers. The coordinate system is organized as follows. The coordinate axis "$X_2$" is along the symmetry axis "2" (the crystallographic $b$ axis), the coordinate axis "$X_3$" (the crystallographic $c$ axis) is along the normal vector of all the layers, while the coordinate axis "$X_1$" (the crystallographic $a$ axis) is perpendicular to "$X_3$" and $X_2$". Note that the mirror plane (the only symmetry element of the ferroelectric phase) is the crystallographic *ac* plane, spanned by the axes "$X_1$" and "$X_3$" (see **Fig. 1**, left). Hereinafter we regard that $CuInP_2S_6$ is a uniaxial ferroelectric, which polar axis "3" is normal to the film surfaces (see red and blue arrows in the central part of **Fig. 1**). We neglect the small in-plane component of $CuInP_2S_6$ spontaneous polarization directed along $X_1$ axis.

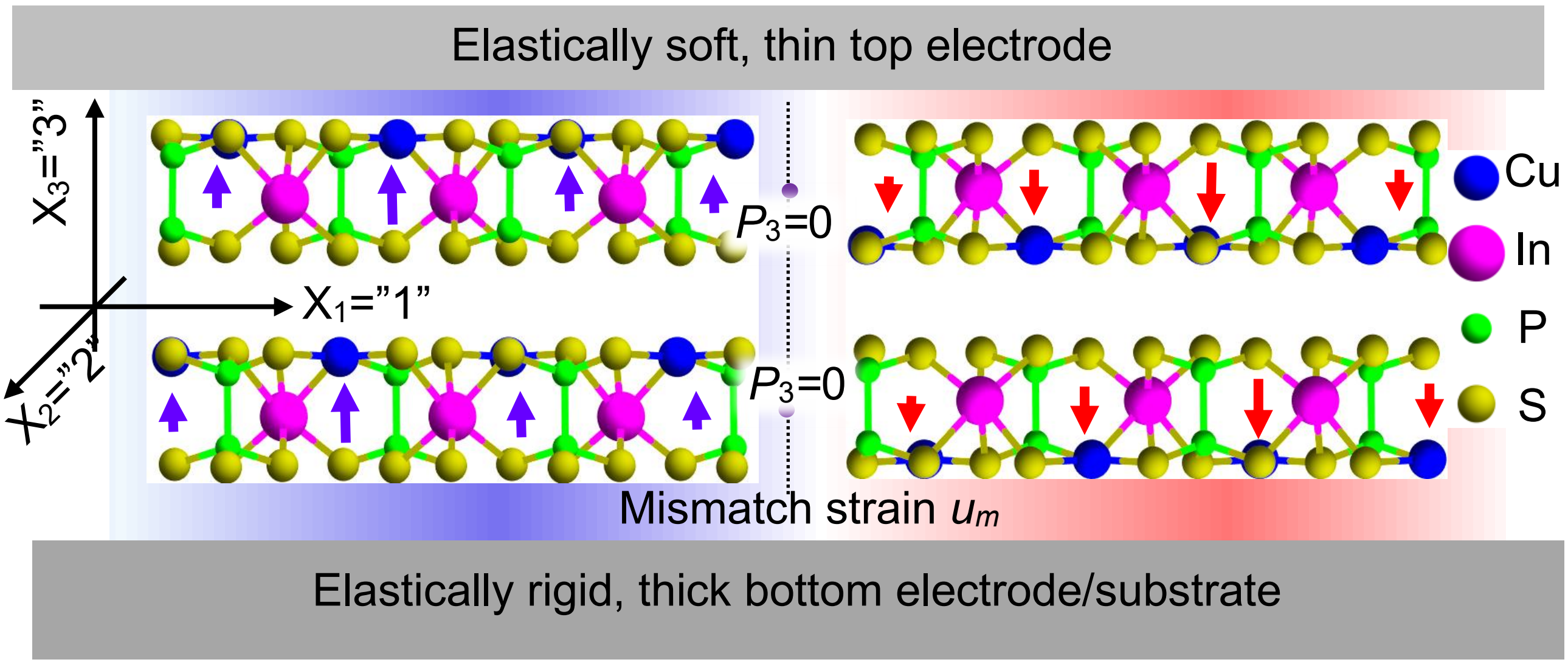

**FIGURE 1.** Schematic of a thin epitaxial $CuInP_2S_6$ film sandwiched between ideally conductive electrodes and clamped on a rigid substrate. The central part shows enlarged $CuInP_2S_6$ layers with crystallographic coordinate system. The polar axis $X_3$ ="3" is normal to the layers, and the axes $X_1$ ="1" and $X_2$ ="2" are in-plane of the layers. Blue and red arrows with varying length illustrate the transverse fluctuations of the ferrielectric polarization, $\delta P_3(x_1, x_2, t)$, with a wavevector $\vec{k}$ along axis $X_1$. The vertical dotted line shows the break between the counter-polarized layers and corresponds to a virtual plane with zero polarization. Central part of the figure is adapted from Ref. [38].

According to the symmetry principle the LGD free energy should be invariant with respect to the symmetry transformations of the paraelectric phase, so that the non-zero tensorial elements may have the even number of indices "2" and/or odd number of "1" and "3", but the total number of both "1" and "3" should be even also. Therefore, one should consider the coupling between the ferrielectric polarization component $P_3$, elastic displacement component $U_3$, and elastic strains $u_{33}$, $u_{23}$ and $u_{13}$.

Using the above approximations, the Lagrange function of $CuInP_2S_6$ film has the form

$$L = \int_t dt \int_{-\infty}^{\infty} dx_1 dx_2 \left[\int_{-h/2}^{h/2} dx_3 \,(F_b - E_k) + F_s\left(x_1, x_2, -\frac{h}{2}\right) + F_s\left(x_1, x_2, \frac{h}{2}\right)\right]. \tag{1}$$

It consists of the kinetic energy with the density $E_k$, bulk and surface free energies with the densities $F_b$ and $F_s$, respectively. The density of kinetic energy,

$$E_k = \frac{\mu}{2}\left(\frac{\partial P_3}{\partial t}\right)^2 + M\frac{\partial P_3}{\partial t}\frac{\partial U_3}{\partial t} + \frac{\rho}{2}\left(\frac{\partial U_3}{\partial t}\right)^2, \tag{2}$$

includes the dynamic flexocoupling with the magnitude $M$ [56]; $\rho$ is the mass density of a material; the coefficient μ is the polarization inertia, which can be expressed via the vacuum dielectric constant $\varepsilon_0$ and the plasma frequency $\omega_p$ as $\mu = \frac{1}{\varepsilon_0 \omega_p^2}$. The bulk and surface densities of the free energy have the following form:

$$F_b = \frac{\alpha}{2}P_3^2 + \frac{\beta}{4}P_3^4 + \frac{\gamma}{6}P_3^6 + \frac{\delta}{8}P_3^8 + g_{3i3j}\frac{\partial P_3}{\partial x_i}\frac{\partial P_3}{\partial x_j} - P_3E_3^e - \frac{1}{2}P_3E_3^d - q_{ij33}u_{ij}P_3^2 - z_{ij3333}u_{ij}P_3^4 -$$
$$w_{ijkl33}u_{ij}u_{kl}P_3^2 + \frac{f_{ij3k}}{2}\left(u_{ij}\frac{\partial P_3}{\partial x_k} - P_3\frac{\partial u_{ij}}{\partial x_k}\right) + \frac{c_{ijkl}}{2}u_{ij}u_{kl} + \frac{v_{ijklmn}}{2}\left(\frac{\partial u_{ij}}{\partial x_k}\right)\left(\frac{\partial u_{lm}}{\partial x_n}\right) - N_3U_3, \quad (3a)$$

$$F_S = \frac{\xi}{2}P_3^2. \quad (3b)$$

The coefficient α obeys the Barret-type expression, $\alpha(T) = \alpha_T T_q\left(\coth\frac{T_q}{T} - \coth\frac{T_q}{T_C}\right)$, where $T_C$ is the Curie temperature, $T_q$ is the quantum vibration temperature, and $T$ is the absolute temperature (in Kelvins). Barret-type expression is valid in a wide temperature range (from law to high temperatures) and reduces to the linear dependence $\alpha(T) = \alpha_T(T - T_C)$ at $T \gg T_q$. All other coefficients in Eq.(3a) are supposed to be temperature independent. The coefficient $\delta$ cannot be negative for the stability of the free energy for arbitrary large $P_3$ values. The coefficients $g_{ijkl}$ are the components of the polarization gradient energy tensor. The values of $g_{ijkl}$ are always positive. Coefficients $f_{ijkl}$ are the components of the static flexocoupling tensor. The coefficients $q_{ijkl}$, $z_{ij3333}$ and $w_{ijkl33}$ are second-order and higher-order electrostriction coupling coefficients [57], respectively. As a rule, the contribution of the term $w_{ijkl33}u_{ij}u_{kl}P_3^2$ is small enough; however, it appeared important for the quantitative description of the sound velocity temperature dependence in $CuInP_2S_6$ [58]. Coefficients $c_{ijkl}$ are the components of elastic stiffness tensor that determines the positive elastic energy. Coefficients $v_{ijklmn}$ are the components of the strain gradient energy that is always positive. $N_3$ is z-component of the external mechanical force bulk density; $E_3^e$ is z-component of external electric field and $E_3^d$ is the depolarization field, which contribution to the free energy is given by the term $\frac{P_3E_3^d}{2}$. The constant $\xi \geq 0$ is the coefficient of polarization surface energy. The surface elastic energy is not considered; that is a usual assumption for ferroics [55].

Considering the Tani mechanism and Khalatnikov relaxation, the explicit form of the LGD-KT equations for elastic displacement component $U_3$ and polarization $P_3$ is:

$$\Lambda\frac{\partial U_3}{\partial t} = -\frac{\delta L}{\delta U_3}, \qquad \Gamma\frac{\partial P_3}{\partial t} = -\frac{\delta L}{\delta P_3}, \quad (4)$$

where Γ and Λ are phenomenological damping constants [38].

The boundary conditions for polarization at the top ($x_3 = h/2$) and bottom ($x_3 = -h/2$) film surfaces of the $CuInP_2S_6$ film can be derived by taking the variation of the Lagrangian $L$ with respect to $P_3$. Performing integration by parts on the polarization-gradient terms isolates the surface contribution, i.e., $\delta L|_{x_3=\pm h/2} = \int dt \int dS \left[\xi P_3 \pm g_{3333}\frac{\partial P_3}{\partial x_3} - \frac{f_{3333}}{2}u_{33}\right]\delta P_3$ . Because $P_3$ is unconstrained (i.e., $\delta P_3$ can take arbitrary values) at the film surfaces ($x_3 = \pm h/2$), the terms in the square bracket need to be zero for letting $\delta L|_{x_3=\pm h/2} = 0$, yielding the so-called mixed-type boundary condition,

$$-\xi P_3 - g_{3333}\frac{\partial P_3}{\partial x_3} - \frac{f_{3333}}{2}u_{33}\Big|_{x_3=h/2} = 0, \quad \xi P_3 - g_{3333}\frac{\partial P_3}{\partial x_3} - \frac{f_{3333}}{2}u_{33}\Big|_{x_3=-h/2} = 0. \tag{5a}$$

For the sake of analytical solution derivation, we regard that $\xi$ is negligibly small in Eq.(3b), which corresponds to the so-called natural boundary conditions [55].

The boundary conditions for elastic displacement $U_i$, strains $u_{ij}$ and stresses $\sigma_{ij}$ components at the clamped bottom surface ($x_3 = -h/2$) and free top surface ($x_3 = h/2$) of the $CuInP_2S_6$ film have the form

$$U_i|_{x_3=-h/2} = 0, \quad u_{11} = u_{22}|_{x_3=-h/2} = u_m, \quad \left(v_{i33lmn}\frac{\partial u_{lm}}{\partial x_n} - \frac{f_{i333}}{2}P_3\right)\Big|_{x_3=-h/2} = 0, \tag{5b}$$

$$\sigma_{3i}|_{x_3=h/2} = 0, \qquad \left(v_{i33lmn}\frac{\partial u_{lm}}{\partial x_n} - \frac{f_{i333}}{2}P_3\right)\Big|_{x_3=h/2} = 0, \tag{5c}$$

where the subscript $i = 1, 2, 3$ (see Supplemental Materials for detail).

Hereinafter the “effective” mismatch strain $u_m^*$ is introduced as:

$$u_m^* = u_m\left(1 - \exp\left[-\frac{h_d}{h}\right]\right), \tag{5f}$$

where $u_m$ are the components of the “seeding” biaxial mismatch strain, $h_d$ is the critical thickness of dislocation appearance [59]. The relaxation of mismatch strains, that inevitably happens in thin films with thickness $h > h_d$, leads to decrease in the mismatch-driven electrostriction coupling between electric polarization and elastic strains.

Electric field $\vec{E}$ satisfies the wave-type equation,

$$\Delta \vec{E} - \frac{\chi\varepsilon_b}{c_0^2}\frac{\partial^2}{\partial t^2}\vec{E} = \chi_0\chi\frac{\partial^2}{\partial t^2}\vec{P} - \frac{1}{\varepsilon_0\varepsilon_b}\boldsymbol{\nabla}\mathrm{div}\,\vec{P}, \tag{6a}$$

where $\varepsilon_0$ and $\chi_0$ are the universal dielectric and magnetic constants, respectively, $\chi$ is the relative magnetic permittivity and $\varepsilon_b$ is the relative dielectric permittivity of background [60], $c_0 = \frac{1}{\sqrt{\chi_0\varepsilon_0}}$ is the speed of light in vacuum, and electric displacement $\vec{D} = \varepsilon_0\varepsilon_b\vec{E} + \vec{P}$. The boundary conditions for an electric field, which correspond to the electroded single-domain $CuInP_2S_6$ thin film of thickness $h$, have the form:

$$E_3^d\Big|_{x_3=h/2} = 0, \qquad E_3^d\Big|_{x_3=-h/2} = 0. \tag{6b}$$

The depolarization field that contains static and dynamic components produced by the inhomogeneity and motion of bound polarization charges. The Fourier image of the depolarization field $E_i^d$ is derived in Supplement S1 of Supplementary Materials [61]; it has the form:

$$\tilde{E}_i^d = -\frac{k_i\left(\vec{k}\cdot\vec{\tilde{P}}\right)-\chi\varepsilon_b\frac{\omega^2}{c_0^2}\tilde{P}_i}{\varepsilon_0\varepsilon_b\left(k^2-\chi\varepsilon_b\frac{\omega^2}{c_0^2}\right)}. \quad (6c)$$

Here $\tilde{P}_i$ is the Fourier image of the polarization component $P_i$, $\vec{k}$ is the wave vector and $\omega$ is the frequency.

The Fourier representation of Eqs. (4), linearized with respect to the fluctuations, $\widetilde{\delta P}_i$ and $\widetilde{\delta U}_i$, allow calculating the generalized susceptibility of the film to electric fields and elastic forces (see e.g., Refs. [38, 54] for details). The characteristic equation for the frequency dispersion, $\omega(\vec{k})$, of the soft phonons was derived from the singularity of the generalized susceptibility in the Fourier $\{\vec{k},\omega\}$-space (see Supplementary Materials in Refs. [38, 54] for details).

To derive analytical expressions, below we consider purely transverse fluctuations of polarization and displacement in the epitaxial $CuInP_2S_6$ film, namely $\widetilde{\delta P}_3(k_1,k_2,\omega)$ and $\widetilde{\delta U}_3(k_1,k_2,\omega)$. This approximation does not consider the propagation of mixed TA – LA phonons in the monoclinic plane due to the mixing of the out-of-plane spontaneous polarization $P_3$ with the small in-plane component of the spontaneous polarization $P_1$. Rigorously speaking, the assumption of purely transverse fluctuations is well-grounded for uniaxial ferroelectrics only, where the longitudinal fluctuations of spontaneous polarization are suppressed by the static depolarization field [38, 54], but the dynamic depolarization field exists anyway. Since $\frac{\partial P_3}{\partial x_3}=0$ and $\frac{\partial U_3}{\partial x_3}=0$ in this case, the condition of transverse fluctuations is consistent with the boundary conditions (5a) only if $\xi\to 0$. The case $\xi=0$ corresponds in the realistic and practically important natural boundary conditions for polarization and strain variations.

From Eq. (6), the static depolarization field is absent in the case, when the scalar product $\vec{k}\cdot\vec{P}=0$, which corresponds to the purely transverse fluctuations of polarization $\widetilde{\delta P}_3(k_1,k_2,\omega)$. Importantly, the dynamic depolarization field, $\tilde{E}_3^d(k_1,k_2,\omega)\cong-\frac{\chi\varepsilon_b\frac{\omega^2}{c_0^2}\widetilde{\delta P}_3(k_1,k_2,\omega)}{\varepsilon_0\varepsilon_b\left(k^2-\chi\varepsilon_b\frac{\omega^2}{c_0^2}\right)}$, which is induced by the transverse fluctuations $\widetilde{\delta P}_3(k_1,k_2,\omega)$, should influence the dispersion law of phonons and ferrons. The magnitude of $\tilde{E}_3^d$ is relatively small for a near-unity relative magnetic permeability (i.e., $\chi\cong 1$, which is reasonable for a non-magnetic material) at all wavevectors above $10^{-4}$ $nm^{-1}$ (see estimates in Supplement S1 [61]). As an example, the wavevector of the transverse optical (TO) phonon in $CuInP_2S_6$, $k_1$ or $k_2$, varies approximately from 0.25 $nm^{-1}$ to 0.35 $nm^{-1}$ within the temporal frequency range of 47.8 GHz to 74.8 GHz (see Supplement S1). Therefore, it is generally acceptable to omit the effect of $\tilde{E}_3^d$ on the dispersion relation herein. However, we note that $\tilde{E}_3^d$ influences the phonon dispersion relation even in bulk $CuInP_2S_6$ at wavevectors below $10^{-4}$ $nm^{-1}$ (see **Figs. S1** and **S2** in Supplement S4 [61]). Transverse components $\tilde{E}_{1,2}^d$, which emerge inevitably in thin films (see Ref. [52]

and estimates in Supplement S1 [61]), do not influence the dispersion law of phonons and ferrons, because we only consider the variation of polarization $P_3$ (parallel to the polar $c$ axis) in this work.

### 3. The Influence Elastic Strains and Flexocoupling on Dispersion of Soft Phonons

The condition of transverse fluctuations leads to $\vec{k}\cdot\overrightarrow{\delta P}=0$, and so the dispersion relation $\omega(\vec{k})$ for the case of bulk-like phonons with strain-renormalized and thickness-renormalized parameters can be found from the 6-th order equation:

$$\alpha_s+\hat{g}\vec{k}^2-i\Gamma\omega-\mu\omega^2-\frac{\chi_0\chi\varepsilon_0\varepsilon_b\omega^2}{\varepsilon_0\varepsilon_b(k^2-\chi_0\chi\varepsilon_0\varepsilon_b\omega^2)}-\frac{\left(\hat{f}\vec{k}^2-M\omega^2\right)^2+4P_S^2\left(\hat{q}\vec{k}+2P_S^2\hat{z}\vec{k}\right)^2}{\hat{v}\vec{k}^4+\hat{c}\vec{k}^2+c\frac{\epsilon(h)}{\phi\, h^2}-i\Lambda\omega-\rho\omega^2}=0. \quad (7a)$$

Hereinafter is the transverse wave vector $\vec{k}=\{k_1,k_2\}$ and $k^2=k_1^2+k_2^2$. The thickness-dependent function $\epsilon(h)$ is introduced

$$\epsilon(h)=1+\frac{5r_c(\zeta-\phi)(\xi+\phi)}{2\phi^2kh}, \quad (7b)$$

where the constants $r_c=\sqrt{1-\frac{c_{44}}{c_{11}}}$, $c=c_{44}\approx c_{55}$, $\phi=\frac{16-4\pi}{\pi^3}$, $\zeta=\frac{4}{\pi^2}$, and $\xi=\frac{4\pi-12}{\pi^2}$ (see **Supplement S4** [61] for details). The value $\alpha_s$, introduced in Eq.(7a), is the temperature-dependent function of spontaneous polarization $P_s$:

$$\alpha_s=\alpha^*+3\beta^*P_s^2+5\gamma^*P_s^4+7\delta^*P_s^6. \quad (8a)$$

The magnitude of $P_s$ can be found from the 7-th order algebraic equation

$$\alpha^*P_s+\beta^*P_s^3+\gamma^*P_s^5+\delta^*P_s^7=0. \quad (8b)$$

The coefficients $\alpha^*$, $\beta^*$, $\gamma^*$ and $\delta^*$ are renormalized by the clamping to substrate and effective mismatch strain $u_m^*$ in the following way:

$$\alpha^*(T,u_m^*)=\alpha_T T_q\left(\coth\frac{T_q}{T}-\coth\frac{T_q}{T_C}\right)-2u_m^*\left(q_{13}+q_{23}-\frac{c_{13}+c_{23}}{c_{33}}q_{33}\right), \quad (8c)$$

$$\beta^*(T,u_m^*)=\beta-2\frac{q_{33}^2}{c_{33}}-4u_m^*\left[z_{133}+z_{233}-\frac{c_{13}+c_{23}}{c_{33}}z_{333}\right], \quad (8d)$$

$$\gamma^*=\gamma-6\frac{q_{33}z_{333}}{c_{33}}, \qquad \delta^*=\delta-4\frac{z_{333}^2}{c_{33}}. \quad (8e)$$

We used Voight notations in expressions (8c)-(8e) and omitted the terms proportional to the parameter $w_{ijklmn}$ and its higher powers. These terms, which are included in much more cumbersome Eqs.(S.19), lead to the 12-th powers of polarization in the renormalized free energy density and increase drastically the mathematical complexity of results [62]. Note that the renormalization of the coefficients $\alpha^*$, $\beta^*$, $\gamma^*$ and $\delta^*$ exist even in the case $u_m^*=0$, because it is determined by the film clamping to a rigid substrate.

The $\vec{k}$-dependent function $\hat{g}\vec{k}^2\equiv g_{3i3j}k_ik_j$, introduced in Eq.(7), is the convolution of the 2D wavevector $\vec{k}$ with the polarization gradient tensor $\hat{g}$, μ is the polarization inertia, $\hat{f}\vec{k}^2\equiv f_{3i3j}k_ik_j$ is the convolution of $\vec{k}$ with the flexocoupling tensor $\hat{f}$, $M$ is the dynamic flexocoupling constant, $\hat{q}\vec{k}\equiv$ $q_{i333}k_i$ is the convolution of $\vec{k}$ with the second-order electrostriction tensor $\hat{q}$, $\hat{z}\vec{k} \equiv z_{i33}k_i$ is the convolution of $\vec{k}$ with the higher-order electrostriction tensor $\hat{z}$, $\hat{v}\vec{k}^4 \equiv v_{3ij3lm}k_i k_j k_l k_m$ is the convolution of $\vec{k}$ with the higher-order strain gradient tensor $\hat{v}$, $\hat{c}\vec{k}^2 \equiv c_{3i3j}k_i k_j$ is the convolution of $\vec{k}$ with the elastic stiffness tensor $\hat{c}$ and ρ is the mass density of a material. The explicit form of all convolutions is listed in Supplement S2 for the case of the 2/m symmetry. Material parameters of a bulk $CuInP_2S_6$ are listed in **Tables S1-S2** [61]; they are collected from Refs. [8-13, 15, 16, 19, 58, 63]).

Note that the contribution of dynamic term $\frac{\chi_0\chi\varepsilon_0\varepsilon_b\omega^2}{\varepsilon_0\varepsilon_b(k^2-\chi_0\chi\varepsilon_0\varepsilon_b\omega^2)}$ in Eq. (7) to phonon dispersion is negligibly small for $CuInP_2S_6$ at all wavevectors above $10^{-4}$ $nm^{-1}$ (for $\chi \cong 1$). The dispersion of "far" infra-red polaritons can be noticeable in the range $10^{-6}$ – $10^{-4}$ $nm^{-1}$, as shown in **Figs. S1** and **S2** (see **Supplement S5** in Supplemental Materials [61]).

The applicability ranges of the dispersion relation $\omega(\vec{k})$ describing bulk-like phonons with thickness-renormalized and strain-renormalized parameters, namely Eqs.(7)-(8), is discussed in **Supplement S4** [61], where we presented calculation details for elastic waves propagation in thin films with elastic boundary conditions (5). We obtained that the dispersion relation (7) is valid in the central part of the film excluding the vicinity of the film-substrate interfaces. Performed FEM confirms that solution (7) for elastic waves have sufficient accuracy in the range of $x_3$ satisfying the inequality $k\left(x_3+\frac{h}{2}\right) \gg \frac{\pi}{2}$ and $k\left(x_3-\frac{h}{2}\right) \geq \frac{\pi}{2}$, where $\vec{k}=\{k_1,k_2\}$ is the transverse wave vector of phonons and/or ferrons. For the transverse wavevectors $k \geq 3\frac{\pi}{h}$, the approximation (7) is valid entire the film depth excluding the immediate vicinity of the film-substrate interfaces. Since the peculiarities of phonon and ferron spectra, considered in the work, are within the range $1\text{nm}^{-1} \geq k \geq 0.1\text{nm}^{-1}$, the condition $kh \geq 3\pi$ corresponds to the film thicknesses above $9.5 - 95$ nm. Surface-like phonon excitations, which are reminiscent of well-known Rayleigh waves, propagate near the freestanding surface of the film, while elastic waves with transverse polarization are strongly suppressed near the clamped surface. The Love-type waves with longitudinal polarization are not considered in the present work due to the strong suppression of electric polarization by emerging depolarization field.

For negligibly small damping and dynamic contribution of the depolarization field (i.e., after neglecting the terms $i\Gamma\omega$, $i\Lambda\omega$ and $\frac{\chi_0\chi\varepsilon_0\varepsilon_b\omega^2}{\varepsilon_0\varepsilon_b(k^2-\chi_0\chi\varepsilon_0\varepsilon_b\omega^2)}$, Eq. (7) that reduces to the biquadratic equation with respect to $\omega^2$. In this case the solution of Eq.(7) was found in Refs. [38, 54]:

$$\omega_{O,A}^2(\vec{k}) = \frac{1}{2(\mu\rho-M^2)}\left[C(\vec{k}) \pm \sqrt{C^2(\vec{k}) - 4(\mu\rho-M^2)B(\vec{k})}\right], \tag{9a}$$

where the functions $C(\vec{k})$ and $B(\vec{k})$ are introduced

$$C(\vec{k}) = \alpha_S\rho + \Gamma\Lambda + \mu\left(\hat{v}\vec{k}^4 + \hat{c}\vec{k}^2 + c\frac{\epsilon(h)}{h^2\phi}\right) - 2\hat{f}\vec{k}^2 M + \hat{g}\vec{k}^2\rho + \mu\hat{v}\vec{k}^4, \tag{9b}$$

$$B(\vec{k}) = (\alpha_S + \hat{g}\vec{k}^2)\left(\hat{v}\vec{k}^4 + \hat{c}\vec{k}^2 + c\frac{\epsilon(h)}{h^2\phi}\right) - 4P_S^2\left(\hat{q}\vec{k} + 2\hat{z}\vec{k}P_S^2\right)^2 - \left(\hat{f}\vec{k}^2\right)^2. \quad (9c)$$

The dispersion relation (9a) contains soft optical (O) and acoustic (A) phonons, corresponding to the signs "+" and "–" before the radical, respectively. In Eqs.(9b)-(9c) the values $\alpha_S$ and $P_S$ are determined by effective mismatch strain $u_m^*$ and temperature $T$. The frequency of optical phonons is always positive, but this is not the case for acoustic phonons, because the condition $4(\mu\rho - M^2)B(\vec{k}) \leq 0$ is possible. The condition $B(\vec{k}) = 0$ is equivalent to the condition $\mathrm{Re}[\omega_A(\vec{k})] = 0$, which determines a dynamic instability of spatially uniform FI phases induced by the flexo-strain effects in a definite range of $u_m^*$, $\hat{f}$ and $T$. Hereinafter we put $f_{55} = f_{44} \equiv f$ for the 2/m parent symmetry and consider films with thicknesses $h$ more than 9.5 — 95 nm to use approximate expressions (7)-(9) with negligibly small thickness-dependent correction $c\frac{\epsilon(h)}{\phi\, h^2}$.

The dependence of the spontaneous polarization $P_s$ on temperature $T$ and mismatch strain $u_m^*$ is shown in **Fig. 2(a)**. The dependence was calculated by a conventional numerical minimization of the LGD free energy (3), as described in Ref. [10]. However, in this work we use slightly different values of electrostriction tensor coefficients $q_{53}$ and $z_{533}$, which better fit available experimental data (see **Table S1** for details). Due to fact, the phase diagram, shown in **Fig. 2(a)**, looks slightly different from those shown in Ref. [10].

Results for phonon dispersion, presented below, correspond to the thermodynamically absolutely-stable states, which are the ferrielectric state 1 (FI1) with larger spontaneous polarization, or ferrielectric state 2 (FI2) with smaller spontaneous polarization, or the paraelectric (PE) phase without spontaneous polarization at fixed temperature $T$ and effective mismatch strain $u_m^*$. There are two regions of phase coexistence associated with first-order phase transitions. The first region, designated as FI1+PE in **Figs. 2(a),** is relatively large. It corresponds to the coexistence of the PE phase and the FI1 state at temperatures above approximately 300 K and exists for compressive mismatch strains. Its potential relief transforms continuously from the single-well PE phase to the three-well FI1 phase with the change in $u_m^*$ (see **Fig. S7(a)** in Suppl. Mat. [61]). The second region, located inside a small triangle limited by the bicritical end point (BEP), the critical end point (CEP) and the crossing of the PE-FI2 boundary with the FI1+PE coexistence boundary (white star), is relatively small. It corresponds to the coexistence of the FI1 and FI2 states at temperatures around 300 K and exists for tensile mismatch strains. Its potential relief transforms continuously from the two-well PE phase to the four-well mixed phase, and then to the three-well FI2 phase with the change in $u_m^*$ (see **Fig. S7(b)** in Suppl. Mat. [61]). The phonon dispersion relations are different in the stable and metastable states, but the system inevitably "falls" with time into the state corresponding to the deepest potential well due to small (but finite) damping. The situation of the almost equal-depth or "quasi-flat" wells is particularly interesting and deserves a separate study that is beyond the scope of the work.

The dependence of the soft optical phonon frequency $\omega_O$ on temperature $T$ and effective mismatch strain $u_m^*$ calculated at $\vec{k} = 0$ and fixed flexoelectric coefficient $f = 4$ V is shown in **Fig. 2(b)**. The dependence of the acoustic phonon frequency $\omega_O$ on temperature $T$ and effective mismatch strain $u_m^*$ calculated at $k_1 = 0.25$ nm$^{-1}$, $k_2 = 0$ and fixed flexoelectric coefficient $f = 4$ V is shown in **Fig. 2(c)**. As anticipated, the dependences of $\omega_O$ and $\omega_A$ on the mismatch strain $u_m^*$ and temperature $T$ correlate completely with the dependence of $P_s$, reflecting the fact that the dispersion of both soft optical and acoustic phonons are determined by the structure of mismatch-induced phase transitions between the FI1, PE and FI2 phases (compare color scales in **Figs. 2(a), (b)** and **(c)**).

Indeed, the region of the PE phase is stable (or metastable) at temperatures $T > 300$ K in **Figs. 2(a)-2(c)**. The first-order strain-induced transition of the ferrielectric state FI1 (a reddish region with a large amplitude of $P_s$) to the ferrielectric state FI2 (a green-blue region with a small amplitude of $P_s$) occurs at $T < 300$ K. The PE–FI2 phase transition is of the second order at tensile strains $u_m^* > 0$, and the PE–FI1 phase transition is of the first order at compressive strains $u_m^* < 0$ [10]. The first order PE–FI1 transition determines the coexistence of the FI1 state and the PE phase for tensile strains (marked as the "FI1+PE" region in **Figs. 2(a)-2(c)**). The second order PE–FI2 phase transition line terminates in the critical end point (CEP, shown by a white circle in **Figs. 2(a)-2(c)**); and the first order FI1–FI2 phase terminates in the bicritical end point (BEP, shown by a black circle in **Figs. 2(a)-2(c)**).

The anomalous feature of **Figs. 2(a)-2(c)** is that the large-polarization state FI1, and therefore higher frequencies of optical and acoustic phonons, exist at compressive strains $u_m^* < 0$ and do not vanish for tensile strains $u_m^* > 0$. At $u_m^* > 0$ the large-polarization FI1 state transforms continuously to the small-polarization state FI2, which in turn transforms continuously to the PE phase (see the dotted line in **Figs. 2(a)-2(c)**).

The color maps of $P_s$, $\omega_O$ and $\omega_A$, shown in **Figs. 2(a)-2(c)**, are principally different from the corresponding dependencies expected in the most strained films of uniaxial and/or multiaxial ferroelectrics, where the out-of-plane polarization is absent or very small at tensile strains ($u_m^* > 0$), and the region of the out-of-plane spontaneous polarization emerges and significantly increases for compressive strains ($u_m^* < 0$) [64]. The effect can be explained as follows. The polarization switching in thin strained films of classical ferroelectric with electrostriction coefficients $q_{33} > 0$, $q_{23} < 0$, and $q_{13} < 0$ leads to a strong increase of the out-of-plane spontaneous polarization for tensile strains, and their strong decrease or disappearance for compressive strains [10, 64]. The strain effect on the ferrielectric polarization of $CuInP_2S_6$ films is "inverted" in comparison with the films of classical ferroelectrics due to the opposite signs of $q_{33} < 0$, $q_{23} > 0$, and $q_{13} > 0$, as well as due to the strongly negative and temperature-dependent nonlinear electrostriction coupling coefficients $z_{i33} < 0$ [10].

Dispersion of soft optical and acoustic phonons, $\omega_O(k_1)$ and $\omega_A(k_1)$, calculated for a mechanically free bulk $CuInP_2S_6$, is shown in **Fig. 2(d)** at higher ($T$ =305 K) and lower ($T$ =100 K) temperatures. It is seen that $\omega_O(k_1)$ decreases in the center of the Brillouin zone (Γ-point, $k = 0$) with increase in temperature, that corresponds to the well-known "softening" of the optical phonons under approaching the FI – PE phase transition point. The slope of the $\omega_A(k_1)$ near the Γ-point is temperature independent within used approximations. At the same time, the minimal distance $\Delta\omega_{OA}$ between the $\omega_O(k_1)$ and $\omega_A(k_1)$ curves increases with decrease in temperature. This happens because $\Delta\omega_{OA}$ is largely determined by the renormalized coefficient $\alpha_s$, electrostriction, static and dynamic flexoelectric coupling constants, which relative contribution to $\Delta\omega_{OA}$ increases with the decrease in temperature. The value of $\Delta\omega_{OA}$ can be roughly estimated as $\Delta\omega_{OA} < \frac{1}{\mu\rho - M^2}\sqrt{C^2(\vec{k}) - 4(\mu\rho - M^2)B(\vec{k})} < \frac{\alpha_s\rho}{\mu\rho - M^2}$. According to the analysis of eigenvectors (corresponding to atom displacements), under the linear interaction (coupling) of modes of the same symmetry, not only the "repulsion" of their frequencies occurs in the "collision" point (marked by dotted ellipse in **Fig. 2(d))**, but also the exchange of the shape of the eigenvectors. This is realized fully when the coupling constants are reduced to zero – the acoustic and optical phonon branches intersect without exchanging the shape of the eigenvectors. In a real crystalline $CuInP_2S_6$, the soft optical phonon branch also interacts linearly with the nearest higher-frequency optical branch, which enhances its optical character.

Dispersion of soft optical and acoustic phonons, $\omega_O(k_1)$ and $\omega_A(k_1)$, calculated for strained $CuInP_2S_6$ films at $k_2 = 0$, lower ($T$ =100 K) and higher ($T$ =305 K) temperatures, are shown in **Fig. 2(e)** and **2(f)**, respectively. Effective mismatch strain $u_m^*$ changes from -0.5% to +0.5% for different curves; and the flexoelectric coefficient is fixed, $f = 4$ V. Compressive strains ($u_m^* < 0$) increases strongly the plateau-like values of $\omega_O(k_1)$ and $\omega_A(k_1)$ (compare blue and green curves). Tensile strains ($u_m^* > 0$) decrease strongly the values of $\omega_O(k_1)$ at small $k_1$ (compare black, magenta and red dashed curves). Also, tensile strains decrease strongly the maximal values of $\omega_A(k_1)$ and can lead to the disappearance of $\mathrm{Re}[\omega_A(k_1)]$ at wavevectors above the critical value, $k_1 \geq k_{cr}$ (compare black, magenta and red solid curves). The frequency of acoustic phonon becomes purely imaginary at $k_1 > k_{cr}$ (see **Fig. S3** in Supplementary Materials [61]). Zeroing of the acoustic phonon frequency, emerging with increase of tensile strains, indicates a dynamic instability of spatially uniform polar phase, that can be induced by the flexo-strain effects in the considered case.

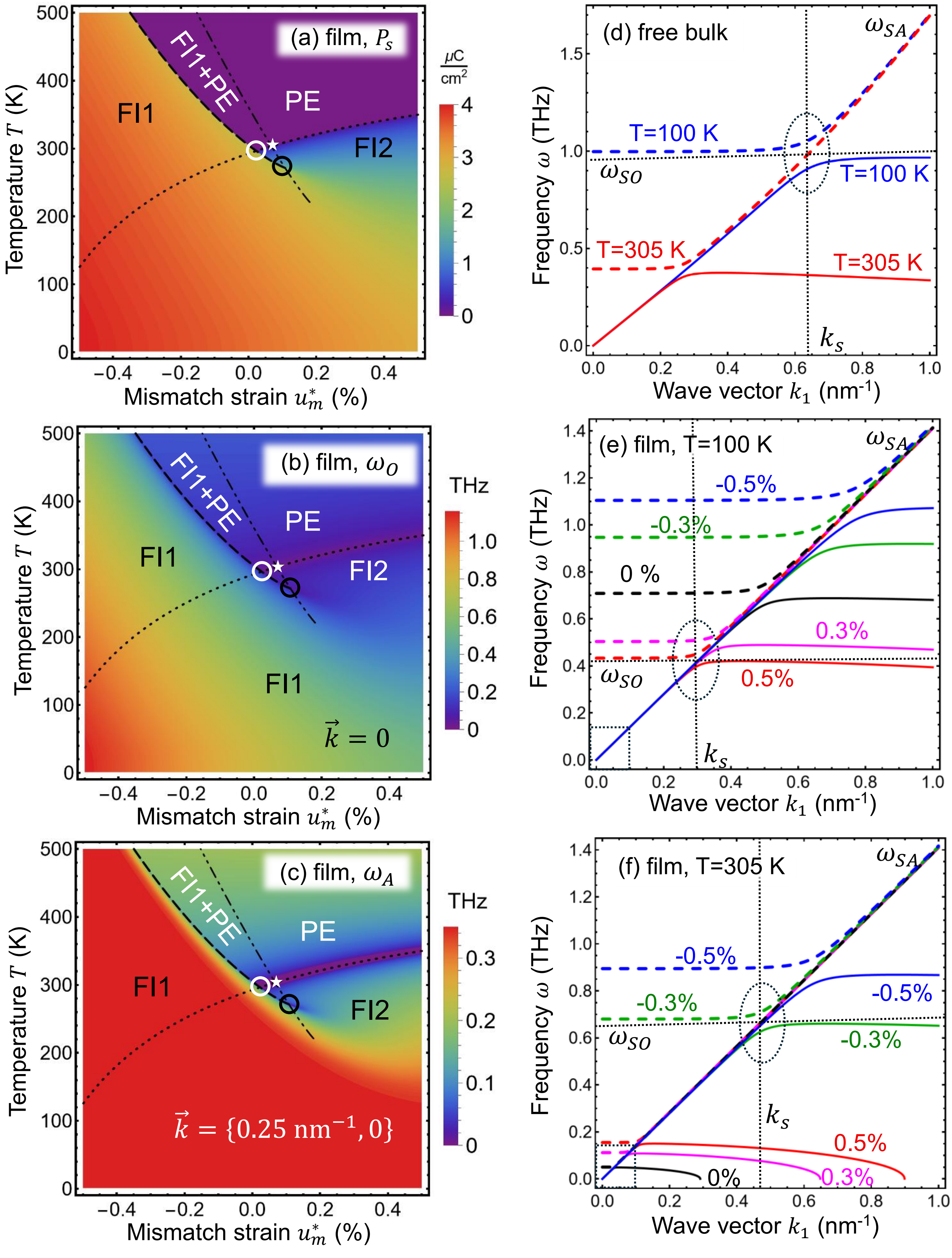

**FIGURE 2**. **(a)** The dependence of the $CuInP_2S_6$ film spontaneous polarization $P_S$ on temperature $T$ and effective mismatch strain $u_m^*$. PE is the paraelectric phase, FI1 and FI2 are the ferrielectric states with large and small $P_S$. The critical and bicritical end points, CEP and BEP, are marked by the white and black circles, respectively. The white star designates the crossing of the PE-FI2 boundary with the FI1+PE coexistence boundary. Color scale is the absolute value of $P_S$ in the deepest potential well of the LGD free energy. The dependence of the soft optical **(b)** and acoustic **(c)** phonon frequency on temperature $T$ and effective mismatch strain $u_m^*$ calculated at $\vec{k} = 0$ **(b)**, $k_1 = 0.25$ nm$^{-1}$ and $k_2 = 0$ **(c)** for strained $CuInP_2S_6$ films. **(d)** Dispersion of soft optical (dashed curves, $\omega_O$) and acoustic (solid curves, $\omega_A$) phonons for a mechanically-free bulk $CuInP_2S_6$ at different temperatures, $T$ =100 K (blue curves) and 305 K (red curves). **(e)-(f)** Dispersion of soft optical (dashed curves) and acoustic (solid curves) phonons calculated for strained $CuInP_2S_6$ films at $k_2 = 0$, $T$ =100 K **(e)** and 305 K **(f)**, and several effective mismatch strains $u_m^* = -0.5\%$ (blue curves), -0.3% (green curves) 0 (black curves), +0.3 % (magenta curves) and +0.5 % (red curves). Black dotted lines and ellipses in the parts **(d)-(f)** show the asymptotic frequency $\omega_{SA}$ and the "collision" point $k_s$ of optical and acoustic modes. The damping coefficients are negligibly small ($\Gamma \to 0$ and $\Lambda \to 0$), the flexoelectric coefficient $f = 4$ V, the relative magnetic permittivity $\chi = 1$ and $h = 100$ nm for all plots. Dotted rectangles mark the k-regions, where Eq.(7) is not applicable. Material parameters of $CuInP_2S_6$ are listed in **Tables S1-S2** [61].

At small wavevectors, the acoustic phonon frequency is weakly dependent on temperature and mismatch strain at wavevectors $\vec{k}$, smaller than the k-point of the optical and acoustic modes maximal proximity, $\vec{k} = \vec{k}_s$. The wavevector $\vec{k}_s$ is the "crossing" or "collision" k-point of the modes, that satisfies the equation $\left(\hat{v}\vec{k}^4 + \hat{c}\vec{k}^2 + c\frac{\epsilon(h)}{\phi\, h^2}\right)\mu = \left(\alpha_s + \hat{g}\vec{k}^2\right)\rho$. The equation follows directly from Eqs.(9), it has an approximate solution $k_s \approx \sqrt{\frac{\alpha_s \rho}{\hat{c}\mu - \hat{g}\rho}} \sim \sqrt{\alpha_s}$, where the $\alpha_s$ is the function of $T$, $u_m^*$ and $P_s(T, u_m^*)$ according to Eq.(8). The magnitude of $\vec{k}_s$ depends strongly on $T$ and $u_m^*$ as shown in **Figs. 2(d) – 2(f)**. The frequency of soft optical phonons (after coupling with the acoustic phonons at $\vec{k} = \vec{k}_s$) become practically independent of $T$ and $u_m^*$ at large wavevectors $k \gg k_s$. This behavior is due to the weak interaction between soft optical and acoustic phonons at both $k \ll k_s$ and $k \gg k_s$. One can derive the asymptotic expressions, $\omega_A^2(k \ll k_s) \to \omega_{SA}^2$, $\omega_O^2(k \gg k_s) \to \omega_{SA}^2$ and $\omega_O^2(k \ll k_s) \cong \omega_{SO}^2$ from Eqs. (9), where the asymptotic frequencies $\omega_{SA}^2(\vec{k}) = \left(\hat{v}\vec{k}^4 + \hat{c}\vec{k}^2 + c\frac{\epsilon(h)}{\phi\, h^2}\right)/\left(\rho - \frac{M^2}{\mu}\right) \approx \left(\hat{v}\vec{k}^4 + \hat{c}\vec{k}^2\right)/\left(\rho - \frac{M^2}{\mu}\right)$ and $\omega_{SO}^2(\vec{k}) = \left(\alpha_s + \hat{g}\vec{k}^2\right)/\left(\mu - \frac{M^2}{\rho}\right)$ are introduced. In fact, $\omega_{SA}$ is the frequency of the bulk acoustic phonons, independent of the flexoelectric coupling $f$, effective mismatch strain $u_m^*$ and temperature $T$, if elastic stiffnesses are considered constant.

The changes in the soft optical phonon frequency with increase in $|u_m^*|$ and/or $T$ is explained by the temperature and strain behavior of the FI1−PE phase transition, that exists at compressive strains only, and its temperature increases with increase in $|u_m^*|$ (see dashed curves in **Figs. 2(a)-2(c)**). The temperature behavior of the FI2 – PE phase transition is the opposite: the transition exists at tensile mismatch strains only, and its temperature increases with increase in $u_m^*$, as does the frequency of the soft optical mode. Therefore, the dispersion curves calculated for $u_m^* < 0$ and $T < 300$ K correspond to the FI1 state, and the dispersion curves for $u_m^* > 0$ and $T > 300$ K correspond to the FI2 state of the strained $CuInP_2S_6$ film. The dispersion curves calculated for the temperatures well below the BEP temperature (270 K) corresponds to the FI1 state of the $CuInP_2S_6$ film for both tensile and compressive mismatch strains.

For relatively small wavevectors (e.g., at $k = 0.25$ nm$^{-1}$), the dependence of the soft optical and acoustic phonon frequency on the effective mismatch strain $u_m^*$ and flexocoupling strength $f$ is largely determined by temperature, as shown in **Figs. 3(a)-3(d)**. Dispersion curves of soft optical ($\omega_O(\vec{k})$, upper curves) and acoustic ($\omega_A(\vec{k})$, lower curves) phonons, calculated for compressive ($u_m^* = -0.5$ %) and tensile ($u_m^* = 0.5$ %) mismatch strains, lower ($T = 100$ K) and higher ($T = 305$ K) temperatures, are shown in **Figs. 3(e)-3(h)** for a set of the flexoelectric coefficient $f$, which vary from 0 (violet curves) to 8 V (red curves) with step size of 1 V for neighboring curves. A common feature of the dispersion curves is the very weak dependence of $\omega_O(\vec{k})$ on $f$ at both small and large wavenumbers. Namely, the difference between the violet and red curves becomes visible in only the k-range near the collision point, $k \cong k_s$. The dependence of $\omega_A$ on $f$ is almost absent at $k < k_s$, it appears at larger $k$ and leads either to slow decrease or to zeroing of $\mathrm{Re}[\omega_A(\vec{k})]$.

For the temperatures much lower than the temperatures of the FI1–PE and the PE–FI2 phase transitions (e.g., for $T \leq 100$ K) and relatively small wavevectors (e.g., for $k \leq 0.3$ nm$^{-1}$) the frequency $\omega_A$ decreases very weakly and monotonically with increase in $f$ from 0 to 8 V, as well as with the change in mismatch from compressive ($u_m^* = -0.5$ %) to tensile ($u_m^* = +0.5$%) strains (see **Fig. 3(a)**). At the same temperatures, the frequency $\omega_O$ decreases relatively strongly and monotonically with the change in $u_m^*$; being almost independent on $f$ (see **Fig. 3(b)**). The behavior of $\omega_A$ at the wavevectors $k \leq k_s$ is largely determined by the asymptotic relation, $\omega_A^2(\vec{k}) \approx \omega_{SA}^2$, where $\omega_{SA}$ is the mismatch-independent frequency of a bulk acoustic phonon (see lower curves in **Figs. 3(e)** and **3(g)**). The behavior of $\omega_O \sim \sqrt{\alpha_s}$ at the wavevectors at $k \leq k_s$ is largely determined by the dependence of the coefficient $\alpha_s$ on the spontaneous polarization $P_s$ and the mismatch strain $u_m^*$ in the FI1 phase at lower temperatures (see **Fig. 2(a)**). Let us mention that $\omega_O$ becomes independent of $T$ and $u_m^*$ only at $k \gg k_s$ (see upper curves in **Figs. 3(e)** and **3(g)**).

For the temperatures close to the temperatures of the FI1–PE and/or the PE–FI2 phase transitions (e.g., for $T = 305$ K) the frequency $\omega_A$ is almost independent on $f$ and $u_m^*$ for tensile strains $u_m^* < -0.05$ % and relatively small $k = 0.25$ nm$^{-1}$ (see the left side of **Fig. 3(c)**). This is because the condition $k \ll k_s$ is still valid for $u_m^* < -0.05$ %, and thus the asymptotic relation $\omega_A^2(\vec{k}) \approx \omega_{SA}^2$ is valid for the compressive strains. For tensile strains ($u_m^* > 0$) and $T = 305$ K the dependence of $\omega_A$ on $f$ and $u_m^*$ is nontrivial (see the right side of **Fig. 3(c)**), because the chosen k-value is close to collision $k_s$-point of the optical and acoustic phonons (see e.g., **Fig. 3(h)**). The critical value of the wavevector $k_{cr}$, that corresponds to the condition $\omega_A(k_{cr}) = 0$, depends strongly on $f$ and $u_m^*$ (see the dark-violet region in **Figs. 3(c)**, where $\mathrm{Re}[\omega_A(\vec{k})] = 0$)). The frequency $\omega_A$ becomes purely imaginary inside the dark-violet region (see **Fig. S3** [61] for details). Outside the dark-violet region the frequency $\omega_A$ depends significantly on the $f$ and $u_m^*$, reflecting the strain-induced transition of the $CuInP_2S_6$ film from the FI1 to the PE phase, and then from the PE to the FI2 phase (see **Fig. 2(a)**). The boundary between large, zero and small values of $\mathrm{Re}[\omega_A]$ becomes slightly more diffuse with the increase in $f$ from 0 to 8 V (see **Figs. 4(c)**).

Generally speaking, the existence and magnitude of the critical wavevector $k_{cr}$ is largely determined by the mismatch strain $u_m^*$, flexoelectric coefficient $f$ and temperature $T$ (compare **Figs. 3(f)** and **3(h)**). The acoustic phonon frequency is a weakly anisotropic function of the wavevector components $k_1$ and $k_2$ at lower temperatures (see insets in **Figs. 3(e)** and **3(g)**). Significant anisotropy appears with increase in temperature (see insets in **Figs. 3(f)** and **3(h)**). More information about the anisotropy of the acoustic phonons is shown **Fig. S4** [61].

The behavior of $\omega_O$ at $T = 305$ K and $k \leq k_s$ is determined by the strain-induced FI1–PE–FI2 phase transitions of the $CuInP_2S_6$ film (compare **Figs. 3(f)** and **3(h)**). The frequency of soft optical phonon is almost isotropic function of the wavevector components $k_1$ and $k_2$ entire studied k-range and temperature range, and therefore it is not shown in **Fig. 3**.

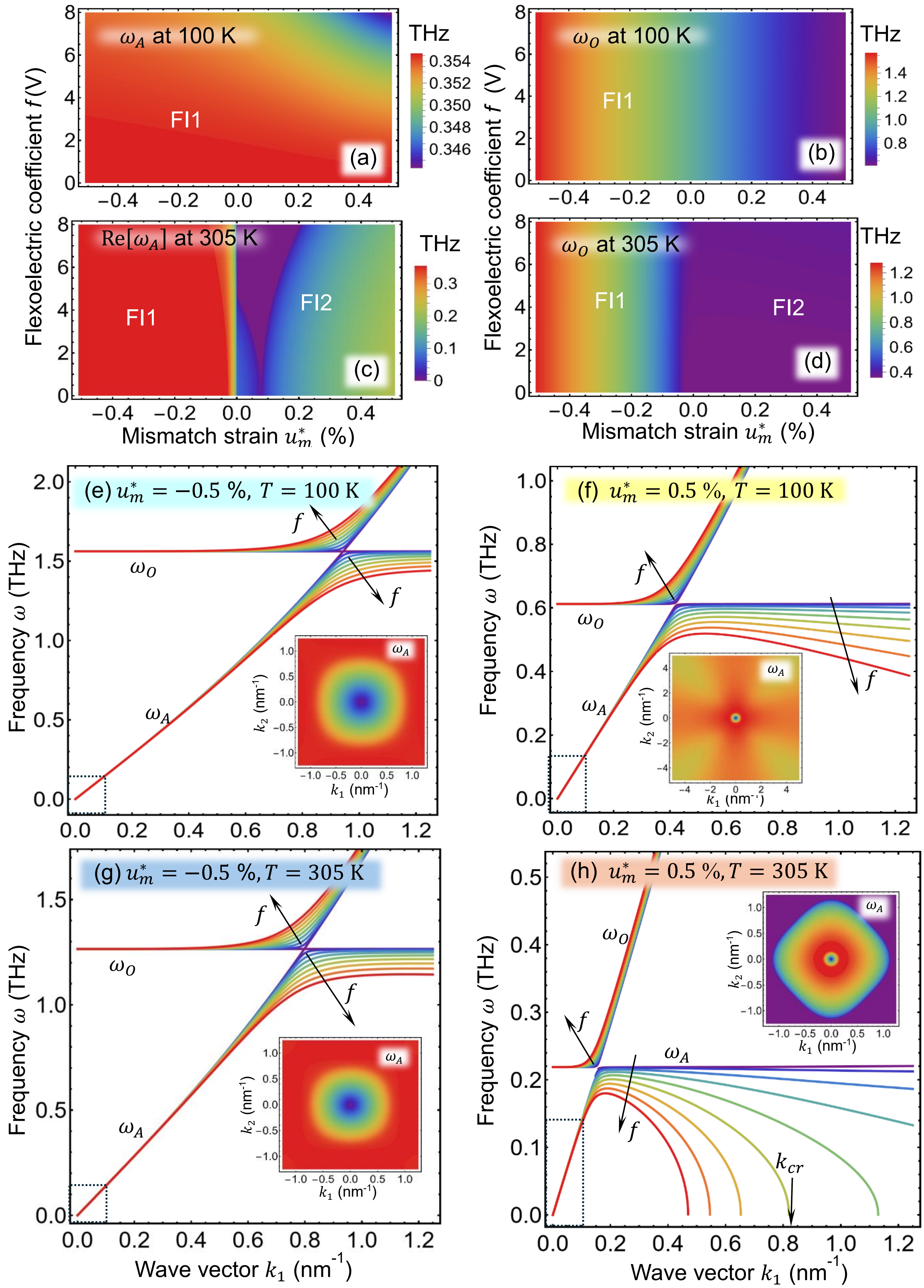

**FIGURE 3.** The acoustic **(a, c)** and soft optical **(b, d)** phonon frequencies, $\omega_A$ and $\omega_O$, as the function of effective mismatch strain $u_m^*$ and flexoelectric coefficient $f$ calculated for $T = 100$ K **(a, b)** and 305 K **(c, d)**

at the wavevector components $k_1 = 0.25$ nm$^{-1}$ and $k_2 = 0$. Dispersion of soft optical ($\omega_O$, upper curves) and acoustic ($\omega_A$, lower curves) phonons calculated for $k_2 = 0$, effective mismatch strains $u_m^* = -0.5$ % **(e, g)** and +0.5% **(f, h)**, $T = 100$ K **(e, f)** and 305 K **(g, h)**, and several values of the flexoelectric coefficient $f$, which vary from 0 (violet curves) to 8 V (red curves) with step size of 1 V for neighboring curves. Insets in the panels **(e)-(h)** show the real part of the acoustic phonon frequency $\omega_A$ as the function of the wavevector components $k_1$ and $k_2$, calculated for $f$= 4 V and $h$ =100 nm. Dotted rectangles mark the k-regions, where Eq.(7) is not applicable. The damping is absent ($\Gamma = 0, \Lambda = 0$) for all plots; material parameters of $CuInP_2S_6$ are listed in **Tables S1-S2** [61].

### 4. The Influence of Elastic Strains and Flexocoupling on the Dispersion of Ferrons

Following Tang et al. [41], we consider the case when the mechanical force and the electric field in Eqs. (3) are Langevin-type noise fields, which obey the fluctuation-dissipation theorem for fluctuation dynamics [65]. Assuming that the damping constant is small and using the perturbation theory, in Ref. [38] we derived the second-order correction for the polarization response to the Langevin-type electric noise:

$$\langle p\rangle \approx \sum_{m=1}^{\infty}\int_{-\infty}^{\infty}\frac{d^2\vec{k}}{(2\pi)^2}\left[\coth\left(\frac{\hbar\omega_O(\vec{k})}{2k_BT}\right)\delta p_O(\vec{k}) + \coth\left(\frac{\hbar\omega_A(\vec{k})}{2k_BT}\right)\delta p_A(\vec{k})\right]. \tag{10}$$

Here $\vec{k} = \{k_1, k_2\}$, $\delta p_O(\vec{k})$ and $\delta p_A(\vec{k})$ are the spectral densities of optical and acoustic ferrons, the eigen frequencies $\omega_O(\vec{k})$ and $\omega_A(\vec{k})$ are given by Eq. (8) at zero damping constants $\Gamma$ and $\Lambda$.

Approximate analytical expressions for $\delta p_O(\vec{k})$ and $\delta p_A(\vec{k})$ were derived in Ref. [54]. Under the absence of external field, they are:

$$\delta p_O(\vec{k}) \approx \frac{-\hbar}{2(\mu - M^2/\rho)}\frac{\eta}{\alpha_S}\left|\frac{\omega_{SA}^2(\vec{k})-\omega_O^2(\vec{k})}{\left(\omega_A^2(\vec{k})-\omega_O^2(\vec{k})\right)\omega_O(\vec{k})}\right|, \tag{11a}$$

$$\delta p_A(\vec{k}) \approx \begin{cases} \frac{-\hbar}{2(\mu - M^2/\rho)}\frac{\eta}{\alpha_S}\left|\frac{\omega_{SA}^2(\vec{k})-\omega_A^2(\vec{k})}{\left(\omega_O^2(\vec{k})-\omega_A^2(\vec{k})\right)\omega_A(\vec{k})}\right|, & \omega_A^2(\vec{k}) > 0, \\ \frac{-\hbar}{2(\mu - M^2/\rho)}\frac{\eta}{\alpha_S}\left|\frac{\omega_{SA}^2(\vec{k})-\omega_A^2(\vec{k})}{\left(\omega_O^2(\vec{k})-\omega_A^2(\vec{k})\right)\omega_A(\vec{k})}\right|\frac{\Gamma}{\sqrt{\Gamma^2-4\mu^2\omega_A^2(\vec{k})}}, & \omega_A^2(\vec{k}) \leq 0. \end{cases} \tag{11b}$$

Here $\omega_{SA}^2(\vec{k}) = \left(\hat{v}\vec{k}^4 + \hat{c}\vec{k}^2 + c\frac{\epsilon(h)}{\phi\, h^2}\right)/\left(\rho - \frac{M^2}{\mu}\right) \approx \left(\hat{v}\vec{k}^4 + \hat{c}\vec{k}^2\right)/\left(\rho - \frac{M^2}{\mu}\right)$ is the frequency of the acoustic phonons in the stress-free homogeneous bulk $CuInP_2S_6$; that is virtually independent of the flexoelectric coupling $f$, effective mismatch strain $u_m^*$ and temperature $T$. The thickness-dependent correction $c\frac{\epsilon(h)}{\phi\, h^2}$ is small for $kh > 3\pi$ that corresponds to the thicknesses $h$ above 9.5 – 95 nm in the actual range of $1\ \mathrm{nm}^{-1} \geq k \geq 0.1\mathrm{nm}^{-1}$ (see **Supplement S4** [61]). The parameter $\eta$ is determined by spontaneous polarization, namely $\eta = 3\beta^* P_s + 10\gamma^* P_s^3 + 21\delta^* P_s^5$ . Note that $\omega_O^2(\vec{k})$ is always positive, but the case $\omega_A^2(\vec{k}) < 0$ is possible for high flexoelectric coefficients (see e.g., **Fig. 3**) and/or for high electric fields close to the coercive field (the latter case is considered in Ref. [54]).

Note $\delta p_A(\vec{k}) \to 0$ at $\Gamma \to 0$ and $\omega_A^2(\vec{k}) < 0$, being consistent with impossibility to quantize the elementary oscillator with imaginary frequency $\omega_A(\vec{k})$ due to its damping. The expression (11b) for $\delta p_A(\vec{k})$ is mathematically rigorous at $\omega_A^2(\vec{k}) \leq 0$, and corresponding integration in the complex plane was performed in Ref. [54]. The ferron concept, proposed by Bauer et al. [41], requires very small (but nonzero) damping constant $\Gamma$ in the amplitude of the Langevin-type noise. The magnitude $\Gamma$ is not included in the final expression (11b) for the ferron density when $\omega_A^2(\vec{k}) > 0$ (see mathematical details in Refs. [41, 54]) allowing us to set $\Gamma \to 0$ for all plots in this work. The physical explanation of the result consists in the current impossibility of quantizing a harmonic oscillator with finite damping. The divergency of $\delta p_A(\vec{k})$ at $\omega_A^2(\vec{k}) \to +0$ is also a mathematically rigorous result; however, it may indicate the inapplicability of the perturbation approach [41, 54] in the immediate vicinity of the threshold $\omega_A^2(\vec{k}) = 0$.

For relatively big wavevectors (e.g., at $k = 0.5$ nm$^{-1}$), the dependence of the acoustic and optical ferron spectral densities, $\delta p_A$ and $\delta p_O$, on the effective mismatch strain $u_m^*$ and flexocoupling strength $f$ is largely determined by temperature, as shown in **Figs. 4(a)-4(d)**.

For the temperatures much lower than the temperatures of the FI1–PE and the PE–FI2 phase transitions (e.g., for $T \leq 100$ K) and relatively small wavevectors (e.g., for $k \leq 0.5$ nm$^{-1}$) the absolute value of $\delta p_A$ increases relatively strongly and monotonically under the change of mismatch from compressive ($u_m^* = -0.5$ %) to tensile ($u_m^* = +0.5$%) strains (see **Fig. 4(a)**). The spectral density $\delta p_O$ changes relatively strongly and nonmonotonically with the change in $u_m^*$ from -0.5 % to +0.5 % at low temperatures and $k \leq 0.5$ nm$^{-1}$ (see **Fig. 4(b)**). The boundary between large and small values of $\delta p_A$ and $\delta p_O$ becomes much more diffuse with the increase in $f$ from 0 to 8 V (see **Figs. 4(a)** and **4(b)**).

For the temperatures close to the temperatures of the FI1–PE and/or the PE–FI2 phase transitions (e.g., for $T = 305$ K) the frequency $\delta p_A$ is very small and almost independent on $f$ and $u_m^*$ for tensile strains $u_m^* < -0.2$ % and $k \sim 0.5$ nm$^{-1}$ (see the left side of **Fig. 4(c)**). This is because the condition $k < k_s$ is still valid for $u_m^* < -0.2$ %, and thus the asymptotic relation $\omega_A^2(\vec{k}) \approx \omega_{SA}^2$ is valid for the compressive strains. For tensile strains ($u_m^* > 0$), $k \sim 0.5$ nm$^{-1}$ and $T = 305$ K the dependence of $\delta p_A$ on $u_m^*$ and $f$ is nontrivial (see the right side of **Fig. 4(c)**), because the chosen k-value is close to collision point of the optical and acoustic phonons at $T = 305$ K (see e.g., **Fig. 3(h)**). The spectral density $\delta p_A$ diverges at the boundary of the blank region, because $\omega_A = 0$ at the boundary and $\delta p_A(\vec{k}) \sim \frac{1}{\omega_A(\vec{k})}$ according to Eq.(11b). The spectral density $\delta p_A \equiv 0$ inside the blank region according to Eq.(11b), because $\omega_A^2(\vec{k}) < 0$ in the region. Outside the blank region the density $\delta p_A$ depends significantly on the $f$ and $u_m^*$, reflecting the strain-induced transition of the $CuInP_2S_6$ film from the FI1 to the PE phase, and then from the PE to the FI2 phase (see **Fig. 2(a)**). The spectral density $\delta p_O$ nonmonotonically depends on the strain $u_m^*$ at $T = 305$ K, namely it becomes larger, then smaller and then vanishes under the change in the change in $u_m^*$ from -0.5 % to +0.5 % (see **Fig. 4(d)**). The boundary between smaller, larger and almost zero values of $\delta p_O$ becomes slightly more diffuse with the increase in $f$ from 0 to 8 V (see **Figs. 4(d)**).

Dispersion curves of the $\delta p_A(\vec{k})$ and $\delta p_O(\vec{k})$, calculated for $k_2 = 0$, fixed flexoelectric coefficient $f = 4$ V, lower ($T = 100$ K) and higher ($T = 305$ K) temperatures, are shown in **Figs. 4(e)-4(h)** for a set of effective mismatch strains $u_m^*$, which vary from -0.5 % (violet curves) to +0.5 % (red curves) with step size of 0.1 % for neighboring curves. A common feature of the dispersion curves is the following. The spectral density $\delta p_O$ is very small at $k \gg k_S$, since the asymptotic equality $\omega_O^2(\vec{k}) \to \omega_{SA}^2(\vec{k})$ is valid in the k-range, as it follows from Eq. (11a). The spectral density $\delta p_A$ is very small at $k \gg k_S$, since the asymptotic equality $\omega_A^2(\vec{k}) \to \omega_{SA}^2(\vec{k})$ is valid in the k-range, as it follows from Eq. (11b). This behavior is due to the weak interaction between the optical and acoustic phonon modes at small and large $\vec{k}$ (see previous section for details). The divergency of $\delta p_A$ appears at $\vec{k} = \vec{k}_{cr}$ due to zeroing of Re$[\omega_A(\vec{k})]$ at the critical k-value, because $\delta p_A(\vec{k}) \sim \frac{1}{\omega_A(\vec{k})}$ according to Eq.(11b) (see **Fig. 4(g)**). The behavior of $\delta p_O(\vec{k})$ at $T = 305$ K is determined by the strain-induced FI1–PE–FI2 phase transitions of the $CuInP_2S_6$ film (compare **Figs. 4(h)** and **3(h)**).

The spectral density acoustic ferrons $\delta p_A(\vec{k})$ is a weakly anisotropic function of the wavevector components $k_1$ and $k_2$ at lower temperatures (see inset in **Figs. 4(e)**). Significant anisotropy appears with increase in temperature (see inset in **Figs. 4(g)**). More information about the anisotropy of acoustic ferrons at lower and higher temperatures can be obtained **Figs. S5** and **S6** [61]. The spectral density of optical ferrons $\delta p_O(\vec{k})$ is weakly anisotropic function of the wavevector components $k_1$ and $k_2$ in the wide temperature and mismatch strain ranges (see insets in **Figs. 4(f)** and **4(h)**).

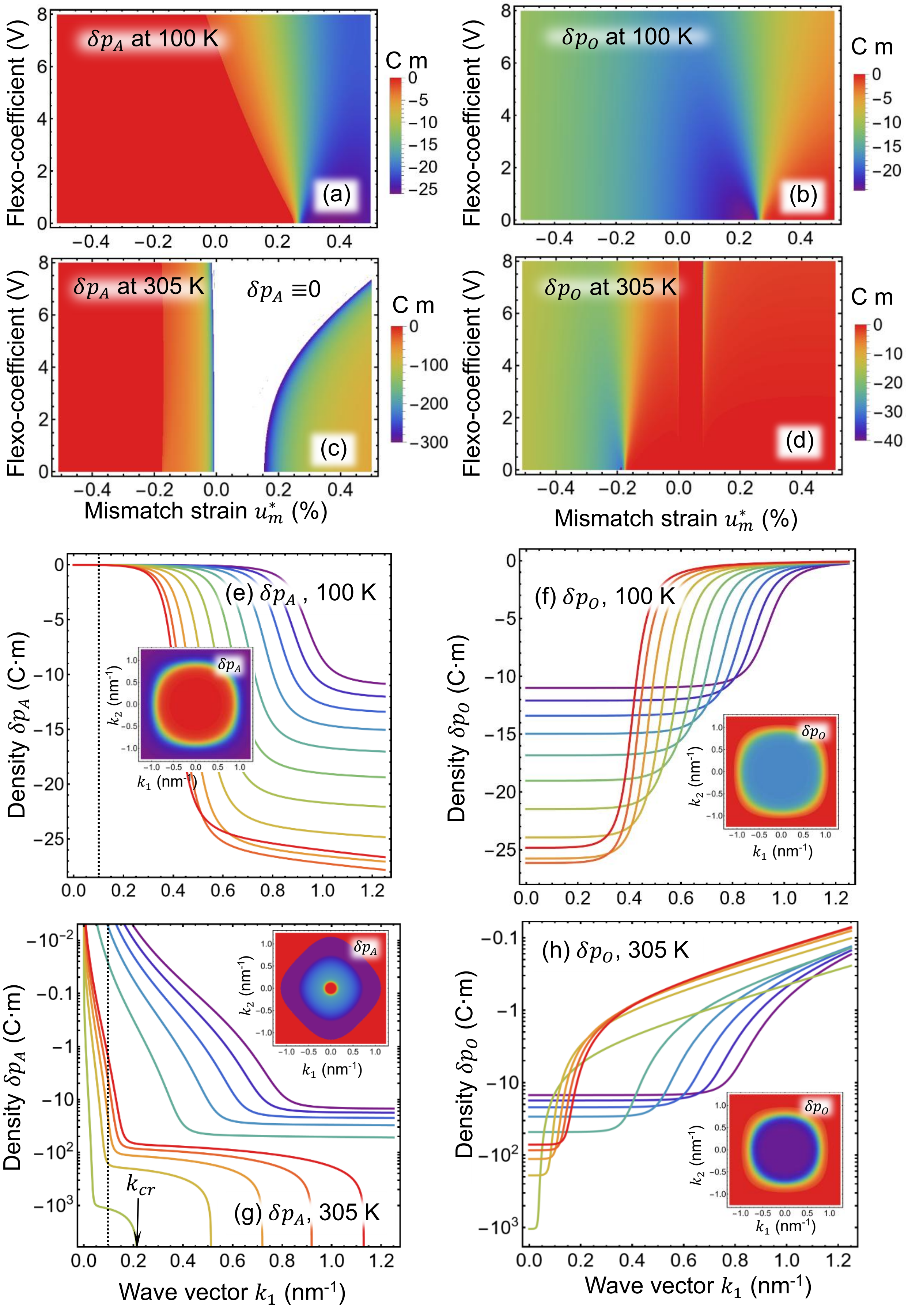

**FIGURE 4.** The spectral density of the acoustic $\delta p_A$ **(a, c)** and optical $\delta p_O$ **(b, d)** ferrons (in $10^{-32}$ C·m) as the function of effective mismatch strain $u_m^*$ and flexoelectric coefficient $f$ calculated for $T = 100$ K **(a, b)** and 305 K **(c, d)** at the wavevector components $k_1 = 0.5$ nm$^{-1}$ and $k_2 = 0$. The spectral density of the acoustic **(e,**

**g)** and optical **(f, h)** ferrons calculated for $k_2 = 0$, flexoelectric coefficient $f = 4$ V, $T = 100$ K **(e, f)** and 305 K **(g, h)**, and several values of the effective mismatch strains $u_m^*$, which vary from -0.5 % (violet curves) to +0.5 % (red curves) with the step size of 0.1 % for neighboring curves. Insets in the panels **(e)-(h)** show the density of ferrons as the function of the wavevector components $k_1$ and $k_2$, calculated for $f$= 4 V and $u_m^*$ =0.5 %. Blank region in part (c) corresponds to $\delta p_A = 0$. Vertical dotted lines mark the $k$-range, where the thickness-dependent corrections for $\delta p_A$ are significant. The damping is absent ($\Gamma = 0, \Lambda = 0$) and $h = 100$ nm for all plots; material parameters of $CuInP_2S_6$ are listed in **Tables S1-S2** [61].

Dispersion of soft optical and acoustic ferron spectral densities, $\delta p_A(\vec{k})$ and $\delta p_O(\vec{k})$, calculated for a set of the flexoelectric coefficients $f$, which vary from 0 to 8 V, compressive ($u_m^* = -0.5$ %) and tensile ($u_m^* = +0.5$ %) effective mismatch strains are shown in **Fig. 5(a)** and **5(b)**, respectively. The color maps of the $\delta p_A(\vec{k})$ and $\delta p_O(\vec{k})$ as a function of $k_1$ and $f$ calculated for $k_2 = 0$, compressive and tensile mismatch strains, are shown in **Figs. 5(c)-5(f)**, respectively. All plots in **Fig. 5** are calculated at $T = 305$ K to show significant flexo-strain-induced changes in the ferron spectra, which appear at temperatures close to the mismatch strain-induced FI1–PE–FI2 phase transitions (shown in **Fig. 2(a)**). At low temperatures the flexo-strain-induced changes are much less pronounced, and so the temperature range is not shown in the main text.

Large red regions in **Figs. 5(c)-5(f)** show the regions where the amplitude of optical or acoustic ferrons is close to zero. Blue regions in **Figs. 5(c)-5(f)** are the regions where the amplitude of optical or acoustic ferrons is maximal. It is seen that these red and blue regions are interchanged with respect to the sign of effective mismatch strain. The "interchange" can be explained by several factors. Primary, it is the strain-induced changes in the spontaneous polarization across the FI1–PE–FI2 phase transitions and weak interaction of the soft optical and acoustic phonon modes entire the k-range except for the vicinity of collision point $k = k_s$, which position is determined by the effective mismatch strain $u_m^*$. Secondary, it is the conditions when the acoustic phonon frequency becomes imaginary at large wavevectors (see the blank region in **Figs. 5(d)**). Note that the boundary between zero, small and large values of $\delta p_A$ and $\delta p_O$ becomes much more diffuse with the increase in $f$ from 0 to 8 V. At $f = 0$, the relatively sharp boundary between zero and large values of $\delta p_A$ and $\delta p_O$ corresponds to $k_1 \approx 0.8$ nm$^{-1}$ at $u_m^* = -0.5$ % and to $k_1 \approx 0.16$ nm$^{-1}$ at $u_m^* = +0.5$ %. These k-points are collision points of the soft optical and acoustic phonon modes, shown in **Fig. 3(g)** and **3(h)**, respectively.

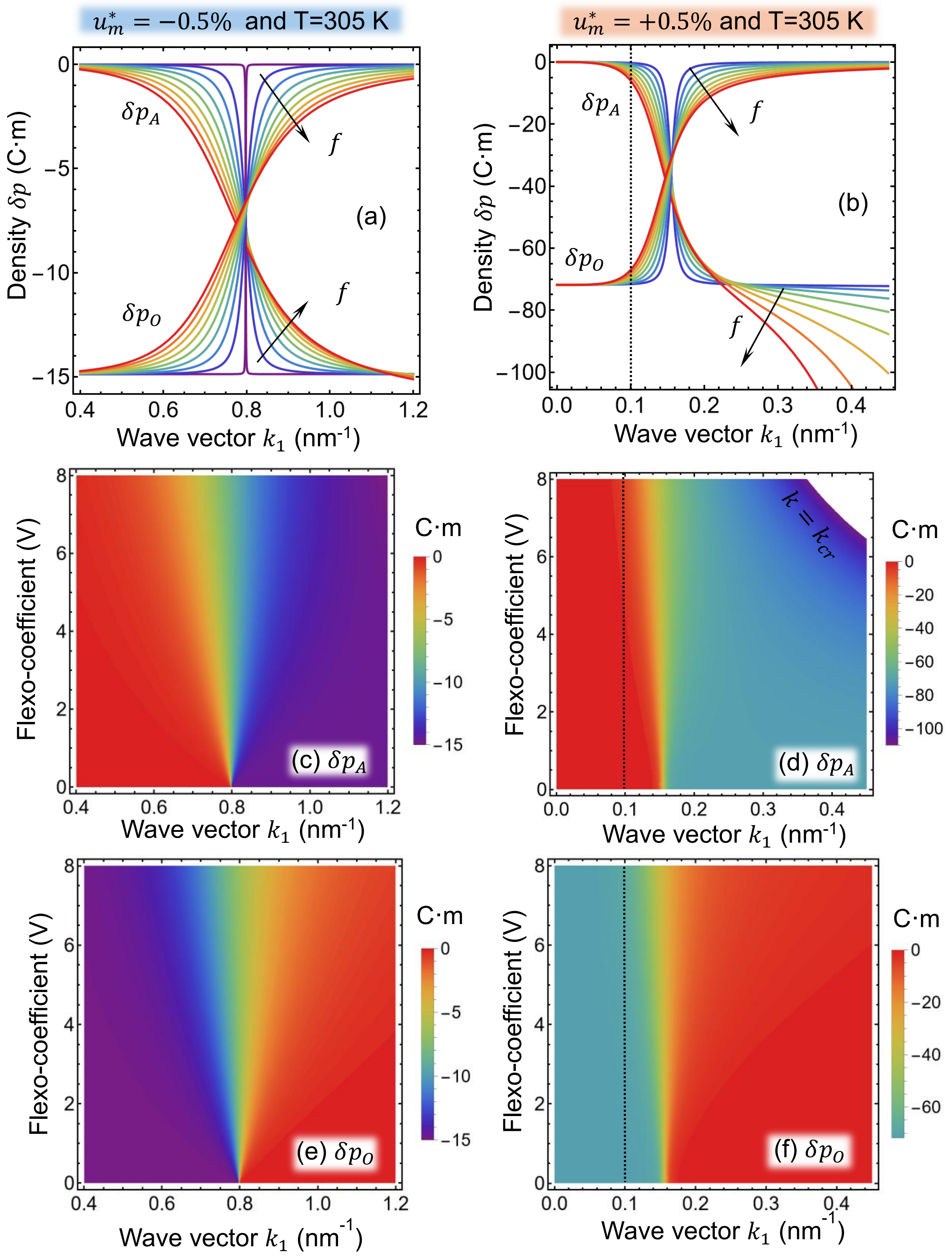

**FIGURE 5.** Dispersion of acoustic and optical ferrons spectral densities $\delta p_A(k_1)$ and $\delta p_O(k_1)$ (in 10$^{-32}$ C·m) calculated for $k_2 = 0$, effective mismatch strains $u_m^* = -0.5$ % **(a)** and +0.5% **(b)**, and several values of the flexoelectric coefficient $f$, which vary from 0 (violet curves) to 8 V (red curves) with step size of 1 V for neighboring curves. The color maps of the $\delta p_A(k_1)$ and $\delta p_O(k_1)$ as a function of the wavenumber $k_1$ and flexoelectric coefficient $f$ calculated for $k_2 = 0$, effective mismatch strains $u_m^* = -0.5$ % **(c, e)** and +0.5 % **(d, f)**. Blank region in part (d) corresponds to $\delta p_A = 0$. Vertical dotted lines mark the $k$-range, where the thickness-

dependent corrections for $\delta p_A$ are significant. The temperature $T = 305$ K and $h = 100$ nm for all plots, material parameters of $CuInP_2S_6$ are listed in **Tables S1-S2** [61].

## 6. Discussion and Conclusions

Using the Landau-Ginzburg-Devonshire approach, in this work we reveal that the dispersion of soft phonons and ferrons is strongly dependent on the sign and magnitude of elastic strains originated from the lattice constants mismatch in thin strained films of van der Waals ferrielectric $CuInP_2S_6$.

In particular, the frequency of acoustic phonons and ferrons tends to zero at nonzero wavevectors $k > k_{cr}$, where the critical value of the wavevector $\vec{k}_{cr}$ is determined by the mismatch strain, temperature and flexocoupling strength. The critical value $\vec{k}_{cr}$, where $\omega_A(\vec{k}_{cr}) = 0$ and $\delta p_A(\vec{k}_{cr})$ diverges, obeys the following equation:

$$\left(\alpha_s + \vec{k}_{cr}\hat{g}\vec{k}_{cr}\right)\left(\vec{k}_{cr}^2\hat{v}\vec{k}_{cr}^2 + \vec{k}_{cr}\hat{c}\vec{k}_{cr}\right) - \left(\vec{k}_{cr}f\vec{k}_{cr}\right)^2 - 4P_s^2\left(\hat{q}\vec{k}_{cr} + 2P_s^2\hat{z}\vec{k}_{cr}\right)^2 = 0. \qquad (12)$$

In equation (12) the parameters $\alpha_s$ and $P_s$ depend on effective mismatch strains $u_m^*$ and temperature $T$. The third term in this equation is proportional to the flexoelectric coefficient $f$.

Considering $CuInP_2S_6$ films with the 2/m symmetry of the parent phase and putting either $k_2 = 0$ or $k_1 = 0$, the analytical expressions for $k_{cr}$ can be derived [38]. The nonzero solutions for $k_{cr}$ have the form:

$$k_{cr1,2}^2(u_m^*, f, T) = \frac{1}{2g_{55}v_{5151}}\left[f_t^2 \pm \sqrt{f_t^4 - 4g_{55}v_{5151}[c_{55}\alpha_s - 4P_s^2(q_{53} + 2P_s^2 z_{533})^2]}\right], \qquad (13a)$$

where $k_2 = 0$ and $f_t^2 = f^2 - v_{5151}\alpha_S - c_{55}g_{55}$, or

$$k_{cr1,2}^2(u_m^*, f, T) = \frac{1}{2g_{44}v_{4242}}\left[f_t^2 \pm \sqrt{f_t^4 - 4g_{44}v_{4242}c_{44}\alpha_s}\right], \qquad (13b)$$

where $k_1 = 0$ and $f_t^2 = f^2 - v_{4242}\alpha_s - c_{44}g_{44}$. Here we used Voigt notations. Both signs "–" (the root $k_{cr1}^2$) and "+" (the root $k_{cr2}^2$) in Eqs.(13) may have physical sense when $f_t^2 > 0$ and the determinant is positive and smaller than $f_t^2$. They difference of the roots, $k_{cr2} - k_{cr1}$, is the "gap" in the spectra of acoustic phonons and ferrons, where the dynamic instability of the spatially uniform phase may appear. The condition of zero determinant corresponds to $k_{cr2} = k_{cr1} = k_0^{cr}$.

The tensor components $v_{ijklnm}$ are very small, as a rule. Under the condition $v_{ijklmn} = 0$, the inequalities $f^2 > c_{55}g_{44}$ and/or $f^2 > c_{55}g_{55}$ should be valid for the appearance of a spatially modulated structure with the threshold period $k_{cr}$. Note that expressions $f^2 = c_{55}g_{44}$ and/or $f^2 = c_{55}g_{55}$ give the maximal values of the static flexoelectric effect coefficient (the "upper thermodynamic limit") established by Yudin et al. [35].

Zeroing of the acoustic phonon frequency, that appears at $\vec{k} = \vec{k}_{cr}$ with increase of tensile strains, indicates a dynamic instability of spatially uniform polar phase induced by the flexo-strain effects. It is seen from **Fig. 6** in which way $\vec{k}_{cr}$ depends on $u_m^*$ and $f$ at fixed $T$. Note that the continuum medium LGD approach is valid quantitatively for the wavelengths exceeding 5 – 10 lattice constants. However, the LGD approach appears valid semi-quantitatively for the wavelengths exceeding 1 – 2 lattice constants, because the correlation lengths, $l_{c1} = \sqrt{\frac{g_{55}}{\alpha_S}}$ and $l_{c2} = \sqrt{\frac{g_{44}}{\alpha_S}}$, are very small (one-half lattice constant or less) entire the actual range of mismatch strains and temperatures for the uniaxial ferrielectric $CuInP_2S_6$ due to anomalously large $\alpha_S$, except for the narrow vicinity of the second order PE-FI2 phase transition.

**Figures 6(a)** and **6(b)** correspond to a low temperature of 100 K; they are calculated for the sign "–" in Eqs. (13a)-(13b). For the temperatures well below the FI-PE transition, the sign "+" led to imaginary values of $k_{cr}^2$ in all the studied region, hence it is not shown in **Fig. 6(a)-(b)**. It is important to note, that from a physical standpoint the blank regions in **Fig. 6(a)-(b)** correspond to the purely imaginary $\vec{k}_{cr}$, i.e. it is the regions of "normal" spatially uniform phase. At low temperatures there is almost no anisotropy between the two k-directions in the inverse space. Indeed, the values at the contour maps in both directions are very similar, as well as the overall distribution of the wave vector, which is almost identical from the left ($k_1$ direction) to right ($k_2$ direction) panels.

**Figures 6(c)-6(f)** correspond to the higher temperature 305 K, which "cross" the FI1-PE-FI2 transition line in **Fig. 2(a)**. These panels are calculated for both signs "+" and "–" in Eqs. (13a)-(13b). The anisotropy between the two k-directions is relatively weak at 305 K for the solution with "–" in Eqs. (13a)-(13b). However, the anisotropy is quite significant the for the "+" sign in Eqs. (13a)-(13b). In particular, the values of the $k_{cr2}$ are slightly higher in the $k_2$-direction. Moreover, the behavior of the $k_{cr2}$ is principally different for compressive (left part) and tensile (right part) mismatch strains. If **Fig. 6(f)**, at approximately zero mismatch strain and at low flexocoefficient values, a blank "split" appears, which loosely corresponds, both temperature and mismatch-induced, to the triple point in **Fig. 2(a)**. Also, it is quite interesting that only the sign "+" gives such behavior. This begs for the question which sign, exactly, gives the correct solutions at higher temperatures.

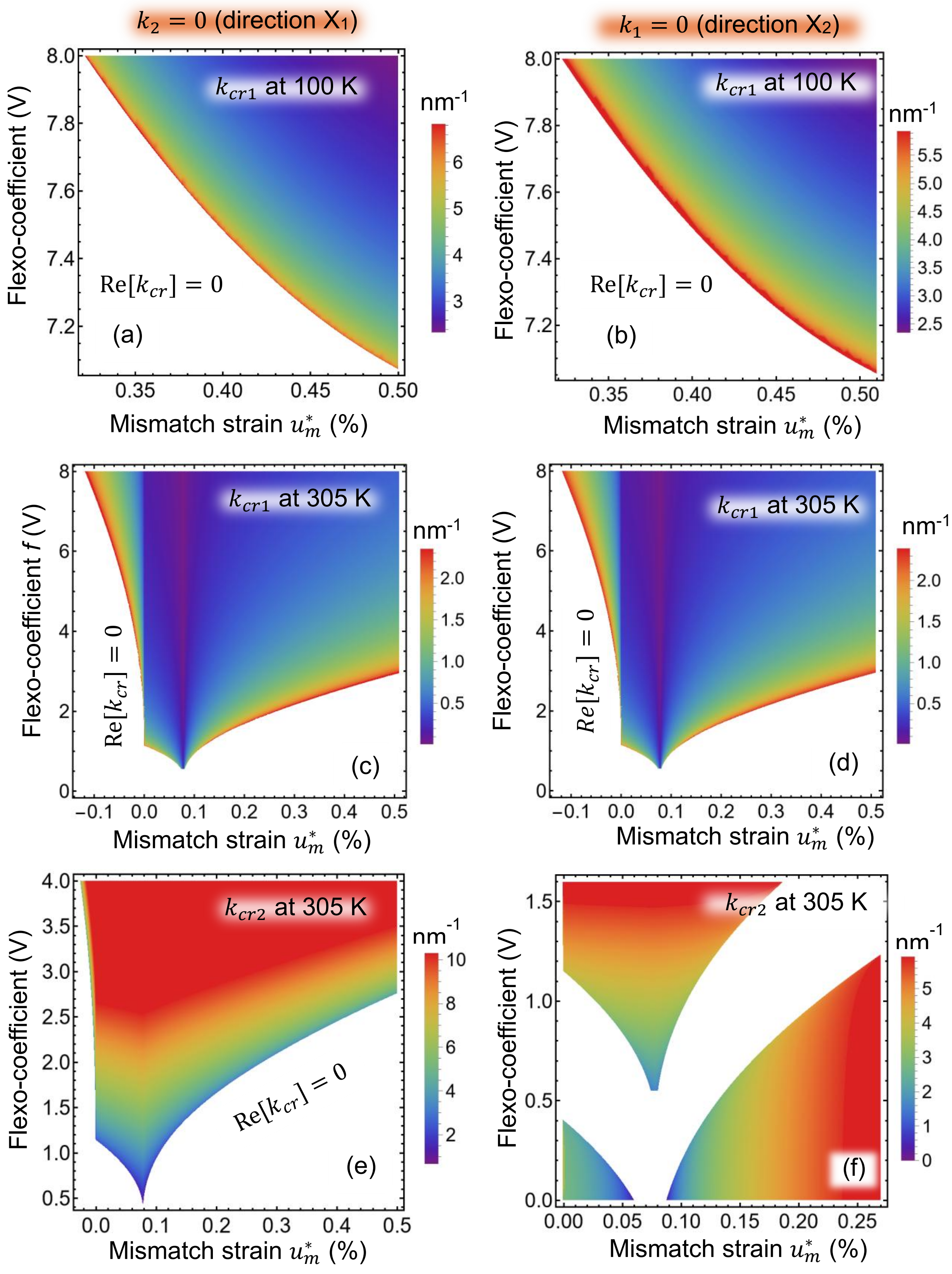

**FIGURE 6.** The critical wavevector $\vec{k}_{cr}$ as the function of effective mismatch strain $u_m^*$ and flexoelectric coefficient $f$ calculated for temperature $T = 100$ K **(a, b)** and 305 K **(c, d, e, f)**, $k_2 = 0$ **(a, c, e)** and $k_1 = 0$ **(b, d, f)**. Blank regions correspond to the purely imaginary $\vec{k}_{cr}$. The limits for $\vec{k}_{cr}$ correspond to the reciprocal lattice constants, which are about $\frac{2\pi}{0.61\text{ nm}} \approx 10\text{ nm}^{-1}$ in the direction $X_1$ and about $\frac{2\pi}{1.06\text{ nm}} \approx 6\text{ nm}^{-1}$ the direction $X_2$. The damping is absent ($\Gamma = 0, \Lambda = 0$) for all plots; material parameters of $CuInP_2S_6$ are listed in **Tables S1-**

**S2** [61].

Note that the predicted features and dynamic instabilities of phonon and ferron dispersion maybe sensitive to the polarization gradient energy and elastic gradient energy coefficients, which order of magnitude is reliably determined from the experimentally observed uncharged domain wall width in $CuInP_2S_6$ [1, 2, 7], but their anisotropy is estimated rather than rigorously determined. However, we would like to underline that all numerical results, presented in the work, are weakly dependent on the anisotropy of the gradient energies, because the results depend on the tensorial convolutions like $\hat{g}\vec{k}^2$ and $\hat{v}\vec{k}^4$. One should change the anisotropy in order of magnitude to achieve noticeable changes in results. Dispersion curves are slightly more sensitive to the magnitude of dynamic flexoelectric constant $M$ (that was determined reliably for $Sn_2P_2S_6$ [1]), however the position of the instability $\omega_A(\vec{k}_{cr}) = 0$ is $M$-independent according to Eq. (12). Obtained results, including the dynamic instability boundary of the acoustic mode, are sensitive to the magnitude of the static flexocoupling constant $f_{44} = f_{55} = f$; and that is why we consider it as a parameter of the theory. The shifts of the acoustic mode instability boundary as a function of the $f$ can be estimated from **Fig. S4** in **Suppl. Materials** [61].

Next interesting question is the possible appearance of incommensurate spatially modulated (ISM) phases in ferroics with intertwined polar and elastic order. It has been shown earlier that the ISM phase emerges and becomes energetically stable when the flexoelectric coupling exceeds the critical value, namely $|f| > f_{cr}$ [37, 66]. Corresponding phase diagram with stable and metastable ISM phases, and with a Lifshitz point is presented in Ref. [66] for a "scalar" 1D case (see Fig. 1(a) therein). Thus, we can suspect that the appearance of ISM phases in $CuInP_2S_6$ films with increase in the flexocoupling strength. Indeed, the dynamic instability of the 2D excitations at $\vec{k} = \vec{k}_{cr}$ may also indicate a possibility of the commensurate-incommensurate phase transition at 305 K for both compressive and tensile strains, which manifests itself most prominently in the $k_2$ direction. In contrast, the dynamic instability may appear in the $k_1$-direction too, but mostly at tensile strains. The increase in the flexocoupling strength makes the commensurate-incommensurate transition possible for small compressive strains in the $k_1$-direction.

To explore the possibility of ISM phase appearance and stability, phase diagrams with stable and metastable ISM states, as the function of temperature and flexoelectric coefficient, are shown in **Fig. 7**. The transition from the paraelectric (PE) phase into the ISM state with polarization $P_3(x_1, x_2) = \delta P_1\, sin(\vec{k}_1^{cr}\vec{x}) + \delta P_2\, sin(\vec{k}_2^{cr}\vec{x})$ occurs at the critical values of the flexo-coefficient $f = f_{cr}$. In the considered 2D case the $f_{cr}$ values are given by expression

$$f_{cr}^2 = v_{5151}\alpha_S + c_{55}g_{55} + 2\sqrt{g_{55}v_{5151}[c_{55}\alpha_s - 4P_s^2(q_{53} + 2P_s^2 z_{533})^2]} \qquad (14a)$$

for the spatial modulation in the direction $X_1$. Thus, $f_{cr}^2 = \left(\sqrt{v_{5151}\alpha_s} + \sqrt{c_{55}g_{55}}\right)^2$ in the PE phase with $P_s^2 = 0$.

For the spatial modulation in the direction $X_2$.

$$f_{cr}^2 = v_{4422}\alpha_s + c_{44}g_{44} + 2\sqrt{g_{44}v_{4242}c_{44}\alpha_s} = \left(\sqrt{v_{4242}\alpha_s} + \sqrt{c_{44}g_{44}}\right)^2 \tag{14b}$$

The function $\alpha_s$ in temperature and mismatch dependent according to Eqs. (8), namely

$$\alpha_s(T, u_m^*) = \alpha(T) - 2u_m^*\left(q_{13} + q_{23} - \frac{c_{13}+c_{23}}{c_{33}}q_{33}\right) + 3\beta^* P_s^2 + 5\gamma^* P_s^4 + 7\delta^* P_s^6. \tag{14c}$$

In the considered case of negligibly small elastic anisotropy ($c_{44} \approx c_{55}$) and equal gradient coefficients ($g_{44} = g_{55}$ and $v_{4422} = v_{5511}$) the PE-ISM boundary is independent of the k-direction, but mismatch-dependent. In the anisotropic case ($c_{44} \neq c_{55}$ and $g_{44} \neq g_{55}$) the transition temperature $T_{IC}$ from the PE to the ISM phase is given by expressions

$$T_{IC} = \begin{cases} T_C + \theta(|f| - f_{cr})^2, & f^2 > f_{cr}^2, \\ \text{absent}, & f^2 < f_{cr}^2, \end{cases} \tag{15}$$

where the parameter $\theta$ and the value of $f_{cr}$ depend on the spatial direction of incommensurate modulation. Under the condition $P_s^2 = 0$, that is valid in the PE phase and at the onset of the ISM phase, the coefficient $\alpha_s = \alpha_T(T_{IC} - T_C)$ and so the critical value is $k_0^{cr} \sim (T_{IC} - T_C)^{1/4}$.

The ISM phase transforms gradually to the homogeneous ferrielectric (FI) phases, namely to FI1 or FI2, or their mixture depending on $u_m^*$ and $T$ values, with decrease in temperature below $T_C$. The coexistence boundary is virtually independent on the anisotropy too, because the term $4P_s^2(q_{53} + 2P_s^2 z_{533})^2$ is negligibly small in comparison with $c_{55}\alpha_s$. Three phases, PE, FI and ISM, coexist in the Lifshitz point (LP). The ISM phase coexists with FI phases at $T < T_{IC}$ and $|f| > f_{cr}$.

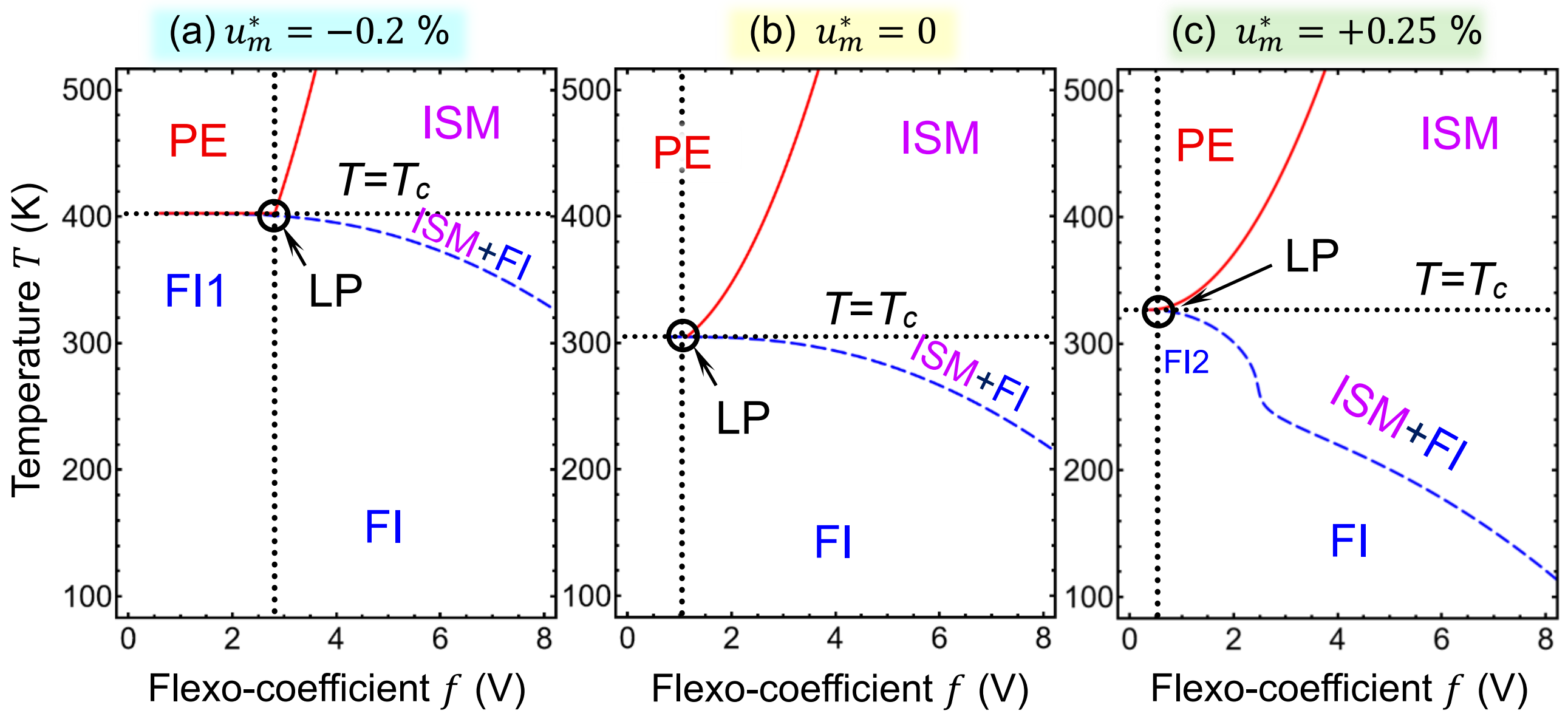

**FIGURE 7.** Phase diagrams as a function of temperature $T$ and absolute value of the flexoelectric coefficient $f$. The diagrams are calculated for $u_m^* = -0.20$ % **(a)**, $u_m^* = 0$ **(b)**, $u_m^* = +0.25$ % **(c)**, and zero damping ($\Gamma = 0, \Lambda = 0$). The FI phase can be FI1 or FI2 states, or their mixture depending on $u_m^*$ and $T$ values. Material

parameters of $CuInP_2S_6$ are listed in **Tables S1-S2** [61].

The equilibrium wave vector $\vec{k}_{eq}$ coincides with the critical value $\vec{k}_{cr}$ at the onset of the ISM phase ($T \approx T_{IC}$), or very close to $\vec{k}_{cr}$ in the so-called "shallow" ISM phase, where the spatial modulation is quasi-harmonic at temperatures slightly lower than $T_{IC}$ [66]. An equilibrium wave vector $\vec{k}_{eq}$ should be calculated numerically in the "deep" SMP (i.e., at temperatures $T_C \leq T \ll T_{IC}$), where the modulation profile becomes strongly anharmonic. However, the magnitude of $\vec{k}_{eq}$ depends on the lateral size of the computation cell (see e.g., results for $S_2P_2S_6$ films in Ref. [67]), and so numerical calculations of $\vec{k}_{eq}$ in the "deep" SMP are not reliable. Only the values of $\vec{k}_{cr}$ are shown in **Fig. 6.** It is necessary to include higher-order terms in the free energy and to evaluate the anisotropy of the gradient and electrostriction tensors for more rigorous determination of the ISM states properties.

Generally speaking, a spatially modulated incommensurate phase may appear in commensurate ferroelectrics, when the frequency of the acoustic phonons becomes zero [66, 68]. Recently, Orenstein et al. [69] observed polarization density waves in $SrTiO_3$ and related their origin to the flexoelectric coupling of ferrons and other modes. Important, that characteristic features of an incommensurate spatially modulated polar phase have been revealed in $CuInP_2S_6$ by Rao et al. [70], who concluded that the incommensurate structure observed at pressures between 12 and 17 GPa is "possibly caused by slight changes in the stacking between layers as well as the continuing decrease in layer spacing due to the compression". As follows from results of this work, the temperature-dependent incommensurate transition may be possible in thin strained $CuInP_2S_6$ films.

According to the analysis of eigenvectors, under the linear coupling of optical and acoustic modes of the same symmetry, not only the repulsion of their frequencies occurs in the collision point, but also the exchange of the shape of the eigenvectors. In result, the dependences of linearly coupled optical and acoustic phonon frequency on the mismatch strain and temperature correlate with the dependence of spontaneous polarization, reflecting the fact that the phonon dispersion is determined by the mismatch-induced phase transitions between the FI1, PE and FI2 phases.

The acoustic phonon frequency is a weakly anisotropic function of the wavevector components at lower temperatures, while significant anisotropy appears with increase in temperature. The frequency of soft optical phonon is almost isotropic function of the wavevector components entire studied wavevector and temperature range. The boundary between large and small values of phonon frequency and ferron amplitude becomes much more diffuse with the increase in the flexocoupling strength. Thus, analytical results, derived in this work, open the way for flexo-strain engineering of soft optical and acoustic phonons and ferrons in thin films of van der Waals ferrielectrics.

**Authors' contribution.** A.N.M. and J.-M.H. generated the research idea. A.N.M. formulated the problem, performed analytical calculations of the phonon and ferron dispersion, and wrote the manuscript draft. E.A.E., Y.Z. and J-M.H. performed calculations of the dynamic depolarization field. E.A.E. and M.Y. wrote the codes and prepared figures. Y.M.V., V.G., L.-Q. C. and J-M.H. worked on the analysis of results and manuscript improvement.

**Acknowledgements.** Authors are very grateful to the Referees for very useful remarks and suggestions. The work is primary supported by the DOE Software Project on "Computational Mesoscale Science and Open Software for Quantum Materials", under Award Number DE-SC0020145 as part of the Computational Materials Sciences Program of US Department of Energy, Office of Science, Basic Energy Sciences. The work of A.N.M. is partially supported by the Target Program of the National Academy of Sciences of Ukraine, Projects No. 5.8/26-П "Energy-saving and environmentally friendly nanoscale ferroics for the development of sensorics, nanoelectronics and spintronics" and 1.4.B/222. The work of E.A.E. is partially supported by the National Academy of Sciences of Ukraine, III-6-26 "Innovative ferroelectric nanomaterials based on silicon-compatible oxides and nitrides of the rare earth and transition metals for strategic requirements of nanoelectronics". Y.M.V. and A.N.M. also acknowledge support from the Horizon Europe Framework Programme (HORIZON-TMA-MSCA-SE), project № 101131229, Piezoelectricity in 2D-materials: materials, modeling, and applications (PIEZO 2D). Results were visualized in Mathematica 15.0 [71].

# SUPPLEMENTAL MATERIALS to
# "Flexo-Strain Engineering of Phonons and Ferrons in Thin Films of Van der Waals Ferrielectrics"

Anna N. Morozovska[1], Eugene A. Eliseev[2], Yujie Zhu[3], Mykola Yelisieiev[4,1], Hanna V. Shevliakova[5], Yulian M. Vysochanskii[6**], Venkatraman Gopalan[7††], Long-Qing Chen[7‡‡], and Jia-Mian Hu[3§§]

[1] Institute of Physics, National Academy of Sciences of Ukraine, 46, pr. Nauky, 03028 Kyiv, Ukraine
[2] Frantsevich Institute for Problems in Materials Science, National Academy of Sciences of Ukraine, Omeliana Pritsaka str., 3, Kyiv, 03142, Ukraine
[3] Department of Materials Science and Engineering, University of Wisconsin-Madison, Madison, WI, 53706, USA
[4] V. Lashkaryov Institute of Semiconductor Physics, National Academy of Sciences of Ukraine, 41, pr. Nauky, 03028 Kyiv, Ukraine
[5] Department of Microelectronics, Igor Sikorsky Kyiv Polytechnic Institute, Kyiv, Ukraine
[6] Institute of Solid-State Physics and Chemistry, Uzhhorod University, 88000 Uzhhorod, Ukraine
[7] Department of Materials Science and Engineering, Pennsylvania State University, University Park, PA 16802, USA

## SUPPLEMENT S1. The "Static" and "Dynamic" Contributions to the Depolarization Field

### S1.A. Calculation details

The influence of the retardation on the internal electric field accompanying the optical phonons/ferrons coupled with the photons. To illustrate this, let us consider Maxwell's equations for the electric field $\vec{E}$, electric displacement $\vec{D}$, magnetic field $\vec{H}$ and magnetic induction $\vec{B}$:

$$\mathrm{div}\,\vec{D} = 0, \quad \mathrm{rot}\,\vec{E} = -\frac{\partial \vec{B}}{\partial t}, \quad \mathrm{div}\,\vec{B} = 0, \quad \mathrm{rot}\,\vec{H} = \frac{\partial \vec{D}}{\partial t}. \tag{S.1}$$

Hereinafter we consider the absence of free charges and the electric currents. Conventional material equations of state have the form

$$\vec{D} = \varepsilon_0 \varepsilon_b \vec{E} + \vec{P}, \qquad \vec{B} = \chi_0 \chi \vec{H}. \tag{S.2}$$

** corresponding author, e-mail: vysochanskii@gmail.com
†† corresponding author, e-mail: vxg8@psu.edu
‡‡ corresponding author, e-mail: lqc3@psu.edu
§§ corresponding author, e-mail: jhu238@wisc.edu

Here $\varepsilon_0$ and $\chi_0$ are the universal dielectric and magnetic constants respectively, $\varepsilon_b$ and $\chi$ are the background dielectric permittivity and relatively magnetic permeability, respectively, and $\vec{P}$ is the polarization as a electric dipole moment of the unit volume. Below we regard that the polarization vector $\vec{P}$ is a given function of the time and spatial coordinates.

Using the curl-operator from both sides of Eq.(S.1) and considering Eq.(S.2), one could easily get the following

$$\mathrm{rot}\,\mathrm{rot}\,\vec{E} = -\chi_0\chi\frac{\partial}{\partial t}\left(\mathrm{rot}\,\vec{H}\right). \tag{S.3}$$

Next, we recall Eq. (S.1d) to get the equation

$$\mathrm{rot}\,\mathrm{rot}\,\vec{E} = -\chi_0\chi\frac{\partial^2}{\partial t^2}\vec{D} \tag{S.4a}$$

It should be noted here that in the general case of the continuous vector field (see e.g., Ref. [72]):

$$\mathrm{rot}\,\mathrm{rot}\,\vec{E} \equiv \mathbf{\nabla}\mathrm{div}\,\vec{E} - \Delta\,\vec{E}, \tag{S.4b}$$

and therefore, one could see from Eqs. (S.4a) and (S.4b):

$$\mathbf{\nabla}\mathrm{div}\,\vec{E} - \Delta\,\vec{E} + \chi_0\chi\frac{\partial^2}{\partial t^2}\vec{D} = 0. \tag{S.4c}$$

Combining Eq.(S.1a) and (S.2a) one gets the following

$$\mathrm{div}\,\vec{E} = -\frac{1}{\varepsilon_0\varepsilon_b}\mathrm{div}\,\vec{P}. \tag{S.5a}$$

Combining Eqs.(S.1), (S.2) and (S.5) one could get the following

$$-\frac{1}{\varepsilon_0\varepsilon_b}\mathbf{\nabla}\mathrm{div}\,\vec{P} - \Delta\,\vec{E} + \chi_0\chi\frac{\partial^2}{\partial t^2}\left(\varepsilon_0\varepsilon_b\vec{E} + \vec{P}\right) = 0 \tag{S.5b}$$

Next, recombing the terms in Eq.(S.5b) one has the equation relating the internal electric field to polarization

$$\Delta\,\vec{E} - \frac{\chi\varepsilon_b}{c_0{}^2}\frac{\partial^2}{\partial t^2}\vec{E} = -\frac{1}{\varepsilon_0\varepsilon_b}\mathbf{\nabla}\mathrm{div}\,\vec{P} + \chi_0\chi\frac{\partial^2}{\partial t^2}\vec{P}. \tag{S.6}$$

Here we introduce the speed of light in vacuum $c_0 = 1/\sqrt{\chi_0\varepsilon_0}$. It is seen that the left-hand side of the Eq.(S.6) is the wave equation, describing the electromagnetic waves travelling with the speed $c_0/\sqrt{\chi\varepsilon_b}$ ("undressed" waves), while the right-hand side represents either the possible source or the disturbance of such waves.

Now let us consider the harmonic waves, i.e. the solution in the form of

$$\vec{E} = \vec{E}_{\boldsymbol{k}}\exp\left(i\vec{k}\vec{x} - i\omega t\right) \tag{S.7a}$$

caused by the polarization wave (phonon or ferron)

$$\vec{P} = \vec{P}_{\boldsymbol{k}}\exp\left(i\vec{k}\vec{x} - i\omega t\right) \tag{S.7b}$$

Using the expressions (S.7) as an ansatz and neglecting "undressed" waves, for which $k^2 = \frac{\chi\varepsilon_b}{c_0^2}\omega^2$, we could get from Eq.(S.6)

$$\vec{E}_{\boldsymbol{k}} = -\frac{\vec{k}\,(\vec{k}\cdot\vec{P}_{\boldsymbol{k}}) - \frac{\chi\varepsilon_b}{c^2}\omega^2\vec{P}_{\boldsymbol{k}}}{\varepsilon_0\varepsilon_b\left(k^2 - \frac{\chi\varepsilon_b}{c^2}\omega^2\right)} \equiv -\frac{\vec{k}\,(\vec{k}\cdot\vec{P}_{\boldsymbol{k}}) - \vec{P}_{\boldsymbol{k}}k^2}{\varepsilon_0\varepsilon_b\left(k^2 - \frac{\chi\varepsilon_b}{c^2}\omega^2\right)} - \frac{\vec{P}_{\boldsymbol{k}}}{\varepsilon_0\varepsilon_b}. \tag{S.8}$$

The solution (S.8) represents so-called internal or "depolarization” electric field. It is seen that one could hardly separate the static ($\omega = 0$) and dynamic ($\omega \neq 0$) contributions to the depolarization field. The expression (S.8) is valid for the infinitely large system. Otherwise, the exact geometrical shape of the system as well as the electric boundary conditions should be considered.

Let us consider the TO-mode with polarization

$$\vec{P} = \vec{e}_1 \exp(i\omega t - ik_2 x_2) P_1(\omega), \tag{S.9}$$

which electric field satisfies the equation, derived from Eq.(S.6):

$$\frac{\partial^2}{\partial x_3^2} E_1 + \left(\frac{\omega^2}{c_b^2} - k_2^2\right) E_1 = -\frac{1}{\varepsilon_0 \varepsilon_b} \frac{\omega^2}{c_b^2} P_1. \tag{S.10}$$

The boundary conditions corresponding to the ideally conducting electrodes (see e.g., Ref. [73]):

$$E_1|_{x_3=\pm h/2} = 0. \tag{S.11}$$

Here $c_b = 1/\sqrt{\chi_0 \varepsilon_0 \chi \varepsilon_b}$ is the speed of light in the material. The solution of the boundary problem (S.10)-(S.11) is [72]:

$$E_1 = \frac{\omega^2}{c_b^2 k_2^2 - \omega^2} \left[1 - \frac{\cosh\left(\tilde{k}(h/2 - x_3)\right)}{\cosh(\tilde{k}h/2)}\right] \frac{P_1(\omega)}{\varepsilon_0 \varepsilon_b}. \tag{S.12}$$

Here we introduced the effective wave vector amplitude $\tilde{k}^2 = k_2^2 - \frac{\omega^2}{c_b^2}$. Thus, the solution represents the pure case of “dynamic” depolarization internal electric field, associated with the TO wave, propagating in the “electroded” layer. This field is caused by the retardation/radiation effects and vanishes in the static limit $\omega \to 0$. Note that free charges on the electrodes are unable to suppress the field (S.12) in the bulk of the layer. In other words, the “pseudo-dynamic” contribution $\sim \frac{\chi \varepsilon_b}{c_0^2} \omega^2 \vec{P}_{\boldsymbol{k}}$ is always present in the depolarization field (regardless of the wave vector orientation) and could hardly be screened by the image charges in the conducting electrodes. The pseudo-dynamic contribution is usually regarded as a small correction due to the relativistic factor $\omega^2/c_b^2$, but it becomes important at optical frequency range.

### S1.B. Qualitative Evaluation of Dynamic Depolarization Field

We further note that the ferroelectric soft mode in $CuInP_2S_6$ can be approximately considered as a transverse optical (TO) phonon, represented by $\delta P_3(k_1,\omega)$ or $\delta P_3(k_2,\omega)$, where where $k_1$ and $k_2$ are the wavenumbers along the $x_1$ ($X_2$ in **Fig. 1**) and $x_2$ ($X_1$ in **Fig. 1**) axis, respectively, and the ‘$\delta$” quantifies the change with respect to $P_3$ at the initial equilibrium state. Below we will show that these TO phonons would not cause significant out-of-plane dynamic depolarization field, i.e., $\delta E_3^{\mathrm{d}}(t) \approx 0$.

Specifically, since we are considering the mode with $k_3 = 0$ in this work (see **Scheme S1**), the dynamical polarization oscillation can be written as $\delta\vec{P} = \left(0,0,\delta P_3^0 e^{\mathrm{i}(k_1 x_1 + k_2 x_2 - \omega t)}\right)$. These TO phonons, $\delta P_3(k_1,\omega)$ or $\delta P_3(k_2,\omega)$, do not induce a variation in the volume bound charge density, i.e., $\delta\rho^{\mathrm{b}} = -\nabla\cdot\delta\vec{P} = -\frac{\partial}{\partial x_3}\left[\delta P_3^0 e^{\mathrm{i}(k_1x_1+k_2x_2-\omega t)}\right] = 0$.

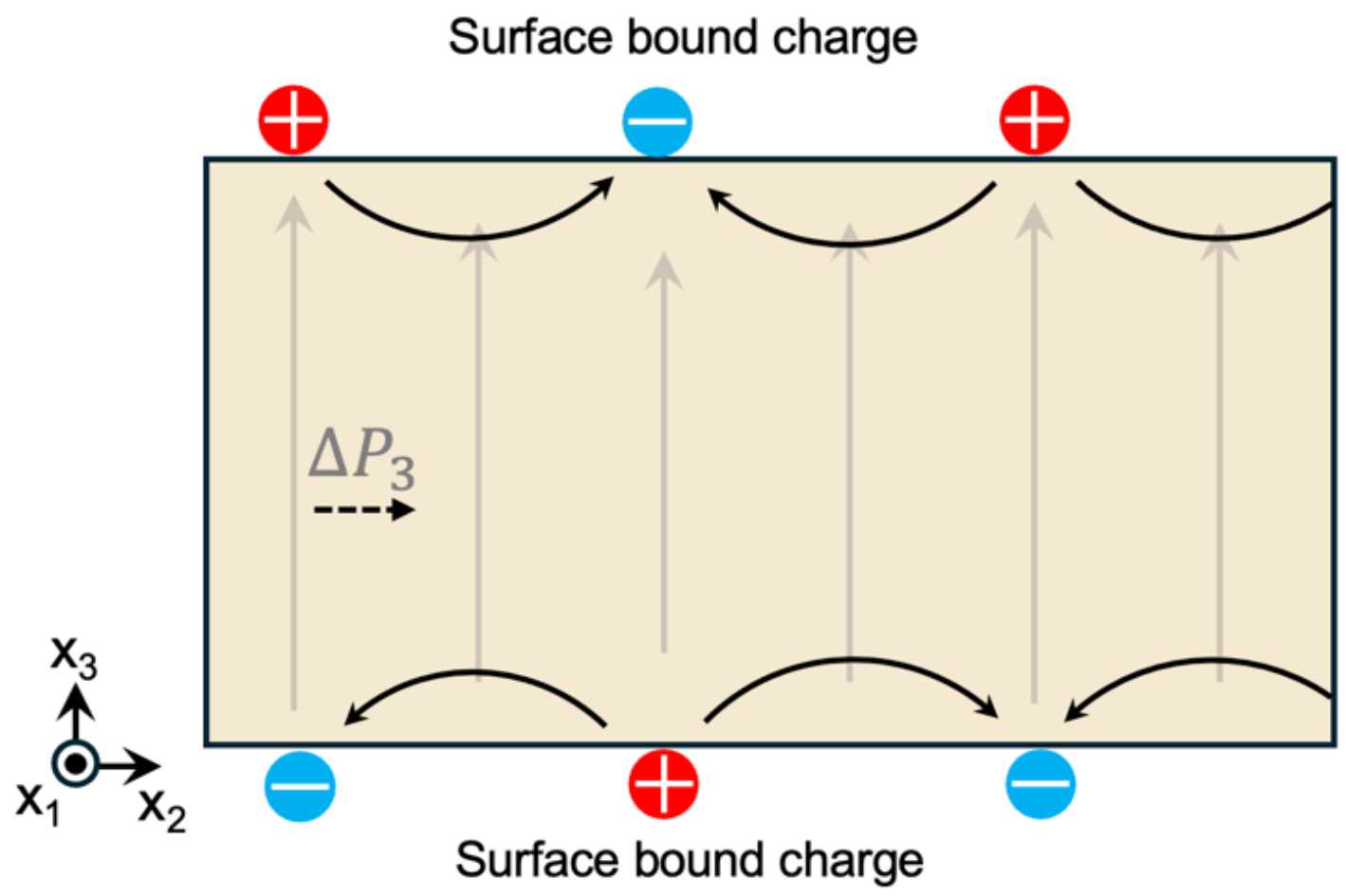

**Scheme S1.** Schematics of the transverse polarization wave $\delta P_3(x_2,t)$, indicated by the vertical gray arrows, where the length of arrows indicate the magnitude of the local $P_3$ at this specific moment; the local dynamical surface bound charge densities, where the positive and negative signs indicate, respectively, an increase and decrease as compared to those at the initial equilibrium state; and the local dynamical depolarization fields, which are confined spatially near the surfaces and have relatively weak components along the $x_3$ axis, as indicated by curved black arrows. The directions of these local fields are reversing periodically. The amplitudes of these local fields decay exponentially along the $x_3$ axis as well.

These TO phonons, however, do induce a dynamic in-plane spatial variation in the surface bound charge density at the top and bottom surfaces of the $CuInP_2S_6$ nanomembrane, which can be written as $\delta\sigma^{\mathrm{b}} = \delta\sigma^{\mathrm{b},0} e^{\mathrm{i}(k_1x_1+k_2x_2-\omega t)}$. In $CuInP_2S_6$, according to Ref. [74], the wavenumber for the TO phonon $k_1$ varies approximately from $2\pi\times0.25$ rad nm$^{-1}$ to $2\pi\times0.35$ rad nm$^{-1}$ within the angular frequency range of $2\pi\times74.8$ rad GHz to $2\pi\times47.8$ rad GHz, corresponding to a wavelength of about 4 nm-2.86 nm. These wavelengths are significantly smaller than the thickness of $CuInP_2S_6$ films considered within continuum LGD approach. Under this condition, as sketched in **Scheme S1**, the surface bound charges tend to form dipolar pairs at the same surface. The resultant dynamic depolarization field is confined near the surface with primarily in-plane components, and decays exponentially into the bulk region of the membrane. It is therefore reasonable to assume $\delta E_3^{\mathrm{d}}(t) \approx 0$ and that the thickness average of $\delta E_1^{\mathrm{d}}(t)$ and $\delta E_2^{\mathrm{d}}(t)$ are negligible.

## SUPPLEMENT S2. The Lagrange Function

Paraelectric and ferroelectric phases of $CuInP_2S_6$ have the point symmetry $2/m$ and $m$, respectively. It was shown that in V-d-W ferroelectric $CuInP_2S_6$, the axis 2 lies in the plane of layers, with spontaneous polarization pointed along the normal vector to the layers. The coordinate system is organized as follows, the coordinate axis "$X_2$" is along the symmetry axis "2", the coordinate axis "$X_3$" is along the normal vector of all the layers, while the coordinate axis "$X_1$" is perpendicular to "$X_3$" and $X_2$". Note that the mirror plane (the only symmetry element of the ferroelectric phase) is parallel to the axes "$X_1$" and "$X_3$".

According to the symmetry principle the Landau free energy should be invariant with respect to the symmetry transformations of the paraelectric phase, so that the non-zero tensorial elements may have the even number of indices "2" and/or odd number of "1" and "3", but the total number of both "1" and "3" should be even also. Hereinafter we regard that $CuInP_2S_6$ is a uniaxial ferroelectric, which polar axis is "3", and so consider the coupling between the polarization component $P_3$, elastic displacement component $U_3$, and corresponding strains $u_3$, $u_4$ and $u_5$ (in Voigt notations).

Using LGD theory and scalar approximation in the considered one-component 1D case, the Lagrange function $L = \int_t dt \int_{-\infty}^{\infty} dx\,(F - K)$ consists of the kinetic energy K and free energy $F$ of ferroelectric. The density of kinetic energy,

$$K = \frac{\mu}{2}\left(\frac{\partial P_3}{\partial t}\right)^2 + M\frac{\partial P_3}{\partial t}\frac{\partial U_3}{\partial t} + \frac{\rho}{2}\left(\frac{\partial U_3}{\partial t}\right)^2, \tag{S.13}$$

includes the dynamic flexocoupling with the magnitude $M$; $\rho$ is the mass density of a material; the coefficient $\mu$ is the polarization inertia, which can be expressed via the vacuum dielectric constant $\varepsilon_0$ and the plasma frequency $\omega_p$ as $\mu = \frac{1}{\varepsilon_0\omega_p^2}$.

The bulk density of the free energy $F$ that depends on polarization component $P_3$ and strain component $u$, and their gradients, has the following form:

$$F = \frac{\alpha}{2}P_3^2 + \frac{\beta}{4}P_3^4 + \frac{\gamma}{6}P_3^6 + \frac{\delta}{8}P_3^8 + \frac{1}{2}g_{3i3j}\frac{\partial P_3}{\partial x_i}\frac{\partial P_3}{\partial x_j} - P_3E_3^e - \frac{1}{2}P_3E_3^d - q_{ij33}u_{ij}P_3^2 - z_{ij3333}u_{ij}P_3^4 +$$
$$\frac{f_{ij3k}}{2}\left(u_{ij}\frac{\partial P_3}{\partial x_k} - P_3\frac{\partial u_{ij}}{\partial x_k}\right) + \frac{c_{ijkl}}{2}u_{ij}u_{kl} + \frac{v_{ijklmn}}{2}\left(\frac{\partial u_{ij}}{\partial x_k}\right)\left(\frac{\partial u_{lm}}{\partial x_n}\right) - N_iU_i \tag{S.14}$$

Note that $q_{43} \equiv 0$ and $z_{433} \equiv 0$ for the considered symmetry group. According to Landau theory, the coefficient $\alpha$ linearly depends on the temperature $T$ for proper ferroelectrics, $\alpha(T) = \alpha_T(T - T_C)$, which is valid well above quantum temperatures. The Barret-type expression, $\alpha(T) = \alpha_T T_q\left(\coth\frac{T_q}{T} - \coth\frac{T_q}{T_C}\right)$, where $T_C$ is the Curie temperature and $T_q$ is the quantum vibration temperature, is valid in a wide temperature range (from low to high temperatures). All other coefficients in Eq.(S.14) are supposed to be temperature independent. The coefficient $\delta \geq 0$ for the stability of the free energy for all $P$ values. The gradient coefficients $g_{ij}$ determines the magnitude of the gradient energy. Coefficients $f_{ij}$ are the components of the static flexocoupling tensor. Coefficients $c_{ii}$ are elastic stiffness. The coefficients $q_{ij}$ and $z_{ijk}$ are second-order and higher-order electrostriction coupling coefficients, respectively. $N_3$ is z-component of the external mechanical force bulk density; $E_3^{ext}$ is z-component of external electric field. Note that the longitudinal fluctuations of polarization are much smaller due to the depolarization field $E_3^d$, which contribution to the free energy is given by the term $\frac{P_3 E_3^d}{2}$.

## SUPPLEMENT S3. Bulk-like excitations of polar and elastic orders with strain-renormalized parameters

Considering the Tani mechanism and Khalatnikov relaxation, the explicit form of the LGD-KT equations, $\delta L/\delta U_i = 0$ and $\delta L/\delta P_3 = 0$, is:

$$v_{3ij3kl}\frac{\partial^4 U_3}{\partial x_i \partial x_j \partial x_k \partial x_l} + \rho\frac{\partial^2 U_i}{\partial t^2} + \Lambda\frac{\partial U_i}{\partial t} - c_{ijkl}\frac{\partial^2 U_k}{\partial x_j \partial x_l} - f_{ijkl}\frac{\partial^2 P_k}{\partial x_j \partial x_l} + q_{ijkl}\frac{\partial (P_k P_l)}{\partial x_j} + z_{ijklmn}\frac{\partial (P_k P_l P_m P_n)}{\partial x_j} + M\frac{\partial^2 P_i}{\partial t^2} = N_i. \qquad \text{(S.15a)}$$

$$\mu\frac{\partial^2 P_3}{\partial t^2} + \Gamma\frac{\partial P_3}{\partial t} + \alpha P_3 + \beta P_3^3 + \gamma P_3^5 + \delta P_3^7 - g_{3jml}\frac{\partial^2 P_m}{\partial x_j \partial x_l} - f_{kl3j}\frac{\partial U_k}{\partial x_j \partial x_l} - 2q_{ij33}\frac{\partial U_i}{\partial x_j}P_3 - 4z_{ij3333}\frac{\partial U_i}{\partial x_j}P_3^3 + M\frac{\partial^2 U_3}{\partial t^2} = E_3, \qquad \text{(S.15b)}$$

$$\alpha_{22}P_2 - g_{2jml}\frac{\partial^2 P_m}{\partial x_j \partial x_l} - f_{kl2j}\frac{\partial U_k}{\partial x_j \partial x_l} = E_2, \qquad \text{(S.15c)}$$

$$\alpha_{11}P_1 - g_{1jml}\frac{\partial^2 P_m}{\partial x_j \partial x_l} - f_{kl1j}\frac{\partial U_k}{\partial x_j \partial x_l} = E_1. \qquad \text{(S.15d)}$$

where $\Gamma$ and $\Lambda$ are phenomenological damping constants.

Next, let's use the 3D Fourier integral expansions for polarization $P_3$, displacement $U_3$, perturbation electric field $E_3$ and mechanical force density $N_3$:

$$P_3 = P_s + \int d\,\omega \int d^3\vec{k}\, exp\left(i\vec{k}\vec{x} - i\omega t\right)\tilde{P}, \quad U_3 = u_{s3j}x_j + \int d\,\omega \int d^3\vec{k}\, exp\left(i\vec{k}\vec{x} - i\omega t\right)\tilde{U}, \qquad \text{(S.16a)}$$

$$E_3 = \int d\,\omega \int d^3\vec{k}\, exp\left(i\vec{k}\vec{x} - i\omega t\right)\tilde{E}, \qquad N_3 = \int d\,\omega \int d^3\vec{k}\, exp\left(i\vec{k}\vec{x} - i\omega t\right)\tilde{N}. \qquad \text{(S.16b)}$$

In the first approximation, the Fourier representation of linearized Eqs.(S.15) has the form:

$$\left(\hat{v}\vec{k}^4 + \hat{c}\vec{k}^2 - i\Lambda\omega - \rho\omega^2\right)\tilde{U} + \left(\hat{f}\vec{k}^2 - M\omega^2\right)\tilde{P} + 2iP_s\left(\hat{q}\vec{k} + 2P_s^2\hat{z}\vec{k}\right)\tilde{P} = \tilde{N}, \qquad \text{(S.17a)}$$

$$\left(\alpha_S + \hat{g}\vec{k}^2 - i\Gamma\omega - \mu\omega^2 + \frac{k_3^2 - \frac{\chi\varepsilon_b}{c^2}\omega^2}{\varepsilon_0\varepsilon_b\left(k^2 - \frac{\chi\varepsilon_b}{c^2}\omega^2\right)}\right)\tilde{P} + \left(\hat{f}\vec{k}^2 - M\omega^2\right)\tilde{U} - 2iP_s\left(\hat{q}\vec{k} + 2P_s^2\hat{z}\vec{k}\right)\tilde{U} = \tilde{E}. \qquad \text{(S.17b)}$$

Hereinafter we regard that $\vec{k}\cdot\vec{P}_k = 0$ and $\vec{k} = \{k_1, k_2\}$, and introduce the following designations for tensorial convolutions:

$$\hat{c}\vec{k}^2 = c_{3i3j}k_i k_j = c_{55}k_1^2 + c_{44}k_2^2, \qquad \text{(S.18a)}$$

$$\hat{v}\vec{k}^4 = v_{3ij3lm}k_i k_j k_l k_m = v_{5151}k_1^4 + v_{4242}k_2^4, \qquad \text{(S.18b)}$$

$$\hat{f}\vec{k}^2 = f_{3i3j}k_i k_j = f_{55}k_1^2 + f_{44}k_2^2, \qquad \text{(S.18c)}$$

$$\hat{g}\vec{k}^2 = g_{3i3j}k_i k_j = g_{55}k_1^2 + g_{44}k_2^2, \qquad \text{(S.18d)}$$

$$\hat{q}\vec{k} = q_{3i33}k_i = q_{53}k_1, \qquad \hat{z}\vec{k} = z_{i33}k_i = z_{533}k_1, \qquad \text{(S.18e)}$$

$$k^2 = k_1^2 + k_2^2. \tag{S.18f}$$

The temperature-dependent function $\alpha_S$, introduced in Eq. (S.18b), depends on the constant spontaneous polarization:

$$\alpha_S \approx \alpha + 3\beta^*|P_s|^2 + 5\gamma^*|P_s|^4 + 7\delta|P_s|^6. \tag{S.19}$$

Using these approximations, the eigen spectrum $\omega(\vec{k})$, corresponding to the bulk-like phonons with LGD parameters renormalized by mismatch strains, can be found from zero determinant:

$$\det\begin{pmatrix} \hat{v}\vec{k}^4 + \hat{c}\vec{k}^2 - i\Lambda\omega - \rho\omega^2 & \hat{f}\vec{k}^2 - M\omega^2 + 2iP_s\left(\hat{q}\vec{k} + 2P_s^2\hat{z}\vec{k}\right) \\ \hat{f}\vec{k}^2 - M\omega^2 - 2iP_s\left(\hat{q}\vec{k} + 2P_s^2\hat{z}\vec{k}\right) & -i\Gamma\omega - \mu\omega^2 + \frac{k_3^2 - \frac{\chi\varepsilon_b}{c^2}\omega^2}{\varepsilon_0\varepsilon_b\left(k^2 - \frac{\chi\varepsilon_b}{c^2}\omega^2\right)} + \alpha_S + \hat{g}\vec{k}^2 \end{pmatrix} = 0. \tag{S.20}$$

## SUPPLEMENT S4. Excitations of polar and elastic orders in strained films

Let us underline that the determinant (S.20) does not account for the "static" elastic boundary conditions (5c), namely $u_{11}|_{x_3=-h/2} = u_m$, $u_{22}|_{x_3=-h/2} = u_m$, $U_i|_{x_3=-h/2} = 0$, at the film/substrate interface, ($x_3 = -h/2$), in the explicit way. However, the bulk-like phonons with LGD parameters renormalized by mismatch strains can be used as a reasonable approximation in the main part of the film excluding the vicinity of the film-substrate interface due to the following reasons.

The rigorous boundary conditions for elastic displacement at the top ($x_3 = h/2$) and bottom ($x_3 = -h/2$) film surfaces of the $CuInP_2S_6$ film can be derived by taking the variation of the Lagrangian $L$ with respect to $U_3$ and $u_{3j}$ (similarly as described in the main text for the variation $\delta P_3$ unconstrained at the film surfaces). Performing integration by parts on the linear and higher strain-gradient terms isolates the surface contribution, i.e., $\delta L|_{x_3=h/2} = \int dt \int dS \left[\frac{\partial \delta L}{\partial U_3}\delta U_3 + \frac{\partial \delta L}{\partial u_{3j}}\delta u_{3j}\right]$. Because these variations are unconstrained (i.e., can take arbitrary values) at the top film surfaces ($x_3 = h/2$), the terms in the square bracket need to be zero for letting $\delta L|_{x_3=h/2} = 0$, yielding mixed-type elastic boundary conditions [75]:

$$\sigma_{3i} = \left(c_{3ijk}u_{kl} + f_{i333}\frac{\partial P_3}{\partial x_3} - q_{i3}P_3^2 - z_{i3}P_3^4 - v_{i333}\frac{\partial^2 u_{33}}{\partial x_3^2}\right)\Big|_{x_3=+h/2} = 0, \tag{S.21a}$$

$$\left(v_{i33lmn}\frac{\partial u_{lm}}{\partial x_n} - \frac{f_{i333}}{2}P_3\right)\Big|_{x_3=-h/2} = 0, \tag{S.21b}$$

where the subscript $i = 1, 2, 3$. At the bottom surface ($x_3 = -h/2$) the variation of $U_3$ is absent, but it does not mean the absence of variation $u_{3j}$, which yields the conventional and additional boundary conditions:

$$U_i|_{x_3=-h/2} = 0, \quad u_{11} = u_{22}|_{x_3=-h/2} = u_m, \quad \left(v_{i33lmn}\frac{\partial u_{lm}}{\partial x_n} - \frac{f_{i333}}{2}P_3\right)\Big|_{x_3=-h/2} = 0. \tag{S.21c}$$

Here the subscript $i = 1, 2, 3$. Linearization of the boundary conditions (S.21) with respect to the purely transverse polarization fluctuations, $P_3 = P_s + \delta P_3(x_1, x_2)$, yields

$$\delta\sigma_{3i}|_{x_3=h/2}=0,\qquad \delta U_i|_{x_3=-h/2}=0,\quad \delta u_{11}=\delta u_{22}|_{x_3=-h/2}=0, \tag{S.22a}$$

$$v_{i33lmn}\frac{\partial}{\partial x_n}\delta u_{lm}-\frac{f_{i333}}{2}\delta P_3\Big|_{x_3=\pm h/2}=0, \tag{S.22b}$$

For very small higher-order strain gradient coefficients $v_{ijklmn}$, that is a common approximation for most ferroics [ 76 ], "additional" elastic boundary conditions (S.22b) can be omitted for the determination of the elastic waves' properties, which are unperturbed by the polarization variation and flexoelectric coupling. In this Supplement we use the "decoupling" approximation and neglect retardation. Also, below we neglect the flexo-electric coupling with longitudinal polarization, i.e., we regard that $\alpha_{11}P_1\approx E_1$ and $\alpha_{22}P_2\approx E_2$, where $\alpha_{11}$ and $\alpha_{22}$ are inverse dielectric susceptibilities.

Next, we introduce the integral representations

$$\delta U_i(\vec{x},t)=\int d^2\vec{k}\,exp\left(i\vec{k}\vec{x}-i\omega t\right)\widetilde{U}_i\left(x_3,\vec{k},\omega\right), \tag{S.23a}$$

$$\delta P_i(\vec{x},t)=\int d^2\vec{k}\,exp\left(i\vec{k}\vec{x}-i\omega t\right)\tilde{P}_i\left(x_3,\vec{k},\omega\right), \tag{S.23b}$$

$$\delta E_i=\int d^2\vec{k}\,exp\left(i\vec{k}\vec{x}-i\omega t\right)\tilde{E}_i\left(x_3,\vec{k},\omega\right), \tag{S.23c}$$

$$\delta N_i=\int d^2\vec{k}\,exp\left(i\vec{k}\vec{x}-i\omega t\right)\widetilde{N}_i\left(x_3,\vec{k},\omega\right), \tag{S.23d}$$

where $\vec{k}=(k_1,k_2,0)$. Using the integral representations (S.23) we obtained from Eqs. (S.14) and (S.21) the system of linear second-order differential equations with the boundary conditions:

$$(-i\Lambda\omega-\rho\omega^2)\widetilde{U}_i+c_{iljm}k_lk_m\widetilde{U}_j-ik_m\left(c_{imj3}+c_{i3jm}\right)\frac{\partial\widetilde{U}_j}{\partial x_3}-c_{i3j3}\frac{\partial^2\widetilde{U}_j}{\partial x_3\partial x_3}+f_{ijkl}k_lk_j\tilde{P}_j-M\omega^2\tilde{P}_i-$$
$$ik_m\left(f_{imj3}+f_{i3jm}\right)\frac{\partial\tilde{P}_j}{\partial x_3}-f_{i3j3}\frac{\partial^2\tilde{P}_j}{\partial x_3\partial x_3}+2iP_s\left(q_{im3j}k_m+2P_s^2z_{im333j}k_m\right)\tilde{P}_j+2P_sq_{i33j}\frac{\partial P_j}{\partial x_3}+$$
$$4P_s^3z_{i3333j}\frac{\partial P_j}{\partial x_3}=\widetilde{N}_i \tag{S.24a}$$

$$\widetilde{U}_i\Big|_{x_3=-h/2}=0, \tag{S.24b}$$

$$ic_{i3jm}k_m\widetilde{U}_j+c_{i3j3}\frac{\partial\widetilde{U}_j}{\partial x_3}+if_{i33m}k_m\tilde{P}_3+f_{i333}\frac{\partial\tilde{P}_3}{\partial x_3}-(2q_{i333}P_s+4z_{i33333}P_s^3)\tilde{P}_3\Big|_{x_3=h/2}=0, \tag{S.24c}$$

$$\left(\alpha_S-i\Gamma\omega-\mu\omega^2+g_{3l3j}k_lk_j\right)\tilde{P}_3-ik_l(g_{333l}+g_{3l33})\frac{\partial\tilde{P}_3}{\partial x_3}-g_{3333}\frac{\partial^2\tilde{P}_3}{\partial x_3\,\partial x_3}+\left(f_{jmin}k_mk_n-\right.$$
$$\left.\delta_{ij}M\omega^2\right)\widetilde{U}_j-ik_l(f_{kl33}+f_{k33l})\frac{\partial\widetilde{U}_k}{\partial x_3}-f_{k333}\frac{\partial\widetilde{U}_k}{\partial x_3\,\partial x_3}-2iP_s\left(q_{ij33}k_j+2P_s^2z_{i33}k_i\right)\widetilde{U}_i=\tilde{E}_i \tag{S.24c}$$

$$\left(g_{3333}\frac{\partial\tilde{P}_3}{\partial x_3}+\frac{f_{3333}}{2}\frac{\partial\widetilde{U}_3}{\partial x_3}\right)\Big|_{x_3=\pm h/2}=0, \tag{S.24b}$$

$$\frac{1}{\varepsilon_0\varepsilon_b}\left(ik_i\,\tilde{P}_i+\frac{\partial P_3}{\partial x_3}\right)+ik_i\,\tilde{E}_i+\frac{\partial\tilde{E}_3}{\partial x_3}=0, \tag{S.24f}$$

$$\tilde{E}_3\Big|_{x_3=\pm h/2}=0, \tag{S.24g}$$

Since all $x_3$-dependent terms of the Fourier series would produce polarization fluctuations, which are strongly suppressed by the depolarization field, in what follows we limit our consideration by the $x_3$-independent $\tilde{P}_3\left(\vec{k},\omega\right)$ considering that the approximation is valid practically everywhere except in regions near film surfaces. In the case of negligibly small $\frac{\partial \tilde{P}_j}{\partial x_3}$ and $\tilde{N}_i = 0$ Eqs. (S.24a)-(S.24c) can be simplified as:

$$(-i\Lambda\omega - \rho\omega^2)\tilde{U}_i + c_{iljm}k_l k_m \tilde{U}_j - ik_m\left(c_{imj3} + c_{i3jm}\right)\frac{\partial \tilde{U}_j}{\partial x_3} - c_{i3j3}\frac{\partial^2 \tilde{U}_j}{\partial x_3 \partial x_3} = M\omega^2 \tilde{P}_i - f_{ijkl}k_l k_j \tilde{P}_j - 2iP_s\left(q_{im3j}k_m + 2P_s^2 z_{im333j}k_m\right)\tilde{P}_j \quad \text{(S.25a)}$$

$$\tilde{U}_i\Big|_{x_3=-h/2} = 0, \quad \text{(S.25b)}$$

$$ic_{i3jm}k_m\tilde{U}_j + c_{i3j3}\frac{\partial \tilde{U}_j}{\partial x_3} = -if_{i33m}k_m\tilde{P}_3 + (2q_{i333}P_s + 4z_{i33333}P_s^3)\tilde{P}_3\Big|_{x_3=h/2}, \quad \text{(S.25c)}$$

The fully decoupled solution of Eqs. (S.25) is the mixture of well-known Rayleigh elastic waves, which propagate near the freestanding surface of the film, and the bulk-like elastic waves, which propagate everywhere except thin regions near film surfaces. Also, transverse excitations are strongly suppressed near the clamped surface of the film.

Using decoupling approximation, the elastic sub-problem could be reduced to well-known dynamic equation for elastic waves:

$$\rho\frac{\partial^2 U_i}{\partial t^2} - c_{ijkl}\frac{\partial^2 U_k}{\partial x_j \partial x_l} = 0. \quad \text{(S.26)}$$

The ansatz for Rayleigh-like or Lamb-type waves is

$$\vec{U} = \{U_1(t,x_1,x_3), 0, U_3(t,x_1,x_3)\}. \quad \text{(S.27)}$$

In this case the strain components could be written as

$$u_{11} = \frac{\partial U_1}{\partial x_1}, \quad u_{13} = \frac{1}{2}\left(\frac{\partial U_1}{\partial x_3} + \frac{\partial U_3}{\partial x_1}\right), \quad u_{33} = \frac{\partial U_3}{\partial x_3}. \quad \text{(S.28)}$$

Non-zero stress components are

$$\sigma_{11} = c_{11}u_{11} + c_{12}u_{33}, \quad \sigma_{13} = 2c_{44}u_{13}, \quad \sigma_{33} = c_{12}u_{11} + c_{11}u_{33}, \quad \text{(S.29)}$$

since $c_{44} = c_{55}$ for the cubic symmetry. Hereinafter we suggest that the elastic stiffness tensor $c_{44} \approx c_{55}$, that is a reasonable approximation even for $CuInP_2S_6$ with 2/m symmetry (see **Table S1** in **Supplement S5**). Using Eqs.(S.26) and (S.27), the dynamic equations for $\vec{U}$ could be written as f

$$\rho\frac{\partial^2 U_1}{\partial t^2} - c_{11}\frac{\partial^2 U_1}{\partial x_1^2} - c_{44}\frac{\partial^2 U_1}{\partial x_3^2} - (c_{12} + c_{44})\frac{\partial^2 U_3}{\partial x_1 \partial x_3} = 0, \quad \text{(S.30a)}$$

$$\rho\frac{\partial^2 U_3}{\partial t^2} - c_{11}\frac{\partial^2 U_3}{\partial x_3^2} - c_{44}\frac{\partial^2 U_3}{\partial x_1^2} - (c_{12} + c_{44})\frac{\partial^2 U_1}{\partial x_1 \partial x_3} = 0. \quad \text{(S.30b)}$$

Let us look for the solution of Eqs. (S.30) in the form of the waves traveling along the surface of the film/layer:

$$U_1(t,x_1,x_3) = \tilde{U}_1(x_3)\exp(i\{k_1x_1 - \omega t\}), \quad U_3(t,x_1,x_3) = \tilde{U}_3(x_3)\exp(i\{k_1x_1 - \omega t\}) \quad \text{(S.31)}$$

Equations for $\tilde{U}_{1,3}$ could be obtained from Eqs. (S.30) in the following form:

$$[-\omega^2\rho + k_1^2 c_{11}]\tilde{U}_1 - c_{44}\frac{\partial^2 \tilde{U}_1}{\partial x_3^2} - ik_1(c_{12} + c_{44})\frac{\partial \tilde{U}_3}{\partial x_3} = 0, \quad \text{(S.32a)}$$

$$[-\omega^2\rho + k_1^2 c_{44}]\tilde{U}_3 - c_{11}\frac{\partial^2 \tilde{U}_3}{\partial x_3^2} - ik_1(c_{12} + c_{44})\frac{\partial \tilde{U}_1}{\partial x_3} = 0. \quad \text{(S.32b)}$$

Let us look for the partial solutions of Eqs. (S.32) in the following form

$$\widetilde{U}_{1,3}(x_3) = v_{1,3}\exp(\lambda\, x_3) \tag{S.33}$$

Where $\lambda$ is the characteristic decay factor, and $v_{1,3}$ are relative amplitudes of the corresponding displacements. Substitution of the solution (S.33) into Eqs. (S.32) yields the system for the unknown amplitudes $v_{1,3}$:

$$\begin{cases}(-\omega^2\rho + k_1^2 c_{11} - c_{44}\lambda^2)v_1 - ik_1\lambda(c_{12}+c_{44})v_3 = 0\\(-\omega^2\rho + k_1^2 c_{44} - c_{11}\lambda^2)v_3 - ik_1\lambda(c_{12}+c_{44})v_1 = 0\end{cases} \tag{S.34}$$

The nontrivial solution of Eqs. (S.34) is possible when the determinant of the stiffness matrix $\hat{C}$ of the system (S.34) is equal to zero:

$$\hat{C} = \begin{pmatrix} k_1^2 c_{11} - c_{44}\lambda^2 - \omega^2\rho & -ik_1\lambda(c_{12}+c_{44}) \\ -ik_1\lambda(c_{12}+c_{44}) & k_1^2 c_{44} - c_{11}\lambda^2 - \omega^2\rho \end{pmatrix} \tag{S.35a}$$

$$det[\![\hat{C}]\!] = (-\omega^2\rho + k_1^2 c_{11} - c_{44}\lambda^2)(-\omega^2\rho + k_1^2 c_{44} - c_{11}\lambda^2) + (c_{12}+c_{44})k_1^2\lambda^2 = 0 \tag{S.35b}$$

In the case of isotropic materials (i.e. when $c_{12}+c_{44} \equiv c_{11}-c_{44}$) one could the solution of Eqs.(S.35b) is

$$\lambda_1^2 = k_1^2 - \frac{\omega^2\rho}{c_{11}} \equiv q^2 \tag{S.36a}$$

$$\lambda_2^2 = k_1^2 - \frac{\omega^2\rho}{c_{44}} \equiv s^2 \tag{S.36b}$$

Note that the dispersion relations $c_{44}k_1^2 = \omega^2\rho$ ($s = 0$ and $U_3 \neq 0$ only) and $c_{11}k_1^2 = \omega^2\rho$ ($q = 0$ and $U_1 \neq 0$ only) correspond to the shear and compression bulk waves, respectively. Substitution of Eqs.(S.36) to the matrix (S.35a) leads to stiffness matrices

$$\hat{C}_1 = \begin{pmatrix} -q^2 & ik_1 q \\ ik_1 q & k_1^2 \end{pmatrix}(c_{44}-c_{11})^2, \tag{S.37a}$$

$$\hat{C}_2 = \begin{pmatrix} k_1^2 & -ik_1 s \\ -ik_1 s & -s^2 \end{pmatrix}(c_{11}-c_{44})^2. \tag{S.37b}$$

The eigenvectors of the matrices (S.37) determine the structure of the general solution:

$$\widetilde{U}_1(x_3) = i\,k_1\left[a\cosh\left(q\left(x_3+\frac{h}{2}\right)\right) + b\sinh\left(q\left(x_3+\frac{h}{2}\right)\right)\right] - s\left[c\sinh\left(s\left(x_3+\frac{h}{2}\right)\right) + d\cosh\left(s\left(x_3+\frac{h}{2}\right)\right)\right], \tag{S.38a}$$

$$\widetilde{U}_3(x_3) = q\left[\left[a\sinh\left(q\left(x_3+\frac{h}{2}\right)\right) + b\cosh\left(q\left(x_3+\frac{h}{2}\right)\right)\right]\right] + i\,k_1\left[c\cosh\left(s\left(x_3+\frac{h}{2}\right)\right) + d\sinh\left(s\left(x_3+\frac{h}{2}\right)\right)\right]. \tag{S.38b}$$

Here the constants $a$, $b$, $c$ and $d$ should be determined from the boundary conditions. Since the film is clamped at $x_3 = -h/2$ and elastically free at the $x_3 = h/2$ (where $h$ is the film thickness) boundary conditions could written as

$$U_1|_{x_3=-h/2} = 0, \qquad U_3|_{x_3=-h/2} = 0, \qquad \sigma_{13}|_{x_3=h/2} = 0, \qquad \sigma_{33}|_{x_3=h/2} = 0. \tag{S.39}$$

Substitution of the general solution (S.38) into conditions (S.39) gives the following linear system of equations for the constants $a$, $b$, $c$ and $d$:

$$\begin{cases} i\,a\,k_1 - d\,s = 0, \\ i\,c\,k_1 + b\,q = 0, \\ \mathrm{c}_{44}\left(2k_1 q\left(a\sinh(h\,q) + b\cosh(h\,q)\right) + i(k_1^2 + s^2)\left(c\cosh(hs) + d\sinh(hs)\right)\right) = 0, \\ [\mathrm{c}_{11}(q^2 - k_1^2) + 2\mathrm{c}_{44}k_1^2]\left(a\cosh(hq) + b\sinh(hq)\right) + 2i\mathrm{c}_{44}k_1 s\left(c\sinh(hs) + d\cosh(hs)\right) = 0. \end{cases}$$

(S.40a)

The equivalent matrix form of Eq.(.40a) is

$$\widehat{M}\begin{pmatrix} a \\ b \\ c \\ d \end{pmatrix} = \begin{pmatrix} 0 \\ 0 \\ 0 \\ 0 \end{pmatrix}, \qquad \text{(S.40b)}$$

where the matrix $\widehat{M}$ is introduced as

$$\widehat{M} = \begin{pmatrix} i\,k_1 & 0 & 0 & -s \\ 0 & q & i\,k_1 & 0 \\ 2k_1 q\sinh(hq) & 2k_1 q\cosh(hq) & i(k_1^2 + s^2)\cosh(hs) & i(k_1^2 + s^2)\sinh(hs) \\ [\mathrm{c}_{11}(q^2 - k_1^2) + 2\mathrm{c}_{44}k_1^2]\cosh(hq) & [\mathrm{c}_{11}(q^2 - k_1^2) + 2\mathrm{c}_{44}k_1^2]\sinh(hq) & 2i\mathrm{c}_{44}k_1 s\sinh(hs) & 2i\mathrm{c}_{44}k_1 s\cosh(hs) \end{pmatrix}$$

(S.40c)

The non-trivial solution of the system (S.40) is possible when its determinant is zero

$$\begin{aligned} \det\left\|\widehat{M}\right\| = -i\big[&q\,s\left(\mathrm{c}_{11}(k_1^2 - q^2)(k_1^2 + s^2) - 2c_{44}k_1^2(3k_1^2 + s^2)\right)\cosh(hq)\cosh(hs) \\ &- 2qsk_1^2\left(\mathrm{c}_{11}(k_1^2 - q^2) - c_{44}(3k_1^2 + s^2)\right) \\ &- k_1^2\{\mathrm{c}_{11}(k_1^2 - q^2)(k_1^2 + s^2) - 2c_{44}(k_1^4 + (k_1^2 + 2q^2)s^2)\}\sinh(hs)\sinh(hq)\big] = 0 \end{aligned}$$

(S.40d)

Equation (S.40d) has at least two independent roots $q = 0$ and $s = 0$, corresponding to the bulk compression and shear waves, respectively. For any of these cases the general solution (S.38) should be replaced with the combination of constant and linear partial solutions, i.e., one has for the case $s = 0$:

$$\widetilde{U}_1(x_3) = i\,k_1[a\cosh(qx_3) + b\sinh(qx_3)] - d`, \qquad \text{(S.41a)}$$

$$\widetilde{U}_3(x_3) = q[a\sinh(qx_3) + b\cosh(qx_3)] + i\,k_1[c` + d`x_3]. \qquad \text{(S.41b)}$$

However, the substitution of solution (S.41) into the boundary conditions (S.39) gives the system of equations only with the trivial solution for the unknown coefficients $a$, $b$, $c`$ and $d`$, which means that the bulk-like shear waves are not possible. Instead, we could look for the solutions with $s \approx 0$. For the sufficiently large $h$ (see below), thus one could simplify (S.40d) assuming that $q \sim k_1 \gg |s|$ and $qh \gg 1$:

$$\det\left\|\widehat{M}\right\| \approx i\omega^2\rho k_1^2\cosh(hq)\,[-5qs\cdot\cosh(hs) + k_1^2\sinh(hs)] = 0 \qquad \text{(S.42a)}$$

It could be checked that Eq.(S.24a) has no roots at real $s$, therefore it is convenient to introduce a new parameter $\tilde{s} = \sqrt{\frac{\omega^2\rho}{c_{66}} - k_1^2}$, so that $s = i\,\tilde{s}$ and rewrite (S.42a) in the following way

$$\det\|\widehat{M}\| \approx \omega^2\rho k_1^2 \cosh(hq)\,[5q\,\tilde{s}\cos(h\tilde{s}) - k_1^2\sin(h\tilde{s})] = 0 \quad \text{(S.42b)}$$

To find the solution of Eq.(S42b) close to $\tilde{s} \approx 0$ in a symbolic way, let us consider the following Pade-approximations

$$\cos(x) \approx \frac{1-\zeta\,x^2}{1+\phi\,x^2}, \quad \zeta = \frac{4}{\pi^2}, \quad \phi = \frac{16-4\pi}{\pi^3}, \quad \text{(S.43a)}$$

$$\sin(x) \approx x\frac{1-\phi\,x^2}{1+\xi\,x^2}, \quad \xi = \frac{4\pi-12}{\pi^2}. \quad \text{(S.43b)}$$

Pade-approximations (S.43) are valid at the range $x \in [0, \pi/2]$, their substitution into (S.42b) gives

$$5\sqrt{k_1^2 - \frac{\omega^2\rho}{c_{11}}}\,\frac{1-\zeta\,h^2\left(\frac{\omega^2\rho}{c_{44}}-k_1^2\right)}{1+\phi\,h^2\left(\frac{\omega^2\rho}{c_{44}}-k_1^2\right)} - hk_1^2\frac{1-\phi\,h^2\left(\frac{\omega^2\rho}{c_{44}}-k_1^2\right)}{1+\xi\,h^2\left(\frac{\omega^2\rho}{c_{44}}-k_1^2\right)} = 0 \quad \text{(S.44a)}$$

and finally, looking for the solution close to the bulk transversal waves with $k_1^2 \approx \frac{\omega^2\rho}{c_{44}}$, one could rewrite (S.44a) as

$$\frac{5}{k_1 h}\sqrt{1 - \frac{c_{44}}{c_{11}}}\,\frac{1-\zeta\,h^2\left(\frac{\omega^2\rho}{c_{44}}-k_1^2\right)}{1+\phi\,h^2\left(\frac{\omega^2\rho}{c_{44}}-k_1^2\right)} - \frac{1-\phi\,h^2\left(\frac{\omega^2\rho}{c_{44}}-k_1^2\right)}{1+\xi\,h^2\left(\frac{\omega^2\rho}{c_{44}}-k_1^2\right)} = 0. \quad \text{(S.44b)}$$

Under the assumptions listed above, Eq.(S.42b) reduces to the biquadratic equation with respect to $\omega^2$; and its solution gives us the first correction to the assumption $k_1^2 \approx \frac{\omega^2\rho}{c_{44}}$:

$$\frac{\omega^2\rho}{c_{44}} \approx k_1^2 - \frac{5r_c(\xi-\zeta) - \sqrt{4k_1^2h^2\phi^2 + 5r_c[5r_c(\xi+\zeta)^2 - 4\,k_1h(\phi^2+\xi\,\zeta)]}}{2h^2(k_1h\,\phi^2 - 5r_c\xi\,\zeta)} \approx k_1^2 + \frac{1}{h^2\phi} + \frac{5r_c(\zeta-\phi)(\xi+\phi)}{2h^3k_1\phi^3} \quad \text{(S.45a)}$$

Here we introduced the parameter $r_c = \sqrt{1 - \frac{c_{44}}{c_{11}}}$, which depends only on elastic stiffness constants.

The simplified dispersion relation is valid when $k_1 h \geq 5\pi$:

$$\omega^2 \approx \frac{c_{44}}{\rho}\left[k_1^2 + \frac{1}{h^2\phi} + \frac{5r_c(\zeta-\phi)(\xi+\phi)}{2h^3k_1\phi^3}\right], \quad \text{(S.45b)}$$

where the constants $\phi = \frac{16-4\pi}{\pi^3}$, $\zeta = \frac{4}{\pi^2}$ and $\xi = \frac{4\pi-12}{\pi^2}$ according to expressions (S.43).

Using Eq.(S.45), the dispersion equation (7) in the main text acquires the form:

$$\alpha_s + \hat{g}\vec{k}^2 - i\Gamma\omega - \mu\omega^2 - \frac{\chi_0\chi\varepsilon_0\varepsilon_b\omega^2}{\varepsilon_0\varepsilon_b(k^2 - \chi_0\chi\varepsilon_0\varepsilon_b\omega^2)} - \frac{\left(\hat{f}\vec{k}^2 - M\omega^2\right)^2 + 4P_S^2\left(\hat{q}\vec{k} + 2P_S^2\hat{z}\vec{k}\right)^2}{\hat{v}\vec{k}^4 + \hat{c}\vec{k}^2 + c_{44}\left(\frac{1}{h^2\phi} + \frac{5r_c(\zeta-\phi)(\xi+\phi)}{2h^3k_1\phi^3}\right) - i\Lambda\omega - \rho\omega^2} = 0. \quad \text{(S.46)}$$

Here the correction $c_{44}\left(\frac{1}{h^2\phi} + \frac{5q_c(\zeta-\phi)(\xi+\phi)}{2h^3k_1\phi^3}\right)$ differs the equation from the bulk solution, corresponding to zero determinant of the matrix (S.20).

Performed FEM confirms that bulk-like solution for elastic waves have sufficient accuracy in the range of $x_3$ satisfying the inequality $k_\perp\left(x_3 + \frac{h}{2}\right) \gg \frac{\pi}{2}$ and $k_\perp\left(x_3 - \frac{h}{2}\right) \geq \frac{\pi}{2}$, where $\vec{k}_\perp = (k_1, k_2)$ is the transverse wave vector of phonons and/or ferrons. Consequently, the bulk-like approximation for elastic solutions, considered in **Supplement S3**, is valid in the central part of the film excluding the immediate vicinity of the film-substrate interface for the transverse wavevectors $k_{\perp} \geq 3\frac{\pi}{h}$. Since the peculiarities of phonon and ferron spectra considered in the main part of the work, are within the range $0.1\text{nm}^{-1} \leq k_{\perp} \leq 1\text{nm}^{-1}$, the condition $k_{\perp}h \geq 3\pi$ corresponds to the film thickness range above 95 – 9.5 nm. Surface-like phonon excitations, which are reminiscent of well-known Rayleigh waves, propagate near the freestanding surface of the film, while elastic waves with transverse polarization are strongly suppressed near the clamped surface.

We also would like to underline that the results obtained numerically can be explained by earlier analytical calculations performed for a simpler case (see details in Supplemental Materials to Ref. [75]). Another way is to consider elastic waves propagation in the substrate that we defer to further studies.

## SUPPLEMENT S5. The Strain Impact on the Free Energy Coefficients and Material Parameters

At first, we made the Legendre transformation of the free energy in Eq. (S.14), $\tilde{F} = F + u_{ij}\sigma_{ji}$, where $\sigma_{ij}$ are the components of a stress tensor. Using the equation of state relating elastic stresses and strains, the coefficients $\alpha^*$, $\beta^*$, $\gamma^*$, and $\delta^*$ in Eq. (S.14) depend on the mismatch strain, $u_m$, as derived in Ref. [77]:

$$\frac{\alpha^*}{2} = \frac{\alpha}{2} + \frac{u_m(Q_{23}(s_{12}-s_{11})+Q_{13}(s_{12}-s_{22}))}{s_{11}s_{22}-s_{12}^2} - u_m^2\frac{(s_{12}-s_{22})^2W_{113}+(s_{11}-s_{12})^2W_{223}}{2(s_{12}^2-s_{11}s_{22})^2}, \quad \text{(S.47a)}$$

$$\frac{\beta^*}{4} = \frac{\beta}{4} + \frac{Q_{23}^2s_{11}-2Q_{13}Q_{23}s_{12}+Q_{13}^2s_{22}}{2(s_{11}s_{22}-s_{12}^2)} + u_m\left\{\frac{(s_{22}-s_{12})Z_{133}+(s_{11}-s_{12})Z_{233}}{s_{12}^2-s_{11}s_{22}} + \frac{Q_{23}(s_{12}(s_{12}-s_{22})W_{113}+s_{11}(s_{11}-s_{12})W_{223})+Q_{13}(s_{22}(-s_{12}+s_{22})W_{113}+s_{12}(-s_{11}+s_{12})W_{223})}{(s_{12}^2-s_{11}s_{22})^2}\right\}, \quad \text{(S.47b)}$$

$$\frac{\gamma^*}{6} = \frac{\gamma}{6} + \frac{(Q_{13}s_{22}-Q_{23}s_{12})Z_{133}+(Q_{23}s_{11}-Q_{13}s_{12})Z_{233}}{s_{11}s_{22}-s_{12}^2} - \frac{-2Q_{13}Q_{23}s_{12}(s_{22}W_{113}+s_{11}W_{223})+Q_{23}^2(s_{12}^2W_{113}+s_{11}^2W_{223})+Q_{13}^2(s_{22}^2W_{113}+s_{12}^2W_{223})}{2(s_{12}^2-s_{11}s_{22})^2} + u_m\frac{s_{22}^2W_{113}Z_{133}+s_{11}^2W_{223}Z_{233}-s_{12}(s_{22}W_{113}+s_{11}W_{223})(Z_{133}+Z_{233})+s_{12}^2(W_{223}Z_{133}+W_{113}Z_{233})}{(s_{12}^2-s_{11}s_{22})^2}, \quad \text{(S.47c)}$$

$$\frac{\delta^*}{8} = \frac{\delta}{8} + \frac{s_{22}Z_{133}^2-2s_{12}Z_{133}Z_{233}+s_{11}Z_{233}^2}{2(s_{11}s_{22}-s_{12}^2)} + \frac{Q_{23}(s_{12}(s_{22}W_{113}+s_{11}W_{223})Z_{133}-s_{12}^2W_{113}Z_{233}-s_{11}^2W_{223}Z_{233})}{(s_{12}^2-s_{11}s_{22})^2} + \frac{Q_{13}(-s_{22}^2W_{113}Z_{133}+s_{12}s_{22}W_{113}Z_{233}+s_{12}W_{223}(-s_{12}Z_{133}+s_{11}Z_{233}))}{(s_{12}^2-s_{11}s_{22})^2}. \quad \text{(S.47d)}$$

Hereinafter the values $Q_{i3}$, $Z_{i33}$, and $W_{ij3}$ denote the components of a single linear and two nonlinear electrostriction strain tensors in the Voigt notation, respectively. Since $W_{ijk}$ are small enough, we remain only linear terms in $W_{ijk}$ in Eqs.(S.47) and omit all terms proportional to higher powers of the parameter. The higher powers of $W_{ijk}$ lead to the 12-th powers of polarization in the renormalized free energy density (S.20), making the mathematical complexity of results like Ref. [ 78 ]. These cumbersome and very small higher renormalization terms can be considered numerically.

**Table S1.** LGD parameters for a bulk ferrielectric $CuInP_2S_6$ Helmholtz free energy $F(P_3, u_{ij})$ with $P_3$ and $u_{ij}$ as independent variables

| **Parameter (dimensionality)** | **Value** |
|---|---|
| $\varepsilon_b$ | 9 |
| $\alpha_T$ ($C^{-2}\cdot$m J/K) | $1.64067\times10^7$ |
| $T_{C,q}$ (K) | $T_C \cong 292.67$, $T_q \cong 50$ |
| $\beta$ ($C^{-4}\cdot m^5$J) | $6.08\times10^{12}(1.-0.00189\,T+1.867\times10^{-6}T^2)$ |
| $\gamma$ ($C^{-6}\cdot m^9$J) | $-1.597\times10^{16}(1-0.00149\,T+1.558\times10^{-6}\,T^2)$ |
| $\delta$ ($C^{-8}\cdot m^{13}$J) | $1.076\times10^{19}(1.-0.000588\,T+1.552\times10^{-6}T^2)$ |
| $q_{i3}$ (J $C^{-2}\cdot$m) | $q_{13}=2.076\times10^{11}(1-0.00213\,T)$<br>$q_{23}=1.755\times10^{11}(1-0.00212\,T)$<br>$q_{33}=-1.613\times10^{11}(1-0.00187\,T)$<br>$q_{53}=-6.62\times10^{9}$ |
| $z_{i33}$ ($C^{-4}\cdot m^5$J) | $z_{133}=-2.420\times10^{14}(1-0.000427\,T)$<br>$z_{233}=-1.856\times10^{14}(1-0.000494\,T)$<br>$z_{333}=0.4281\times10^{14}(1-0.007890\,T)$<br>$z_{533}=2.17\times10^{12}$ |
| $c_{ij}$ (Pa) | $c_{11}=9.986\times10^{10}$ , $c_{12}=2.901\times10^{10}$, $c_{13}=-0.86\times10^{9}$, $c_{23}=-1.93\times10^{9}$, $c_{22}=10.17\times10^{10}$, $c_{33}=2.802\times10^{10}$, $c_{44}=6.99\times10^{9}$, $c_{55}=6.71\times10^{9}$, $c_{66}=3.756\times10^{10}$ |
| $g_{3i3j}$ (J $m^3/C^2$) | Estimated parameter, which has an order of $10^{-10}$, e.g., $g \cong (0.3 - 2.0)\times10^{-9}$ |
| $f_{ijkl}$ (V) | $f_{44}=f_{55}=f=0-10$ |
| $v_{3ij3kl}$ | $v_{5511}=v_{4422}=3\;10^{-9}$ |
| $\Gamma$ (s m J/$C^2$) | $<10^{-3}$ |
| $\mu$ ($s^2$ m J/$C^2$) | $8*10^{-14}$ |
| $\rho$ (kg/$m^3$) | 3427 |
| $M$ ($s^2$ J/(m $C^2$)) | $10^{-11}$ |

**Table S2**. LGD parameters for a bulk ferrielectric $CuInP_2S_6$ Hibbs free energy $F(P_3, \sigma_{ij})$ with $P_3$ and $\sigma_{ij}$ as independent variables

| **Designation** | **Units** | **Numerical value** |
|---|---|---|
| $\varepsilon_b$ | Dimensionless | 9 |
| $\alpha_T$ | $C^{-2}\cdot$m J/K | $1.64067\times10^7$ |
| $T_C$ | K | 292.67 |
| $\beta$ | $C^{-4}\cdot m^5$J | $3.148\times10^{12}$ |
| $\gamma$ | $C^{-6}\cdot m^9$J | $-1.0776\times10^{16}$ |
| $\delta$ | $C^{-8}\cdot m^{13}$J | $7.6318\times10^{18}$ |
| $Q_{i3}$ | $C^{-2}\cdot m^4$ | $Q_{13}=1.70136-0.00363\,T$, $Q_{23}=1.13424-0.00242\,T$, $Q_{33}=-5.622+0.0105\,T$, $Q_{53}=-0.986$ |
| $Z_{i33}$ | $C^{-4}\cdot m^8$ | $Z_{133}=-2059.65+0.8\,T$, $Z_{233}=-1211.26+0.45\,T$, $Z_{333}=1381.37-12\,T$, $Z_{533}=323$ |
| $s_{ij}$ | $Pa^{-1}$ | $s_{11}=1.092\times10^{-11}$ , $s_{12}=-0.311\times10^{-11}$ , $s_{13}=+0.0120\times10^{-11}$, $s_{22}=1.074\times10^{-11}$, $s_{23}=+0.0644\times10^{-11}$, $s_{33}=3.574\times10^{-11}$, $s_{44}=14.31\times10^{-11}$, $s_{55}=14.90\times10^{-11}$, $s_{66}=2.662\times10^{-11}$ |

## SUPPLEMENT S6. Additional Figures

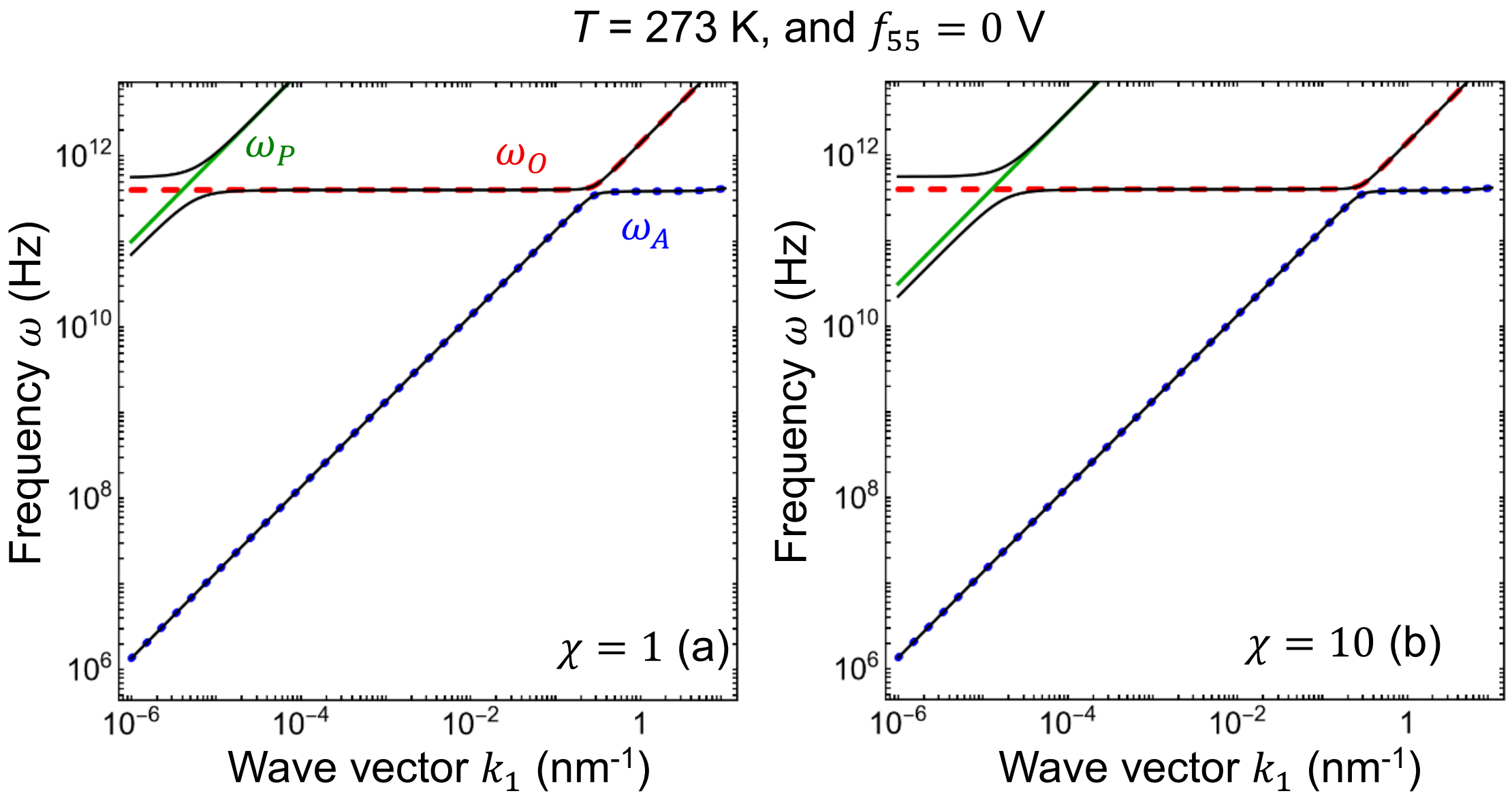

**FIGURE S1.** The dispersion of the different ferron modes (black curves) along with optical, $\omega_O(k_1)$, and acoustic phonon, $\omega_A(k_1)$, modes (see red dashed and blue dotted curves respectively), calculated for $k_2 = 0$, $T$ =273 K, flexoelectric coefficient $f_{55} = 0$ V and the relative magnetic permittivity $\chi = 1$ **(a)** and 10 **(b)**. Green lines show the dispersion of the photons, $\omega_p = c_b k_1$. The damping coefficients are negligibly small for all curves ($\Gamma \to 0$ and $\Lambda \to 0$). The material parameters of $CuInP_2S_6$ are listed in **Tables S1-S2**.

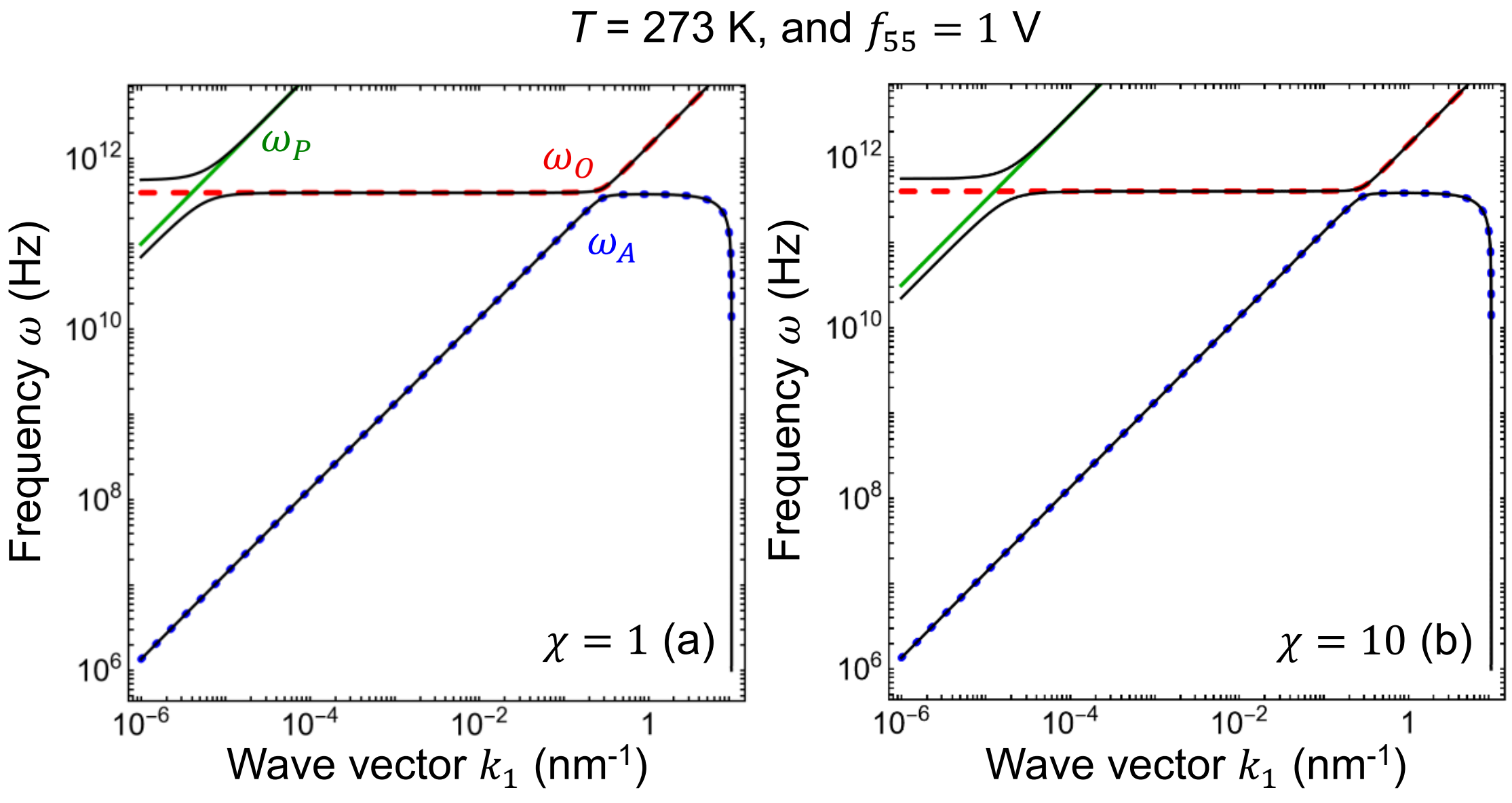

**FIGURE S2.** The dispersion of the ferron modes (black curves) along with optical ($\omega_O(k_1)$, (red dashed curves) and acoustic ($\omega_A(k_1)$, blue dotted curves) phonon modes, calculated for $k_2 = 0$, $T$ =273 K, flexoelectric coefficient $f_{55}$ =1 V and the relative magnetic permittivity $\chi$ = 1 **(a)** and 10 **(b)**. Green lines show the dispersion of the photons, $\omega_p = c_b k_1$. The damping coefficients are negligibly small for all curves ($\Gamma \to 0$ and $\Lambda \to 0$). The material parameters of $CuInP_2S_6$ are listed in **Tables S1-S2**.

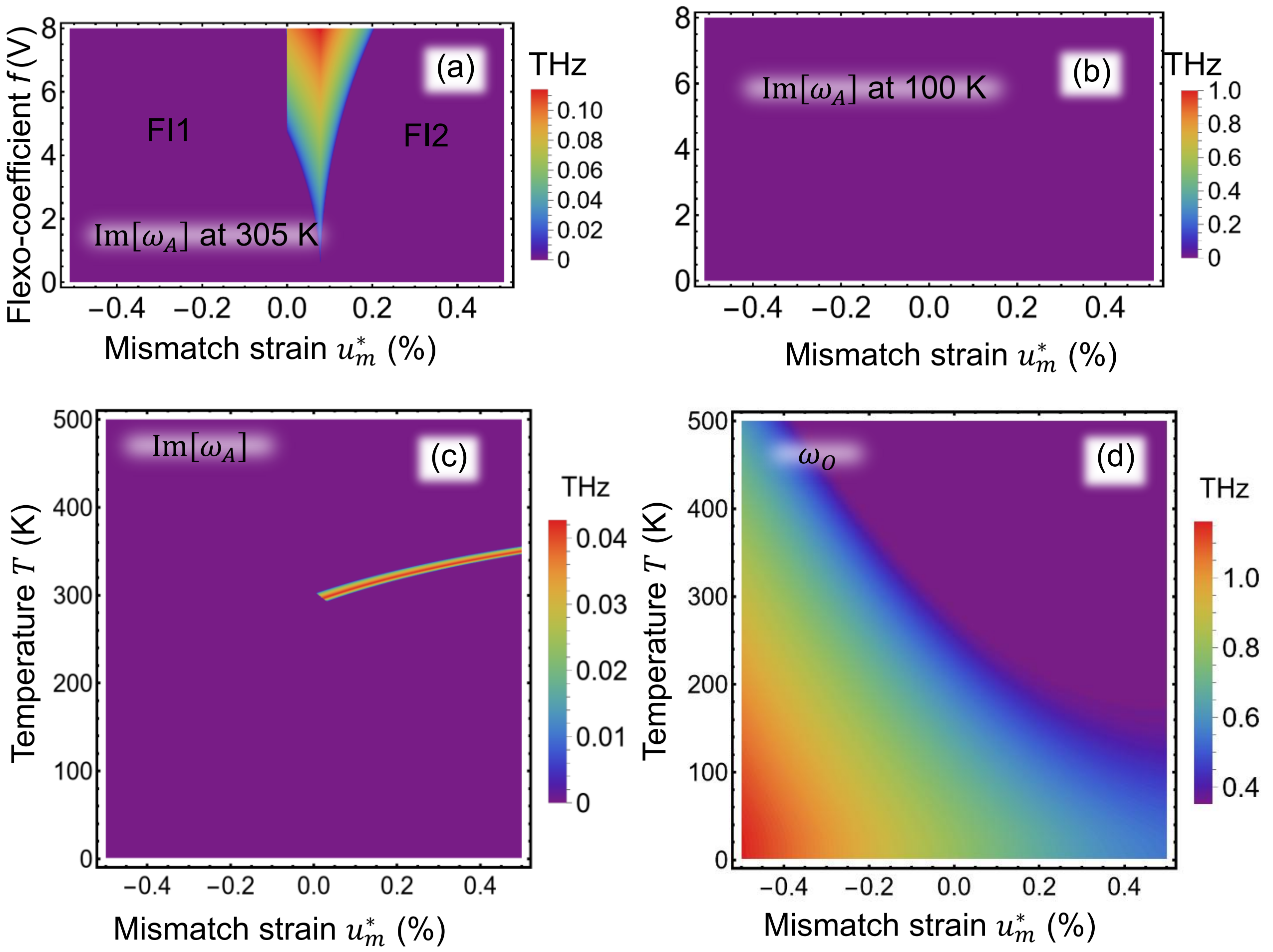

**FIGURE S3.** The imaginary part of the acoustic phonon frequency $\omega_A$ as the function of effective mismatch strain $u_m^*$ and flexoelectric coefficient $f$ calculated for the wavevector components $k_1 = k_2 = 0.25$ nm$^{-1}$. The damping is absent ($\Gamma = 0, \Lambda = 0$), $T = 305$ K **(a)** and $T = 100$ K **(b)**. The dependence of the acoustic phonon frequency imaginary part **(c)** and optical phonon frequency **(d)** on temperature $T$ and effective mismatch strain $u_m^*$ calculated at $k_1 = 0.25$ nm$^{-1}$, $k_2 = 0$, and $f = 4$ V or strained $CuInP_2S_6$ films. Material parameters of $CuInP_2S_6$ are listed in **Tables S1-S2.**

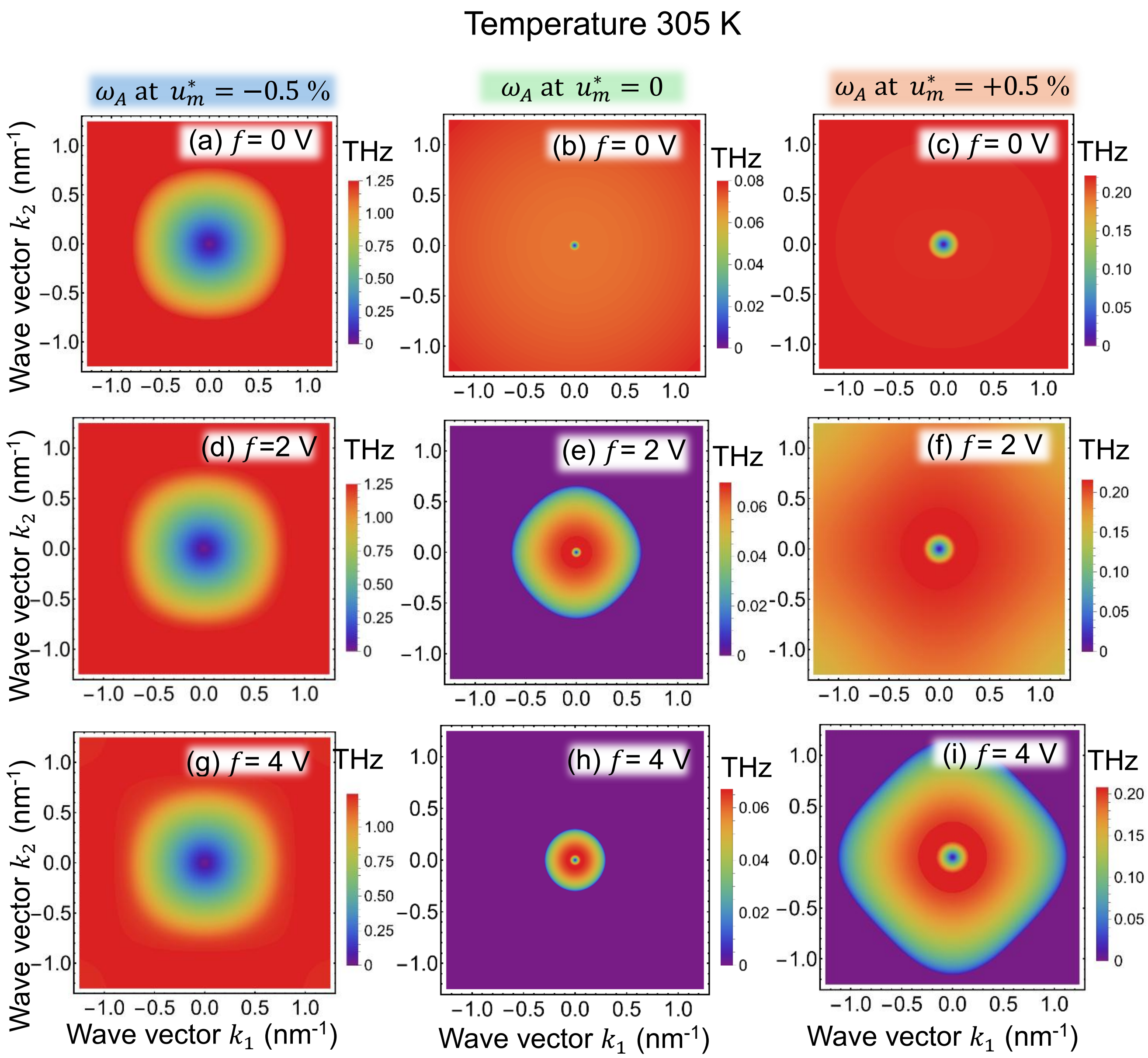

**FIGURE S4.** The real part of the acoustic phonon frequency as the function of the wavevector components $k_1$ and $k_2$, calculated for several values of the flexoelectric coefficient $f$=0 **(a, b, c)**, 2 **(d, e, f)** and 4 V **(g, h, i).** The effective mismatch strain $u_m^*$ is -0.5 % **(a, d, g),** 0 **(b, e, h)** and +0.5 % **(c, f, i)**. The damping is absent ($\Gamma = 0, \Lambda = 0$) and $T = 305$ K for all plots; material parameters of $CuInP_2S_6$ are listed in **Tables S1-S2.**

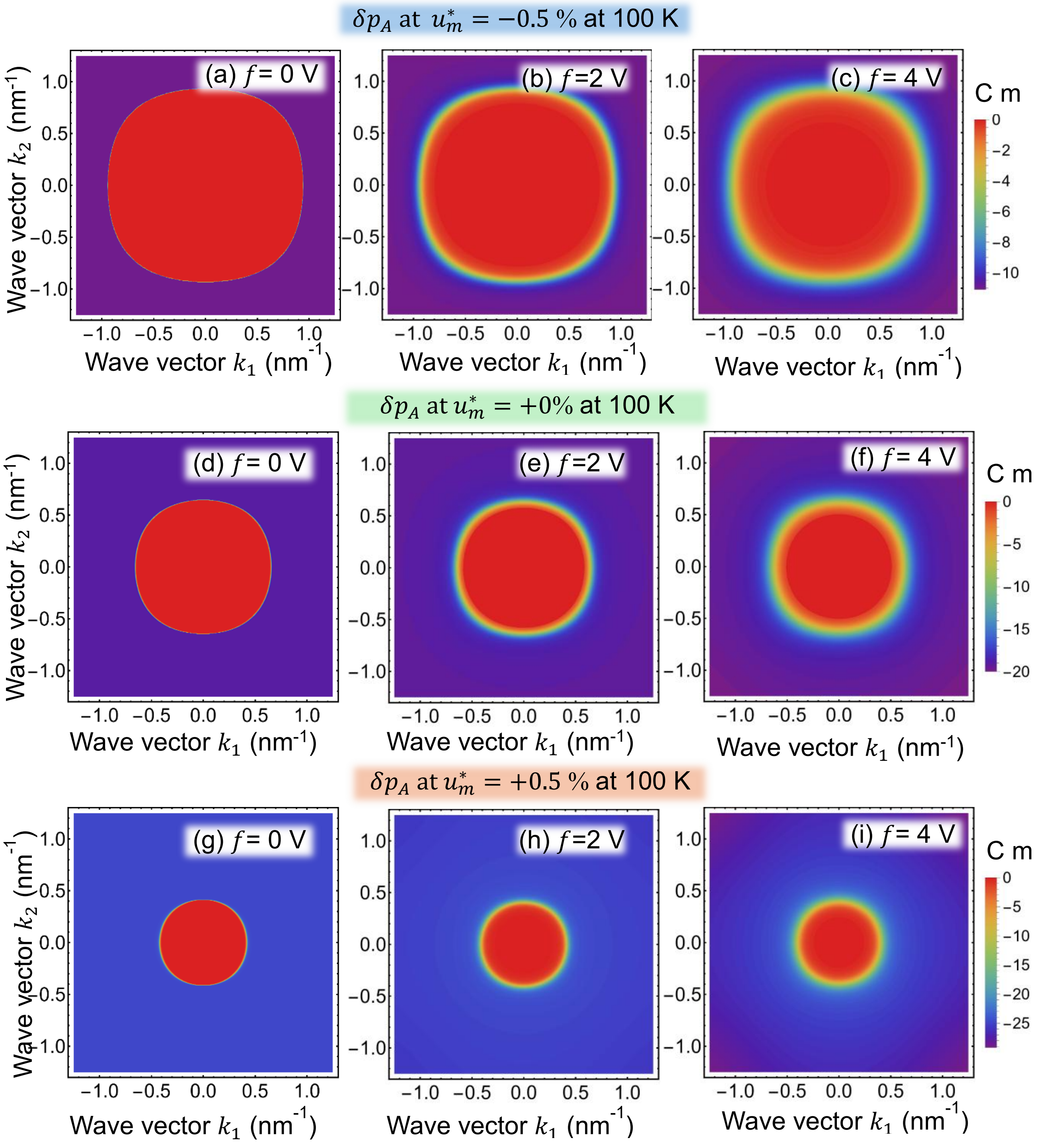

**FIGURE S5.** The spectral density of the acoustic ferrons $\delta p_A$ (in $10^{-32}$ C·m) as the function of the wavevector components $k_1$ and $k_2$, calculated for several values the flexoelectric coefficient $f$=0 **(a, d, j)**, 2 **(b, e, h)** and 4 V **(c, f, i).** The effective mismatch strain $u_m^*$ is -0.5 % **(a, b, c),** 0 % **(d, e, f)** and +0.5 % **(g, h, i)**. The damping is absent ($\Gamma = 0, \Lambda = 0$) and $T = 100$ K for all plots; material parameters of $CuInP_2S_6$ are listed in **Tables S1-S2.**

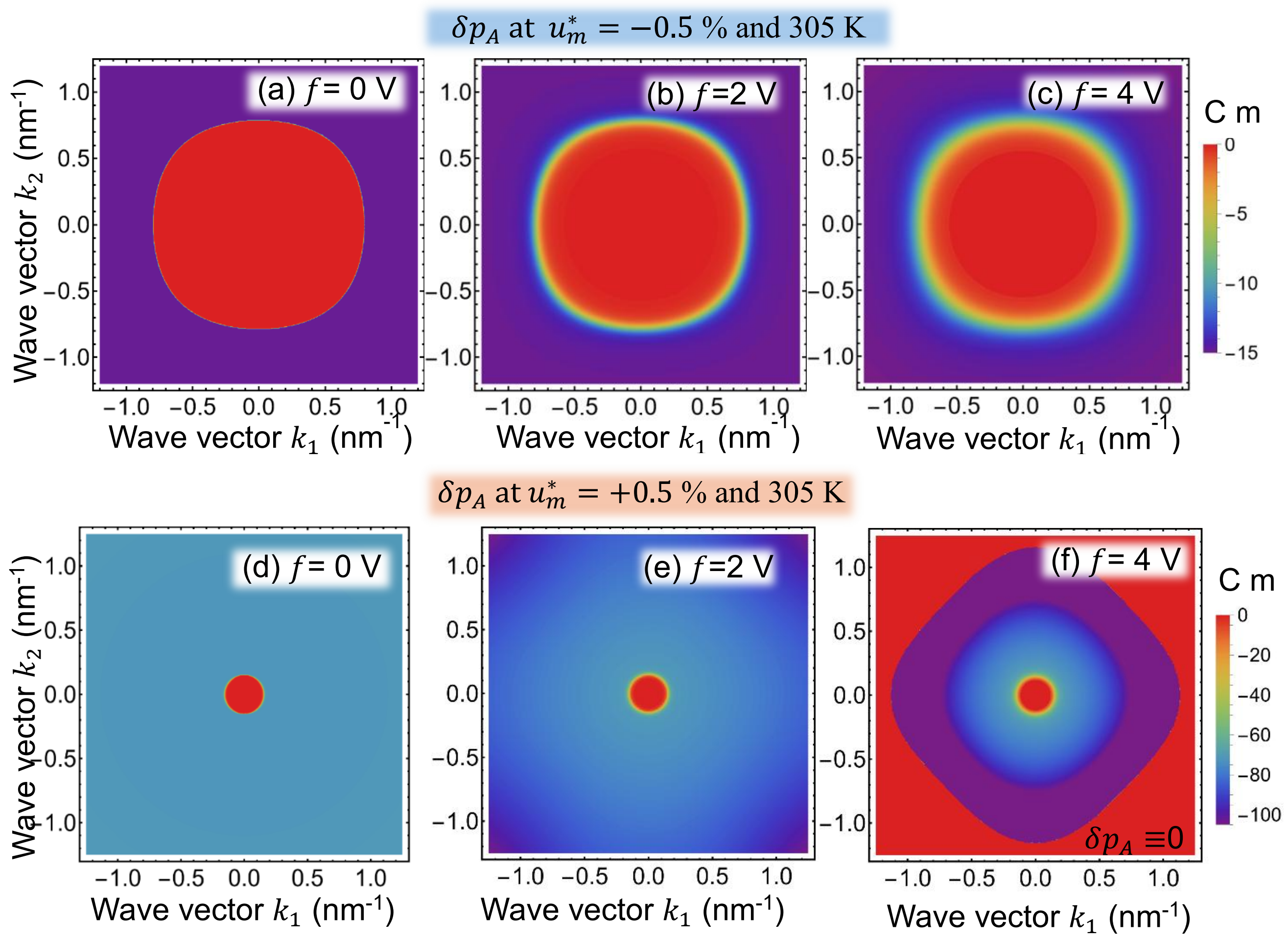

**FIGURE S6.** The spectral density of the acoustic ferrons $\delta p_A$ (in $10^{-32}$ C·m) as the function of the wavevector components $k_1$ and $k_2$, calculated for several values the flexoelectric coefficient $f$=0 **(a, d)**, 2 **(b, e)** and 4 V **(c, f)**. The effective mismatch strain $u_m^*$ is -0.5 % **(a, b, c)** and +0.5 % **(d, e, f)**. The damping is absent ($\Gamma = 0, \Lambda = 0$) and $T = 305$ K for all plots; material parameters of $CuInP_2S_6$ are listed in **Tables S1-S2.**

## SUPPLEMENT S7. Thermodynamically Stable and Metastable States

All results for phonon dispersion, presented in the main text, correspond to the thermodynamically absolutely-stable states, which are the FI1 state with high spontaneous polarization (HP), or the FI2 state with low spontaneous polarization (LP), or the PE phase without spontaneous polarization depending on the temperature $T$, film thickness $h$ and effective mismatch $u_m^*$. There are two regions of phase coexistence associated with first-order phase transitions. The first region, designated as FI1+PE in **Figs. 2(a)-(c),** is relatively large. It corresponds to the coexistence of the PE phase and the FI1 state at temperatures above approximately 300 K and exists for compressive mismatch strains. Corresponding potential relief, shown in **Fig. S7(a)**, continuously changes from the single-well PE phase (at $u_m^* = 0.05$ %) to the three-well FI1 phase (at $u_m^* = -0.15$ %). The second region, located inside a small triangle limited by the BEP, the CEP and the white star (crossing of the PE-FI2 boundary with the FI1+PE coexistence boundary) in **Figs. 2(a)-(c)**, is relatively small. It corresponds to the coexistence of the LP and HP states at temperatures around 300 K for tensile misfit strains.

Corresponding potential relief, shown in **Fig. S7(b)**, continuously changes from the two-well PE phase (at $u_m^* = 0.15$ %) to the four-well mixed phase (at $u_m^* = 0.05$ %), and then to the three-well FI2 phase (at $u_m^* = -0.05$ %).

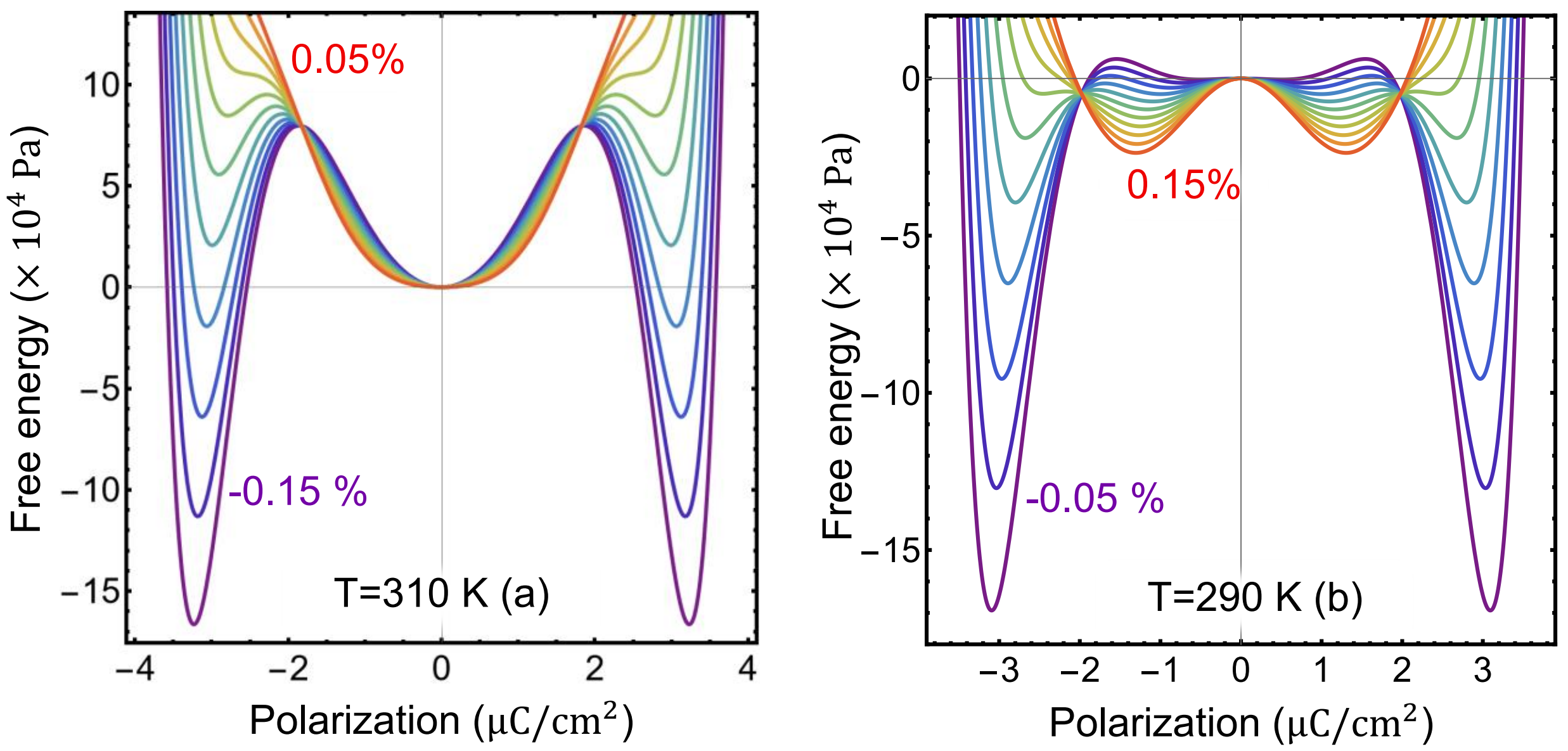

**FIGURE S7.** The free energy as the function of polarization calculated for $T$ =310 K and several values of the effective mismatch strains $u_m^*$, varying from -0.15 % (violet curves) to +0.05 % (red curves) with the step of 0.02 % for neighboring curves **(a)**; T=290 K and several values of the effective mismatch strains $u_m^*$, which vary from -0.05 % (violet curves) to +0.15 % (red curves) with the step of 0.02 % for neighboring curves **(b).** Material parameters of $CuInP_2S_6$ are listed in **Tables S1-S2**.

From mathematical standpoint the phonon dispersion can be calculated in both stable and metastable wells using Eqs. (7)-(8), where the substituted value of $P_s$ corresponds to the HP, LP, or PE state. In result one obtains that the softening of acoustic phonons may occur in both cases of metastable shallow and stable deeper wells. However, this is a formal result related to approximate analytical description; it does not consider the temporal dynamics of the wide fluctuation spectra (e.g., in the form of Langevin-type electric noise, see Eq. (10)).

We performed preliminary simulations to study soft optic and acoustic phonon dispersions and incoherent ferrons dynamics, as small perturbations with the Langevin-type noise spectra near all existing wells, which can include 1, 2, 3 or 4 wells at fixed temperature and mismatch strain. In result we observed the long-living time of perturbations near the deepest potential well(s) and relatively short-living time of the fluctuations near the shallow-well(s). However, we observed also unclassified mixture of quasi-harmonic perturbations in the case when the depths of all four wells are very close (e.g., for $u_m^* = 0.06$ % and $T = 290$ K), and the potential relief transforms into the wide quasi-flat (or flat) well(s).

Thus, the phonon dispersion relations are different in the LP state FI2 and in the HP state FI1 in the case of four wells of significantly different depth, but the system inevitably "falls" into the deepest well with time due to small but finite damping. Exactly this case is already considered in the work. The situation of the almost equal or "quasi-flat" wells is particularly interesting and deserves a separate study. At present we can conclude that predicted softening of acoustic phonons may occur near the deepest well(s), but we cannot make such conclusion for shallow wells, as well as for the degenerated case of quasi-flat wells. These interesting and complicated cases deserve further investigations.